\documentclass [12pt]{article}
\usepackage [cp1251]{inputenc}
\usepackage [english, russian]{babel}
\usepackage {graphicx}
\usepackage {amssymb}
\usepackage {amsmath}
\usepackage {longtable}
\title{Nonlinear dynamo in obliquely rotating stratified electroconductive fluid in an uniformly  magnetic field }
\author{$^{1}$\textbf{M.I. Kopp},$^{1}$\textbf{K.N. Kulik}, $^3$\textbf{A.V. Tur}, $^{1,2}$\textbf{V.V. Yanovsky}}

\begin{document}

\maketitle

$^{1}$ \textit{Institute for Single Crystals, NAS  Ukraine, Nauky Ave. 60, Kharkov 61001, Ukraine}

$^{2}$\textit{V.N. Karazin Kharkiv National University 4 Svobody Sq., Kharkov 61022, Ukraine}

$^{3}$\textit{Universit\'{e} de Toulouse [UPS], CNRS, Institut de Recherche en Astrophysique et Plan\'{e}tologie,
9 avenue du Colonel Roche, BP 44346, 31028 Toulouse Cedex 4, France}

\abstract{We study a new type of large-scale instability, which arises in obliquely rotating stratified electroconductive fluid with an external uniform magnetic field and a small-scale external force having zero helicity. This force gives rise to small-scale oscillations of the velocity with a small Reynolds number. Using the method of multi-scale asymptotic expansions  there are obtained nonlinear equations for vortex and magnetic perturbations in the third order  in Reynolds number. Studied is the linear stage of magneto-vortex dynamo caused by instabilities of  $\alpha$-effect type. Stationary solutions for the equations of nonlinear magneto-vortex dynamo are found by numerical methods in the form of localized chaotic structures.}


\section{Introduction}

As is known, the problems of generation of magnetic fields of planets, stars, galaxies and other cosmic  objects are studied within the framework  of dynamo theory. For the first time, the term \guillemotleft dynamo\guillemotright   in the scope of origin of magnetic fields was proposed by Larmor \cite{1s}. In his opinion, hydrodynamic motion of an electroconductive fluid can  generate a magnetic field by acting as a dynamo.  In the linear theory or kinematic dynamo, where magnetic energy is small  in comparison with the kinetic energy of motion of the medium, magnetic forces hardly influence flow of the medium. By  now   the kinematic theory of dynamo  has been practically built  \cite{2s}-\cite{11s}. In this theory a significant role belongs to rotational motion of cosmic bodies that gives rise to  various  wave (e.g. of Rossby or inertial  waves )  and vortex motions (geostrophic, etc.\cite{12s}-\cite{19s}). In particular, under the influence of the Coriolis force the initial mirror-symmetric turbulence  turns into helical  one characterized by breakdown of  the  mirror symmetry of turbulent fluid motion.  A significant topological characteristic of helical  turbulence is the invariant  $J_{s} =\overline{\vec{v}rot\vec{v}}$, the measure of knottedness of vortex field force lines  \cite{20s}.  In \cite{21s}  it was shown, that  generation of a large-scale field occurs under the action of a turbulent e.m.f. proportional to the mean magnetic field  $\vec{ {\mathcal E}}=\alpha \overline{\vec{H}}$. Coefficient $\alpha$ is proportional to the mean helicity of the velocity field   $\alpha \sim \overline{\vec{v}rot\vec{v}}$  and   has got a definition of  $\alpha $-effect.  The generation properties of helical turbulence were considered  not only in magnetohydrodynamics or in electroconductive media, but also  in conventional hydrodynamics. For the first time the  hypothesis comcerning the ability of  helical turbulence to generate large-scale vortices was  reported in \cite{22s}. It was based on the formal similarity  of the equations of magnetic field induction  $\vec{H}$  and those  for vorticity  $\vec{\omega }=rot\vec{v}$. However, as proved in \cite{22s},  $\alpha $-effect cannot occur in an incompressible turbulent fluid due to the symmetry of the Reynolds tensor of stresses  in the averaged Navier–Stokes equations. Thus, helical turbulence per se  is not sufficient for the onset of hydrodynamic (HD) $\alpha $-effect, other factors are also necessary for symmetry breakdown in  a turbulent flow. As shown in  \cite{24s} and \cite{25s}, such factors are compressibility and temperature gradient in a gravitational field, respectively. The effect of generation of large-scale vortex structures (LSVS) by helical turbulence is called vortex dynamo. The mechanisms of vortex dynamo were worked out with reference to turbulent atmosphere and ocean. In particular, there was  developed the theory of convective vortex dynamo \cite{25s}-\cite{31s}. According to  this theory, helical turbulence gives  rise to large-scale instability leading to the formation of a convective cell interpreted as a huge vortex of tropical cyclone type. Moreover, there are known  many papers devoted to LSVS generation taking into account the effects of rotation  \cite{32s}-\cite{37s}. Just another   $\alpha$-effect is reported in \cite{38s}, where turbulent fluid motion is modelled by means of an external small-scale force $\vec{F}_{0} $. This model  is characterized   by parity violation (at zero helicity: $\vec{F}_{0} rot\vec{F}_{0} =0$). The effect of generation of large-scale disturbances by such a force is called anisotropic kinetic $\alpha$-effect, or AKA-effect \cite{38s}. In the mentioned paper there is considered large-scale instability in an incompressible fluid by means of  the method of asymptotic multi-scale expansions. In this method the Reynolds number $R=\frac{v_{0} t_{0} }{\lambda _{0} } \ll 1$    is used as a  small parameter for small-scale pulsations of the velocity $v_{0} $  caused by the small-scale force.

It is evident, that applicability of kinematic theory of magnetic and vortex dynamos is limited. During rather long period of time amplified  fields  (vortex and magnetic ones) start influencing the flows. In this case the behavior of the magnetic field and the motion of the substance must be considered concordantly, i.e. in the scope of nonlinear theory. The observed magnetic fields of real objects  obviously exist just in nonlinear regime and this testifies to significance of  nonlinear theory  \cite{39s}. In the mentioned  paper the nonlinear theory of magnetic dynamo is  based on generalization of the theory of mean field  (see e.g. \cite{7s} ) taking into account nonlinear effects. However, this theory does not allow us to distinguish strictly from the whole hierarchy of perturbations the principal order at which instability occurs. Therefore, an alternative for construction of a nonlinear dynamo theory is the method of multi-scale asymptotic expansions  \cite{38s}. The use of this method  makes it possible to develop nonlinear theories of vortex dynamo for compressible media  \cite{40s}-\cite{41s}, as well as for convective media with a helical external force  \cite{30s}-\cite{31s}. The asymptotic multi-scale method was used to reveal large-scale instability  in a thermally stratified conductive medium  in the case of helicity  of small-scale  velocity and magnetic fields  \cite{42s}-\cite{43s}. Development of such a large-scale instability in a convective electroconductive medium results in generation of both vortex and magnetic fields. Self-consistent or nonlinear theory of magneto-vortex dynamo in a convective electroconductive medium with small-scale helicity  was built in \cite{43s}. In this work, the possibility of the formation of stationary chaotic large-scale structures in magnetic and vortex fields was shown for the first time. Moreover, there was considered the particular case of the formation of large-scale stationary magnetic structures. These structures were classified as stationary solutions of three types: nonlinear waves, solitons and kinks. Qualitative estimations of the linear stage  \cite{42s} for  solar conditions   made it possible to establish a good agreement of  the characteristic scales and times of  the formed hydrodynamic structures with those of the structures revealed experimentally \cite{44s}.

In the above-mentioned papers helical turbulence was considered  either  to be known, or the problem  of its generation was considered independently \cite{45s}.
\begin{figure}
  \centering
	\includegraphics[height=6 cm, width=6 cm]{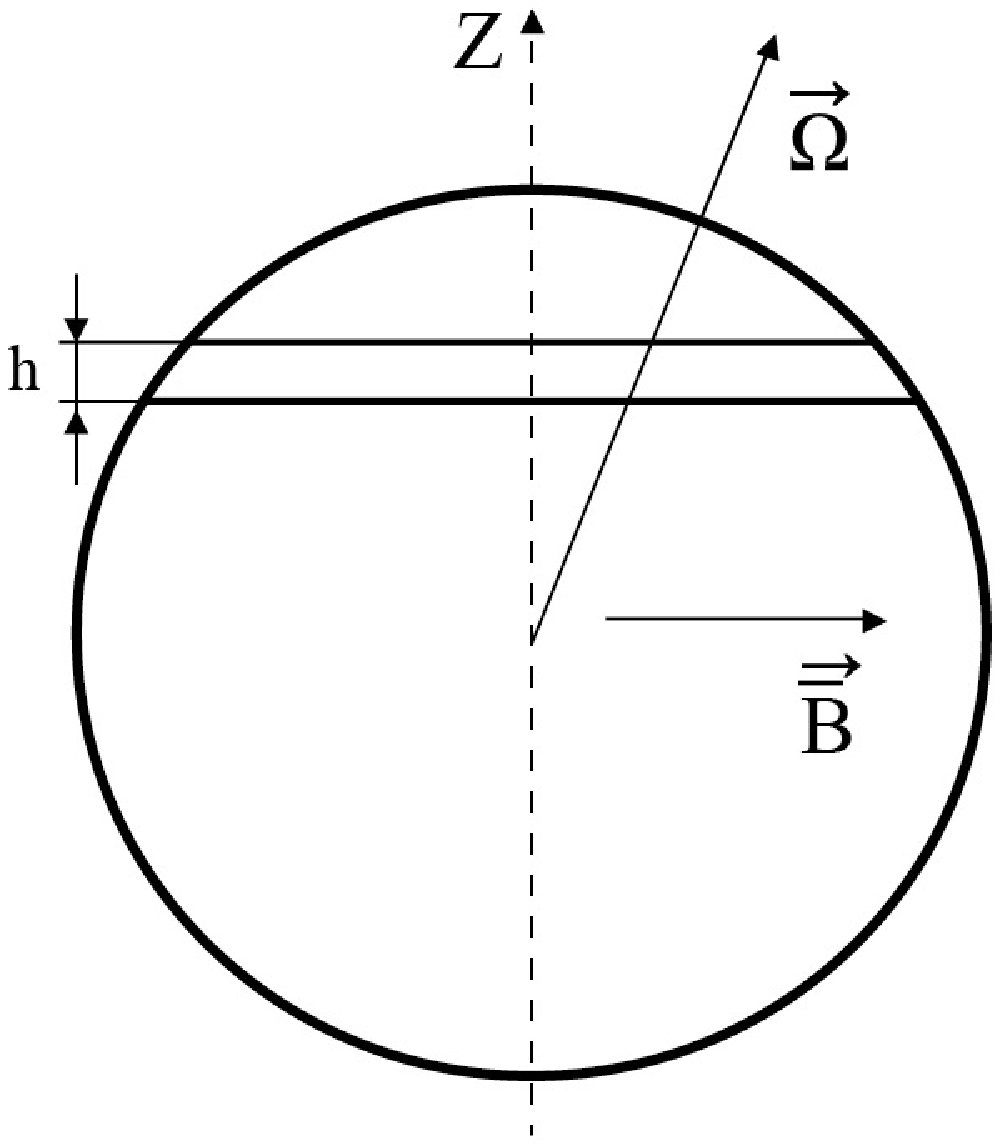}
    \includegraphics[height=6 cm, width=6 cm]{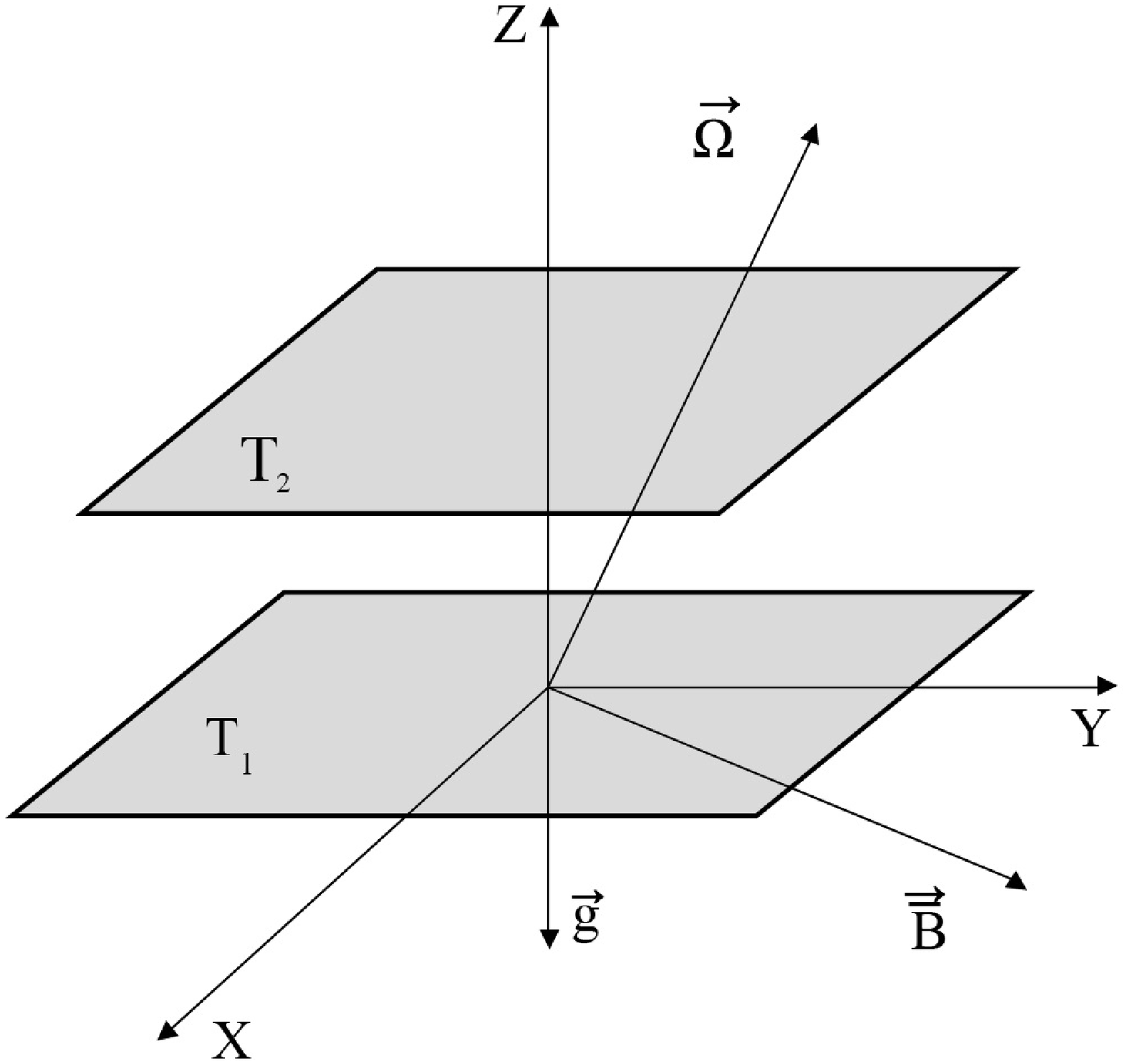}\\
  \caption{ \small   Scheme of a thin layer of rotating electroconductive fluid of an astrophysical object. In the general case the angular velocity  $\vec{\Omega }$ is inclined  to the plane  $(X,Y)$ where the induction vector  $\overline{B}$ of uniform magnetic filed is located. The gradient of equilibrium temperature is directed  vertically  downwards: $T_{1} >T_{2} $ -- heating from below.
}\label{fg1}
\end{figure}
Naturally,   this  promts to study the possibility of generation of large-scale vortex and magnetic fields in rotating  media under the action of a small-scale force with zero helicity $\vec{F}_{0} rot\vec{F}_{0} =0$. Such an example of LSVS generation in a rotating incompressible fluid is reported in  \cite{46s}.  As shown in the said paper,  development of large-scale instability in obliquely rotating fluid gives rise to nonlinear large-scale helical structures of Beltrami vortex type, or to localized kinks with internal helical structure. In \cite{47s} the new HD  $\alpha $-effect revealed in  \cite{46s} was generalized to the case of electroconductive fluid.  This allowed to reveal  the  large-scale instability leading to generation of LSVS and magnetic fields.  Thereat,  the nonlinear stage was shown to be characterized by the  presence of chaotic localized vortex and magneic structures.
As is known  \cite{48s}-\cite{49s}, a large-scale motion caused by nonuniform  heating in a gravitation field (free convection) exists in convective zones of the Sun and other stars, as well as in the core of the Earth and other planets. The convection in which the  rotation axes of the medium and uniform magnetic field coincide with the direction of gravitation vector, was studied in detail in  \cite{49s}. However, for astrophysical problems it is most significant to consider the case when the directions of the rotation axes and of  magnetic fields  are perpendicular, or do not coincide with one another.  The role of azimuthal magnetic field essentially increases for convective fluid layers located in the equatorial region of the rotating object. As known from the theory of magnetic dynamo \cite{2s}-\cite{7s}, the toroidal magnetic field in the Earth crust or  in the solar atmosphere exceeds the poloidal field by an order of magnitude. Starting from this fact, here we will consider generation and nonlinear evolution of vortex and magnetic fields in a rotating stratified electoconductive  fluid in an external uniform magnetic field under the action of the nonhelical force  $\vec{F}_{0} $. Suppose that   the vector of angular rotation velocity  $\vec{\Omega }$  is  deviated from the vertical direction $OZ$, and the vector of the external magnetic field  $\overline{B}$ is located  in the horizontal plane $XOY$ perpendicular to the direction of the gravity force $\vec{g}$ (Fig. \ref{fg1}). Such a geometry of the problem considered here is most suitable for description of dynamo processes in rotating cosmic objects.

The results obtained in the present work may find  application in a number of astrophysical problems.

\begin{figure}
  \centering
    \includegraphics[width=6 cm]{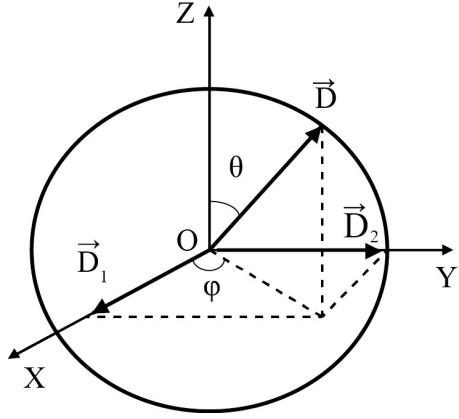}\\
  \caption{\small  For the case, when the external magnetic field $\vec{\overline{B}}=0$, shows the relationship of the Cartesian projections of the rotation parameter $\vec D$   (or the angular velocity vector of rotation $\vec \Omega$ ) with their projections in a spherical coordinate system.}\label{fg2}
\end{figure}

\section{Basic equations and formulation of the problem}

Consider the dynamics of perturbed state of the electroconductive fluid  located in the constant gravitation  $\vec{g}$ and magnetic  $\vec{\bar{B}}$ fields with the constant temperature gradient  $\nabla \overline{T}$ in the system of rotating coordinates:
\begin{equation} \label{eq1}
\frac{\partial v_{i} }{\partial t}+v_{k} \frac{\partial v_{i} }{\partial x_{k} } =\nu \frac{\partial^2 v_{i}}{\partial x_k^2} -\frac{1}{\overline{\rho }} \frac{\partial P}{\partial x_{i} }+2\varepsilon_{ijk} v_{j} { \Omega }_{k}+\frac{\varepsilon_{ijk}\varepsilon_{jml}}{4\pi \overline{\rho}}\frac{\partial B_l}{\partial x_m}(B_k+\overline{B}_k) + g e_{i}\beta \Theta+F_{0}^{i}
\end{equation}
\begin{equation} \label{eq2}
\frac{\partial B_i}{\partial t}=\varepsilon_{ijk}\varepsilon_{knp}\frac{\partial }{\partial x_j}\left(v_n (B_p+\overline{B}_p)\right)+\nu_m \frac{\partial^2 B_i}{\partial x_k^2} \end{equation}
\begin{equation} \label{eq3}
  \frac{\partial \Theta}{\partial t} +v_{k} \frac{\partial \Theta}{\partial x_{k} }-Ae_{k} v_{k} =\chi \frac{\partial^2 \Theta}{\partial x_k^2}
\end{equation}
\begin{equation} \label{eq4}
  \frac{\partial v_{i} }{\partial x_{i} } =\frac{\partial B_{i} }{\partial x_{i} }= 0
\end{equation}
Here  $v_{i} $, $P$, $B_{i} $, $\Theta $ are the perturbances of  the velocity, pressure, magnetic field induction and temperature of the fluid $(i=x,y,z)$; $\overline{B}_{i} =\textrm{const}$, the induction of the external homogeneous magnetic field; $\overline{\rho }$, the equilibrium density of the medium: $\overline{\rho }=const$; $\nu$, $\chi $, the fluid viscosity and thermal conductivity coefficients,  respectively; $\nu_{m} =\frac{c^{2}}{4\pi \sigma_{c}} $, the  magnetic viscosity coefficient; $\sigma_{c} $, the coefficient of electrical conductivity of the medium; $\beta $, the thermal expansion coefficient. The system of magnetic hydrodynamic Eqs.(\ref{eq1})-(\ref{eq4}) is written in  the Boussinesq approximation \cite{48s} and describes the evolution of disturbances  relative to the equilibrium state, specified by the constant temperature gradient $\nabla \overline{T}=-A\vec{e}$ $(A>0)$ and the hydrostatic pressure: $\nabla\left(\overline{P}+\frac{\overline{B}^2}{8\pi}\right)=\overline{\rho }\vec{g}$. Here we neglect the centrifugal forces, since the condition  $g \gg \Omega^2r$ , where  $r$ is the characteristic radius of fluid rotation, is considered to be satisfied.

Now let us formulate the following problem which geometry is shown in Fig. \ref{fg1}. Consider a thin layer (with the thickness $h$ ) of a rotating electoconductive fluid in which the lower and the upper surfaces have the temperatures  $T_{1} $ and  $T_{2} $, respectively, there at  $T_{1} >T_{2} $ , i.e. heating from below. In this case the direction of the  temperature gradient  $\nabla \overline{T}=\vec{A}$  coincides with the one of the gravitation field  $\vec{g}=-g\vec{e}_{z} $, where  $\vec{e}=(0,0,1)$ is the unit vector in the direction of the axis $Z$. The temperature profile  $\overline{T}$ linearly depends on the vertical coordinate  $z$: $\overline{T}(z)=T_{1}-\frac{T_{1}-T_{2} }{h} \cdot z$. The vector of angular rotation velocity $\vec{\Omega }=\left(\Omega_{1} ,\Omega_{2} ,\Omega_{3} \right)$ is considered to be constant (solid-body rotation) and inclined with respect
\begin{figure}
  \centering
  \includegraphics[height=7 cm]{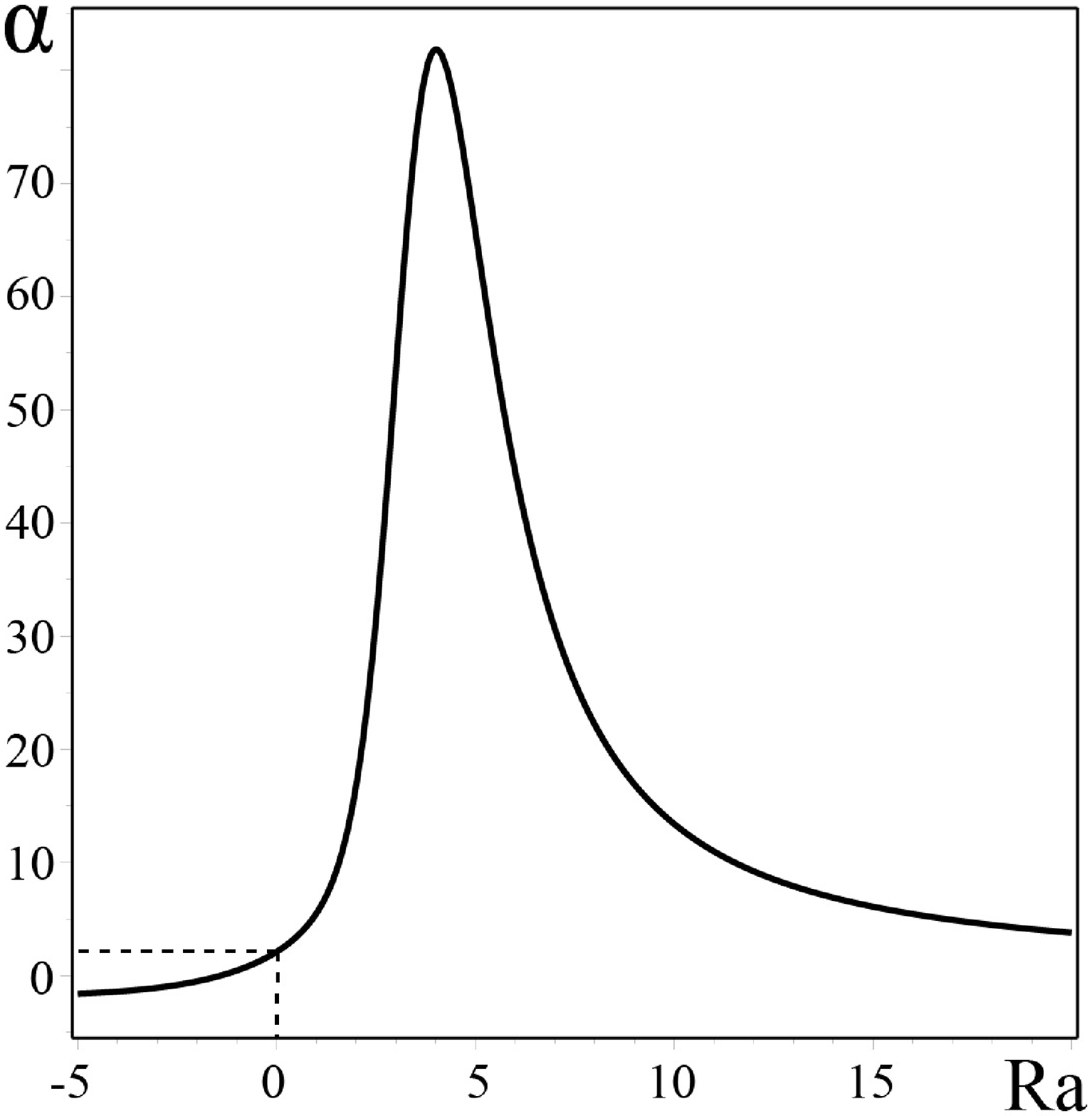}
  \includegraphics[height=7 cm]{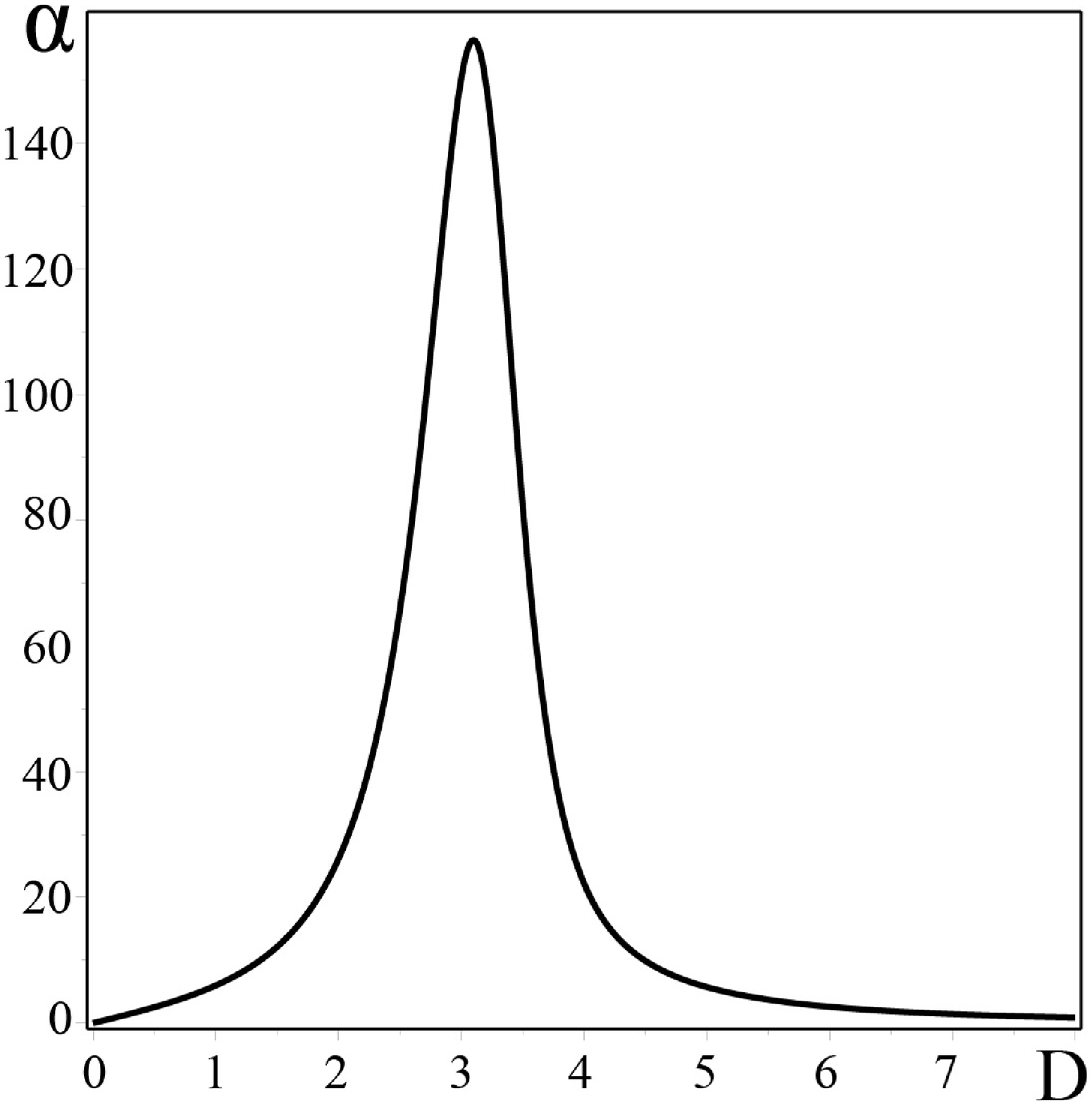}\\
  \caption{\small  On the left the plot of  $\alpha$-effect of parameter stratification of the medium $Ra$ (Rayleigh number), and on the right the plot of the  $\alpha $-effect of the parameter of rotation $D$. } \label{fg3}
\end{figure}
to the plane  $(X,Y)$ where the vector of homogeneous magnetic field  $\vec{\bar{B}}=(\overline{B}_{1} ,\overline{B}_{2} ,0)$ lies. Eq. (\ref{eq1}) contains the external force  $\vec{F}_{0} $, which models the source of external excitation in the medium of small-scale and high-frequency fluctuations of the velocity field  $\vec{v}_{0} $ with the small Reynolds number $R=\frac{v_{0} t_{0} }{\lambda_{0} } \ll 1$.  Here we will consider the non-helical external force  $\vec{F}_{0} $ with the following properties:
\begin{equation} \label{eq5}
div\vec{F}_{0} =0, \; \vec{F}_{0} rot\vec{F}_{0}
=0,\;  rot\vec{F}_{0} \ne 0 , \;    \vec{F}_{0} =f_{0} \vec{F}_{0} \left(\frac{x}{\lambda
_{0} } ;  \frac{t}{t_{0} } \right)
\end{equation}
where  $\lambda_{0} $ is the characteristic scale, $t_{0} $ is the characteristic time, $f_{0} $ is the characteristic amplitude  of the external force. Now choose the external force in a  rotating coordinate system in the form:
\begin{equation} \label{eq6}
F_{0} ^{z} =0,\;\vec{F}_0 =f_{0}\left(\vec{i}cos\phi _{2} +\vec{j}cos\phi _{1}\right), $$
$$ \phi_{1} =\vec{\kappa}_{1}\vec{x}-\omega_{0}t, \; \phi_{2}=\vec{\kappa}_{2} \vec{x}-\omega_{0}t,\;\vec{\kappa}_{1}=\kappa_{0}\left(1,0,0\right),\;\vec{\kappa}_{2}=\kappa_{0} \left(0,1,0\right).
\end{equation}
It is evident that this external force satisfies all the conditions  (\ref{eq5}). Let us consider the dimensionless variables in Eqs. (\ref{eq1})-(\ref{eq4})  which notations   preserve  the ones of the dimensional variables (for convenience):
\[\vec{x} \to \frac{\vec{x}}{\lambda_{0} } ,\quad t\to \frac{t}{t_{0} } ,\; \; \;
\; \vec{v}\to \frac{\vec{v}}{v_{0} } ,\quad \vec{F}_{0} \to \frac{\vec{F}_{0}}{f_{0} } ,\quad \vec{B}\to \frac{\vec{B}}{B_{0}},\quad \overline{\vec{B}}\to \frac{\overline{\vec{B}}}{B_{0}},\quad \Theta \to \frac{\Theta}{\lambda_0 A},\]
\[t_{0} =\frac{\lambda_{0}^{2}}{\nu },\quad  f_{0} =\frac{v_{0} \nu }{\lambda_{0}^{2} }, \quad P\to \frac{P}{P_{0}} ,\quad  P_{0} =\overline{\rho}\frac{\nu v_{0} }{\lambda_{0}}. \]
Here  $v_{0} ,\; B_{0} ,\; P_{0} $ are the characteristic values of small-scale pulsations of the velocity, magnetic field and pressure. In the dimensionless variables Eq.(\ref{eq1})-(\ref{eq3})  will have the form:
\begin{equation} \label{eq7}
\frac{\partial v_{i} }{\partial t}+Rv_{k} \frac{\partial v_{i} }{\partial x_{k} } = \frac{\partial^2 v_{i}}{\partial x_k^2} - \frac{\partial P}{\partial x_{i} }+\varepsilon_{ijk} v_{j}D_{k}+\frac{Q}{RPm}\varepsilon_{ijk}\varepsilon_{jml}\frac{\partial B_l}{\partial x_m}(B_k+\overline{B}_k) + e_{i}\frac{Ra}{RPr} \Theta+F_{0}^{i}
\end{equation}
\begin{equation} \label{eq8}
\frac{\partial B_i}{\partial t}-Pm^{-1}\frac{\partial^2 B_i}{\partial x_k^2} =R\varepsilon_{ijk}\varepsilon_{knp}\frac{\partial }{\partial x_j}\left(v_n (B_p+\overline{B}_p)\right) \end{equation}
\begin{equation} \label{eq9}
  \frac{\partial \Theta}{\partial t}+Rv_{k} \frac{\partial \Theta}{\partial x_{k} }-Re_{k} v_{k} =Pr^{-1} \frac{\partial^2 \Theta}{\partial x_k^2}
\end{equation}
\begin{figure}
  \centering
  \includegraphics[height=7 cm]{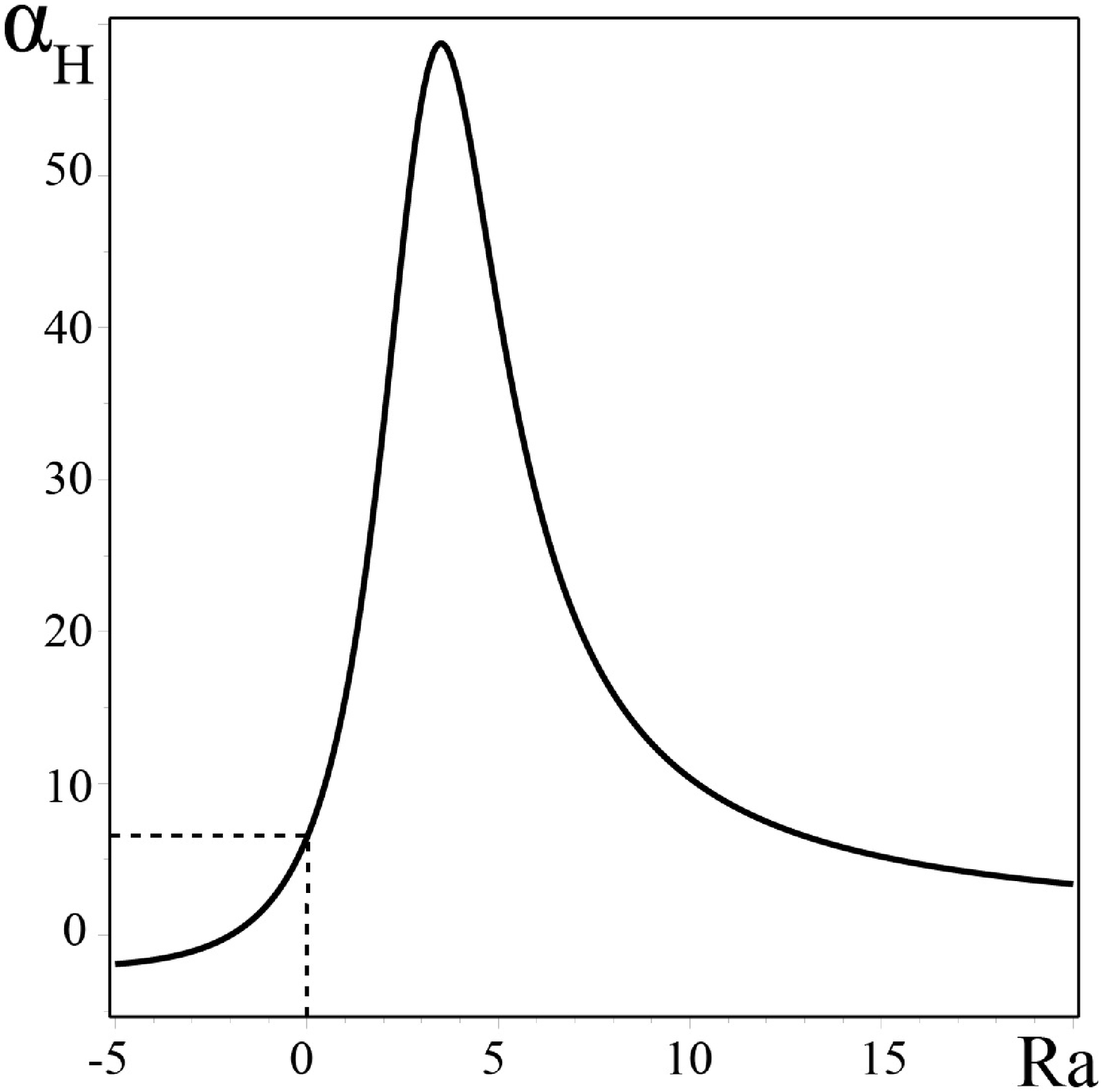}
  \includegraphics[height=7 cm]{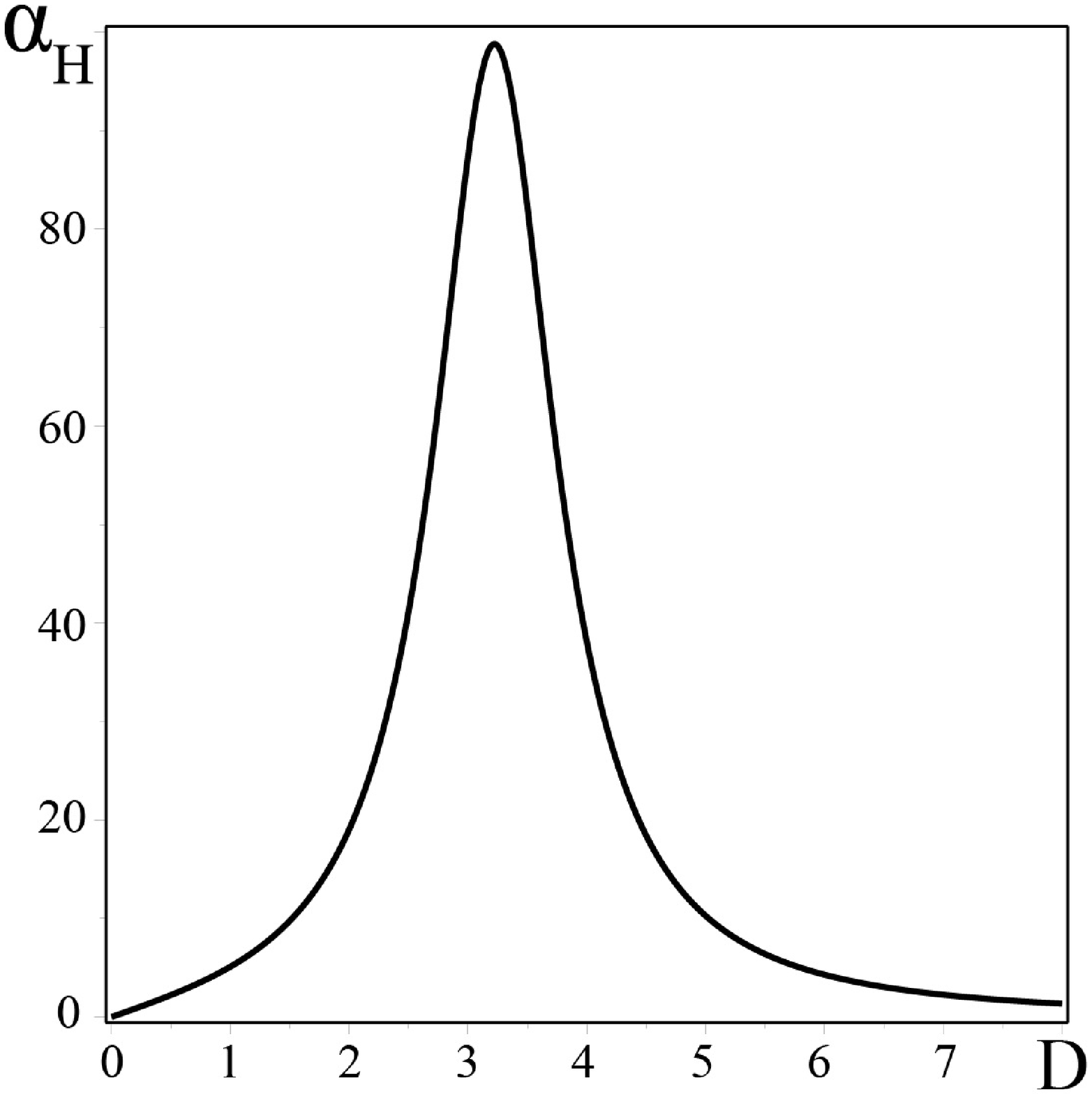}\\
  \caption{ \small  On the left the plot of   $\alpha_H$-effect of parameter stratification of the medium $Ra$ (Rayleigh number), and on the right the plot of the $\alpha_H$-effect of the parameter of rotation $D$.} \label{fg4}
\end{figure}
When going to  new temperature $\Theta \to \Theta/R$ and magnetic field  $B \to B/R$, finally obtain:
\begin{equation} \label{eq10}
\frac{\partial v_{i} }{\partial t}+Rv_{k} \frac{\partial v_{i} }{\partial x_{k} } = \frac{\partial^2 v_{i}}{\partial x_k^2} - \frac{\partial P}{\partial x_{i} }+\varepsilon_{ijk} v_{j}D_{k}+ R\widetilde{Q}\varepsilon_{ijk}\varepsilon_{jml}\frac{\partial B_l}{\partial x_m}B_k+\widetilde{Q}\varepsilon_{ijk}\varepsilon_{jml}\frac{\partial B_l}{\partial x_m}\overline{B}_k+ e_{i}\widetilde{Ra} \Theta+F_{0}^{i}
\end{equation}
\begin{equation} \label{eq11}
\frac{\partial B_i}{\partial t}-Pm^{-1}\frac{\partial^2 B_i}{\partial x_k^2} =R\varepsilon_{ijk}\varepsilon_{knp}\frac{\partial }{\partial x_j}\left(v_n B_p\right)+\varepsilon_{ijk}\varepsilon_{knp}\frac{\partial }{\partial x_j}\left(v_n\overline{B}_p\right) \end{equation}
\begin{equation} \label{eq12}
  \frac{\partial \Theta}{\partial t}-Pr^{-1} \frac{\partial^2 \Theta}{\partial x_k^2}=-Rv_{k} \frac{\partial \Theta}{\partial x_{k} }+e_{k} v_{k}
\end{equation}
\begin{equation} \label{eq13}
  \frac{\partial v_{i} }{\partial x_{i} } =\frac{\partial B_{i} }{\partial x_{i} }= 0
\end{equation}
Here we use the following dimensionless parameters: $\widetilde{Ra}=\frac{Ra}{Pr} $, $Ra=\frac{g\beta A\lambda _{0}^{4} }{\nu \chi } $ is the Rayleigh  number in the scale $\lambda _{0} $; $D_{i} =\frac{2\Omega _{i} \lambda _{0}^{2} }{\nu } $ -- the rotation parameter in the scale  $\lambda _{0} $ ($i=1,2,3$) connected with the Taylor number $Ta_{i} =D_{i}^{2} $; $\widetilde{Q}=\frac{Q}{Pm} $, $Q=\frac{\sigma_{c} B_{0}^{2} \lambda_{0}^{2}}{c^{2}\overline{\rho }\nu } $ -- the Chandrasekhar number; $Pm=\frac{\nu }{\nu_{m} } $ -- the magnetic Prandtl number; $Pr=\frac{\nu }{\chi } $ -- the Prandtl number.

The small parameter of asymptotic expansion is the Reynolds number  $R=\frac{v_{0}t_{0}}{\lambda_{0}} \ll 1$, the parameters $D$, $\widetilde{Q}$ and  $\widetilde{Ra}$ being arbitrary. Due to the presence of the small parameter ($R \ll 1$) in the system of Eqs.(\ref{eq10})-(\ref{eq13}), we can apply  the theory of multi-scale asymptotic expansions  (see e.g. \cite{30s}-\cite{31s}, \cite{38s}). In contrast to the theory of mean field \cite{2s}-\cite{7s}, we can sequentially those in each order by $R$, see the dynamics of perturbations for different space and time scales. In particular, in the zero order of   $R$, small-scale and high-frequency oscillations of the velocity   $\vec{v}_{0} $ are excited by the external force  $\vec{F}_{0} $  acting  at the equillibrium state. Naturally, the dynamics of small-scale fields depends on external factors such as  rotation and stratification of the medium, magnetic and gravitation fields, etc.  Such oscillations are characterized by zero average values, however, nonlinear interactions in some orders of the perturbation theory give rise to the terms which do not vanish at averaging. The method of finding the solvability conditions for multiscale asymptotic  expansion which define the evolution equations for large-scale perturbations will be considered  in more detail in the next section.
\begin{figure}
  \centering
  \includegraphics[height=6 cm]{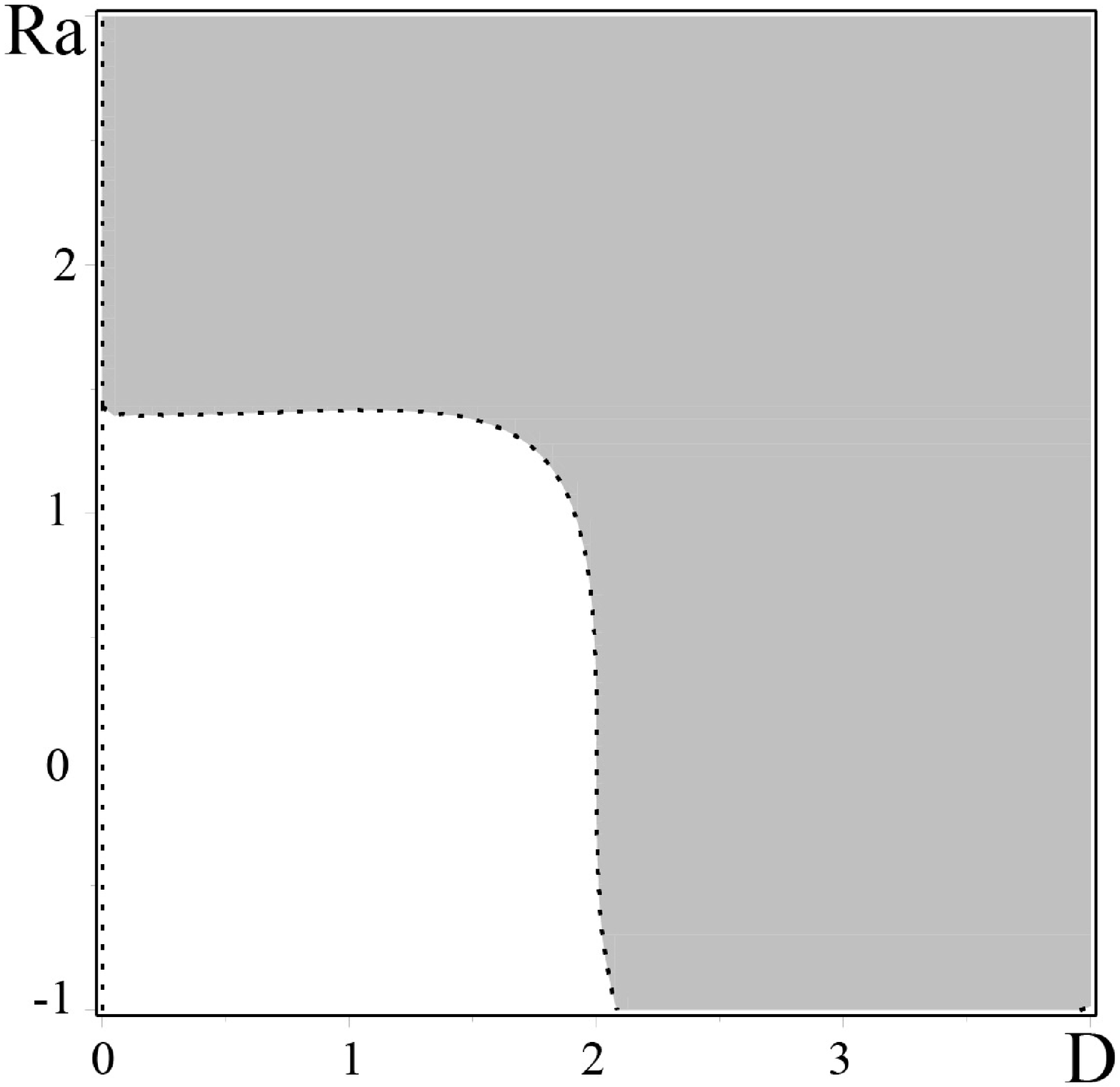}
  \includegraphics[height=6 cm]{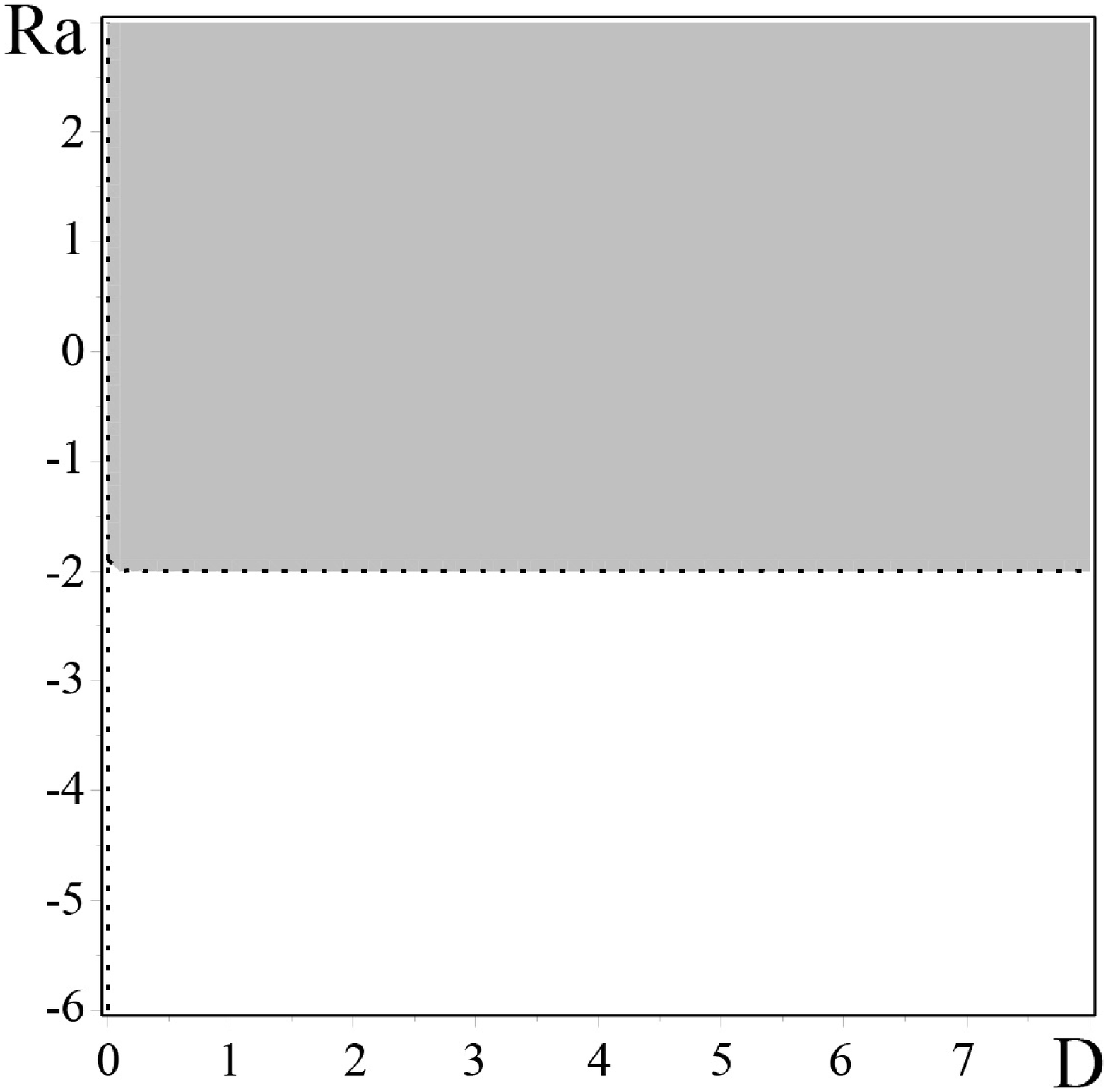}\\
  \caption { \small On the left the plot for $\alpha$  in the plane $(D,Ra)$, where the gray color shows the region corresponding to positive values  $\alpha>0 $  (unstable solutions), and the white negative values $\alpha$. On the right is the plot  for $\alpha_H$  in the plane $(D,Ra)$, where the gray color shows the region corresponding to positive values  $\alpha_H >0 $ (unstable solutions), and the white negative values $\alpha_H$.} \label{fg5}
\end{figure}

\section{Equations for large-scale fields }

In accordance with the method \cite{30s}-\cite{31s},\cite{38s}  of construction of asymptotic equations let us present the derivatives with respect to space and time in Eqs.(\ref{eq10})-(\ref{eq13}) in the form of the asymptotic expansion:
\begin{equation} \label{eq14}
 \frac{\partial }{\partial t} \to \partial _{t}
+R^{4} \partial _{T}, \;\;\; \frac{\partial }{\partial x_{i} } \to \partial _{i} +R^{2}
\nabla _{i}
\end{equation}
where $\partial_{i} $ and $\partial_{t} $ denote the derivatives of/with respect to the fast variables  $x_{0} =\left(\vec{x}_{0} ,t_{0} \right)$, whereas  $\nabla_{i} $ and $\partial_{T} $ are the derivatives of/with respect to the slow variables $X=(\vec{X},T)$. The variables  $x_{0} $ and $X$ may be referred to as small- and large-scale variables, accordingly.  While constructing the nonlinear theory, the variables  $\vec{V}$, $\vec{B}$, $P$  are to be presented in the form of the asymptotic series:
\[\vec{V} \left(\vec{x},t\right)\; \; =\frac{1}{R} \vec{W}_{-1} \left(X\right)+\vec{v}_{0} \left(x_{0} \right)+R\vec{v}_{1} + R^{2} \vec{v}_{2} +R^{3} \vec{v}_{3} +\cdots  \]
\begin{equation} \label{eq15}
 \vec{B}\left(\vec{x},t\right)=\frac{1}{R} \vec{B}_{-1}
\left(X\right)+\vec{B}_{0} \left(x_{0} \right)+R\vec{B}_{1} +R^{2} \vec{B}_{2} +R^{3}
\vec{B}_{3} +\cdots
\end{equation}
\[\Theta(\vec{x},t)=\frac{1}{R} T_{-1} \left(X\right)+T_{0} \left(x_{0} \right)+RT_{1} +R^{2} T_{2} +R^{3} T_{3} +\cdots  \]
\[P(x)=\frac{1}{R^{3} } P_{-3} +\frac{1}{R^{2} } P_{-2} +\frac{1}{R} P_{-1} +P_{0}\left(x_{0} \right)
+R(P_{1} +\overline{P}_{1} \left(X\right))+R^{2} P_{2} +R^{3} P_{3} +\cdots \]
 Let us substitute the expansions  (\ref{eq14})-(\ref{eq15}) into the system of Eqs.(\ref{eq10})-(\ref{eq13}) , then select the terms of the same orders in  $R$ up to the degree  $R^{3}$ inclusively and obtain the equations of multi-scale asymptotic expansion. The algebraic structure of the asymptotic expansion of Eqs.(\ref{eq10})-(\ref{eq13}) of different orders in  $R$ is presented in Appendix I. In the latter it is shown that  the basic secular equations, i.e. those for large-scale fields, are obtained in the order  $R^{3}$
\begin{equation} \label{eq16}
  \partial _{t} W_{-1}^{i} -\nabla_k^2 W_{-1}^{i}+{{\nabla }_{k}}\overline{(v_{0}^{k}v_{0}^{i})}=-\nabla _{i} \overline{P}_{1}
	+\widetilde{Q}\varepsilon _{ijk} \varepsilon _{jml} \left({{\nabla }_{m}}\overline{(B_{0}^{l}B_{0}^{k})}\right)
\end{equation}
\begin{equation} \label{eq17}
 \partial _{t} B_{-1}^{i} -Pm^{-1} \nabla_k^2 B_{-1}^{i}=\varepsilon _{ijk} \varepsilon _{knp} {{\nabla }_{j}}\overline{(v_{0}^{n}B_{0}^{p})}
\end{equation}
\begin{figure}
  \centering
  \includegraphics[width=7 cm, height= 7 cm]{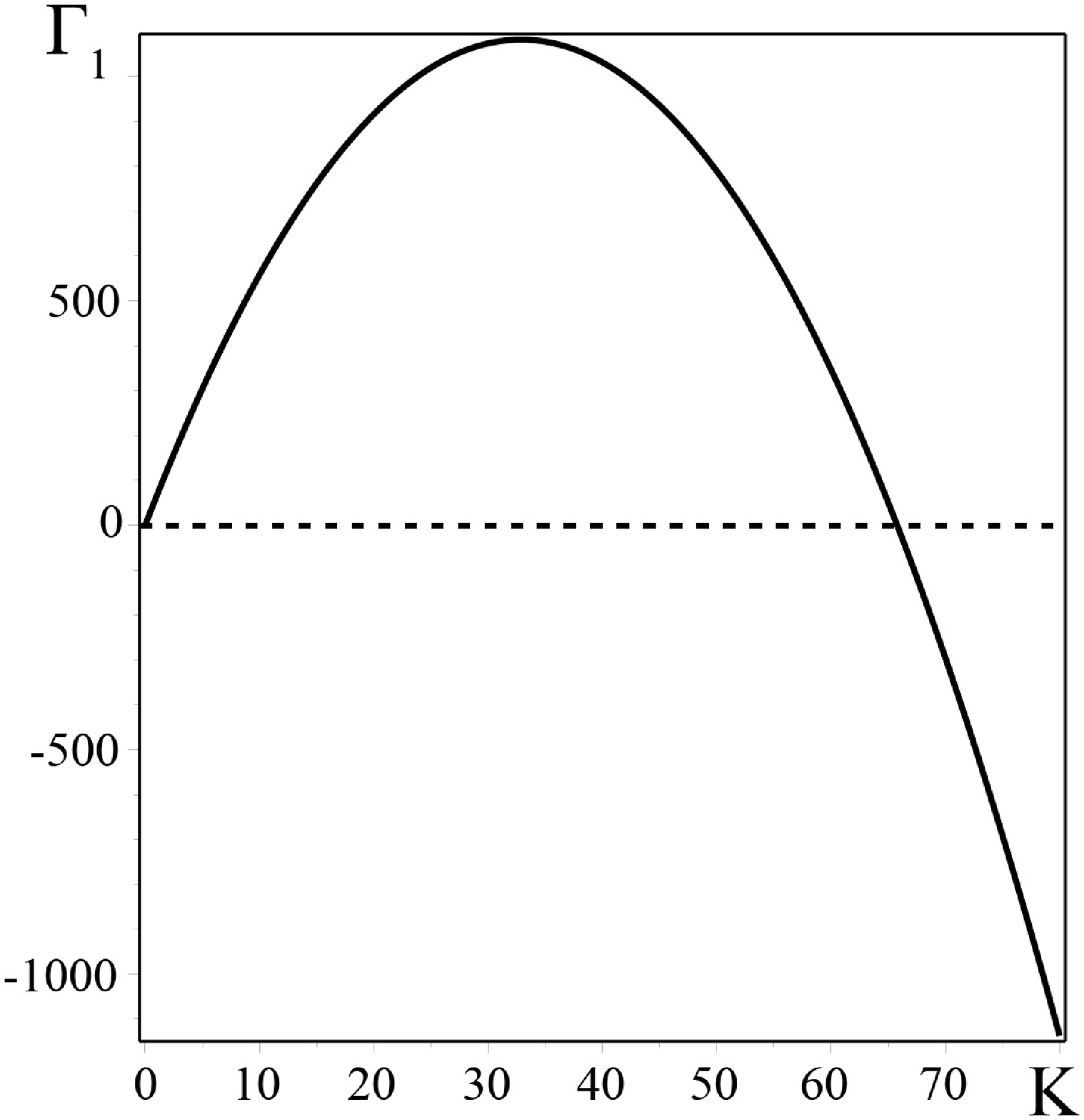}
	\includegraphics[width=7 cm, height= 7 cm]{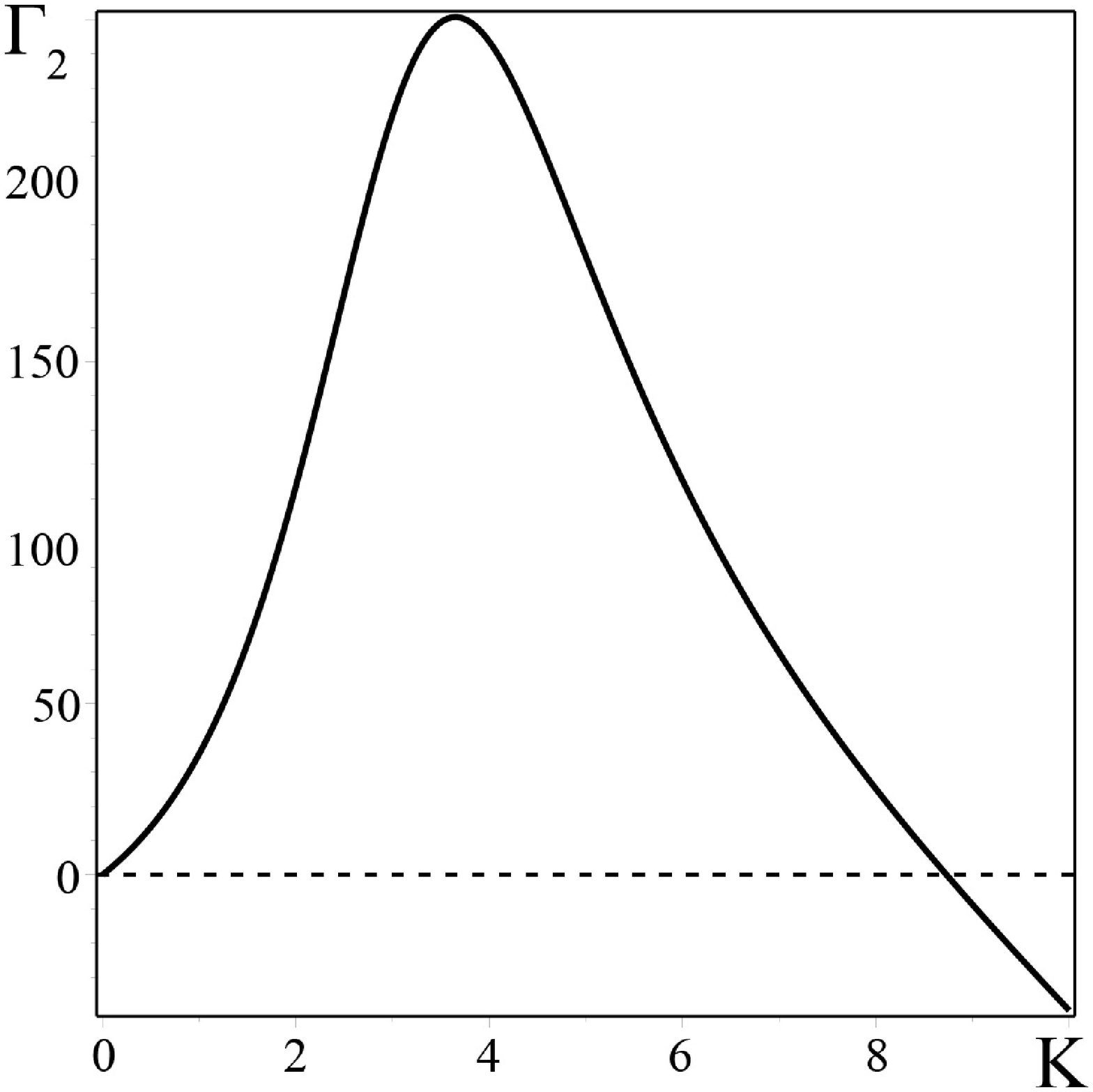}\\
\caption{ \small  On the left is the plot of the dependence of the instability increment  for  $\alpha $-effect on the wave numbers $K$; on the right is a plot  of the dependence of the instability increment for the $\alpha $-effect on the wave numbers $K$. The graphs are constructed for fixed parameters of stratification $Ra=5$ and rotation $D=2.5$.}\label{fg6}
\end{figure}
\begin{equation}\label{eq18}\partial_{T} T_{-1} -Pr^{-1} \nabla_k^2 T_{-1} =-\nabla _{k} \left(\overline{v_{0}^{k} T_{0}
}\right) \end{equation}
Using the  convolution of the tensors  $\varepsilon_{ijk} \varepsilon_{jml} =\delta_{km} \delta_{il} -\delta_{im} \delta_{kl} ,\; \varepsilon_{ijk} \varepsilon_{knp} =\delta_{in} \delta_{jp} -\delta_{ip} \delta_{jn} $ and the denotations  $\vec{W}=\vec{W}_{-1} ,\; \vec{H}=\vec{B}_{-1} $  obtain Eqs.(\ref{eq16})-(\ref{eq17}) in the form:
\begin{equation} \label{eq19}
   \partial _{T} W_{i} -\nabla_k^2 W_{i} +{{\nabla }_{k}}\overline{(v_{0}^{k}v_{0}^{i})}=-\nabla _{i} \overline{P}_{1} +\widetilde{Q}\left( {{\nabla }_{k}}\overline{(B_{0}^{i}B_{0}^{k})}-\frac{{{\nabla }_{i}}}{2}\overline{{{(B_{0}^{k})}^{2}}} \right)
\end{equation}
\begin{equation} \label{eq20}
  \partial _{T} H_{i} -Pm^{-1} \nabla_k^2 H_{i} =
{{\nabla }_{j}}\overline{(v_{0}^{i}B_{0}^{j})}-{{\nabla }_{j}}\overline{(v_{0}^{j}B_{0}^{i})}
\end{equation}
Eqs.(\ref{eq16})-(\ref{eq18}) are supplemented with the secular equations  derived  in Appendix  I:
\[-\nabla_{i} P_{-3} +\varepsilon_{ijk} W_{j} D_{k}+e_{i}\widetilde{Ra}T_{-1} =0,\quad  W_{-1}^{z}=0, \]
\[   W_{-1}^{k} \nabla_{k} W_{-1}^{i} =-\nabla_{i} P_{-1} +\widetilde{Q}\varepsilon_{ijk} \varepsilon_{jml}\left(\nabla_{m} B_{-1}^{l}B_{-1}^{k}+
\nabla_{m}B_{-1}^{l} \overline{B_k} \right), \]
\[\varepsilon_{ijk}\varepsilon_{knp} \left(\nabla_{j} W_{-1}^{n} B_{-1}^{p}+\nabla_{j} W_{-1}^{n}\overline{B_p} \right)=0, \]
\[ W_{-1}^{k} \nabla_{k}T_{-1} =0,\quad \nabla_{i}W_{-1}^{i} =0, \quad \nabla_{i} B_{-1}^{i} =0. \]
To obtain the system of Eqs.(\ref{eq16})-(\ref{eq18}) decribing the evolution of large-scale fields we had to reach the third order of the perturbation theory.
Such a phenomenon is typical of the use of the method of multi-scale expansions.  As seen from Eqs.(\ref{eq16})-(\ref{eq17}), the large-scale temperature  $T_{-1} $ does not influence the dynamics of the large-scale  field of the velocity  $\vec{W}_{-1} $ and the magnetic field $\vec{B}_{-1} $, therefore let us confine ourselves to investigation of Eqs.(\ref{eq16})-(\ref{eq17}). These equations acquire a closed form after calculation of the correlation functions, i.e.  the Reynolds stresses  $\nabla_{k} \overline{(v_{0}^{k} v_{0}^{i} )}$, the Maxwell stresses $\nabla_{k} \overline{(B_{0}^{i} B_{0}^{k} )}$ and the turbulent e.m.f. ${{\mathcal{E}}_{n}}={{\varepsilon }_{nij}}\overline{v_{0}^{i}B_{0}^{j}}$  . Calculation of these correlation functions is essentially simplified due to the \guillemotleft quasi-two-dimensional\guillemotright approximation often used for  description of  large-scale vortex and magnetic fields  in a number of astrophysical and geophysical problems \cite{3s,14s,30s,31s}.
\begin{figure}
  \centering
    \includegraphics[width=6 cm]{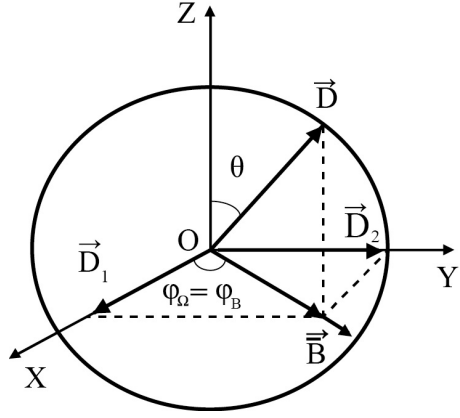}\\
  \caption{\small  The connection of the Cartesian projections of the rotation parameter $\vec D$ (or the angular velocity vector of rotation $\vec \Omega$ ) with their projections in a spherical coordinate system is shown. The direction of the external magnetic field  $\vec{\overline{B}}$  is chosen so that the angles  $\varphi$  of deviation from the axis  $OX$ for the rotation vector  $\varphi_\Omega$ and magnetic field $\varphi_B$ coincide: $\varphi_\Omega =\varphi_B$.}\label{fg7}
\end{figure}
In the framework of this approximation, in the present  study we consider  the large-scale derivative with respect to  $Z$  more preferable, i.e.
\[\qquad \nabla_Z\equiv\frac{\partial }{\partial  Z} \gg \frac{\partial }{\partial X} , \frac{\partial }{\partial Y}, \]
thereat the geometry of large-scale fields has the following form:
\begin{equation} \label{eq21}
   \vec{W}=\left(W_{1} \left(Z\right),\; W_{2}
\left(Z\right),\; 0\right), \vec{H}=\left(H_{1}\left(Z\right),\; H_{2}\left(Z\right),
\; 0\right)
\end{equation}
In the scope of \guillemotleft quasi-two-dimensional\guillemotright  problem the system of Eqs.(\ref{eq14})-(\ref{eq15})   is simplified:
\begin{equation} \label{eq22} \partial_{T} W_{1}-\nabla_Z^2 W_{1} +{{\nabla }_{Z}}\overline{(v_{0}^{z}v_{0}^{x})}=\widetilde{Q}{{\nabla }_{Z}}\overline{(B_{0}^{z}B_{0}^{x})}\end{equation}
\begin{equation} \label{eq23} \partial_{T} W_{2}-\nabla_Z^2 W_{2} +{{\nabla }_{Z}}\overline{(v_{0}^{z}v_{0}^{y})}=\widetilde{Q}{{\nabla }_{Z}}\overline{(B_{0}^{z}B_{0}^{y})}\end{equation}
\begin{equation} \label{eq24} \partial_{T} H_{1}-Pm^{-1} \nabla_Z^2 H_{1} =
{{\nabla }_{Z}}\overline{(v_{0}^{x}B_{0}^{z})}-{{\nabla }_{Z}}\overline{(v_{0}^{z}B_{0}^{x})} \end{equation}
\begin{equation} \label{eq25} \partial_{T} H_{2} -Pm^{-1} \nabla_Z^2 H_{2} ={{\nabla }_{Z}}\overline{(v_{0}^{y}B_{0}^{z})}-{{\nabla }_{Z}}\overline{(v_{0}^{z}B_{0}^{y})} \end{equation}
\begin{equation} \label{eq26} \partial_{T}T_{-1}-Pr^{-1}\nabla_Z^2 {T}_{-1} +\nabla_{{Z}}\left(\overline{{v}_{0}^{{z}}{T}_{0}}\right)=0 \end{equation}

To derive Eqs.\eqref{eq22}-\eqref{eq26}   in a closed form we will use the solutions of the equations for small-scale fields in the zeroth order of $R$ obtained in Appendix  II. Then it is necessary to calculate the correlators contained in the system  \eqref{eq22}-\eqref{eq26}.
The technical aspect of this problem is considered in detail in Appendix  III.  The calculations performed here make it possible to obtain the following  closed equations for large-scaler fields of the velocity  $(W_{1} ,W_{2} )$ and the magnetic fields  $(H_{1} ,H_{2} )$:
\begin{equation} \label{eq27}
 \partial_{T} W_{1}-\nabla_Z^2 W_{1} +\nabla_{Z}\left(\alpha_{\left(2\right)}\cdot \left(1-W_{2} \right)\right)=0
\end{equation}
\begin{equation}\label{eq28}
 \partial_{T} W_{2}-\nabla_Z^2 W_{2} -\nabla_{Z}\left(\alpha_{\left(1\right)}\cdot \left(1-W_{1} \right)\right)=0
\end{equation}
\begin{equation} \label{eq29}
 \partial_{T} H_{1}-Pm^{-1}\nabla_Z^2 H_{1} +\nabla_{Z}\left(\alpha_{H}^{\left(2\right)}\cdot H_{2} \right)=0
\end{equation}
\begin{equation} \label{eq30}
 \partial_{T} H_{2}-Pm^{-1}\nabla_Z^2 H_{2}-\nabla_{Z}\left(\alpha_{H}^{\left(1\right)}\cdot H_{1} \right)=0
\end{equation}
 where the nonlinear coefficients  $\alpha_{\left(1\right)} $, $\alpha_{\left(2\right)}$, $\alpha_{H}^{\left(1\right)}$, $\alpha_{H}^{\left(2\right)}$   have the form :
\[\alpha_{\left(1\right)} =\frac{f_{0}^{2} }{2} \cdot\frac{D_{1} q_{1} Q_{1} \left(1-W_{1} \right)^{-1} }{4
\left(1-W_{1}\right)^{2}q_1^2\widetilde{Q}_1^2 +\left[D_{1}^{2} +W_{1}\left(2-W_{1}\right)+\mu_{1} \right]^{2}+\xi_1 },\]
\begin{figure}
  \centering
  \includegraphics[height=5.5 cm, width=5 cm]{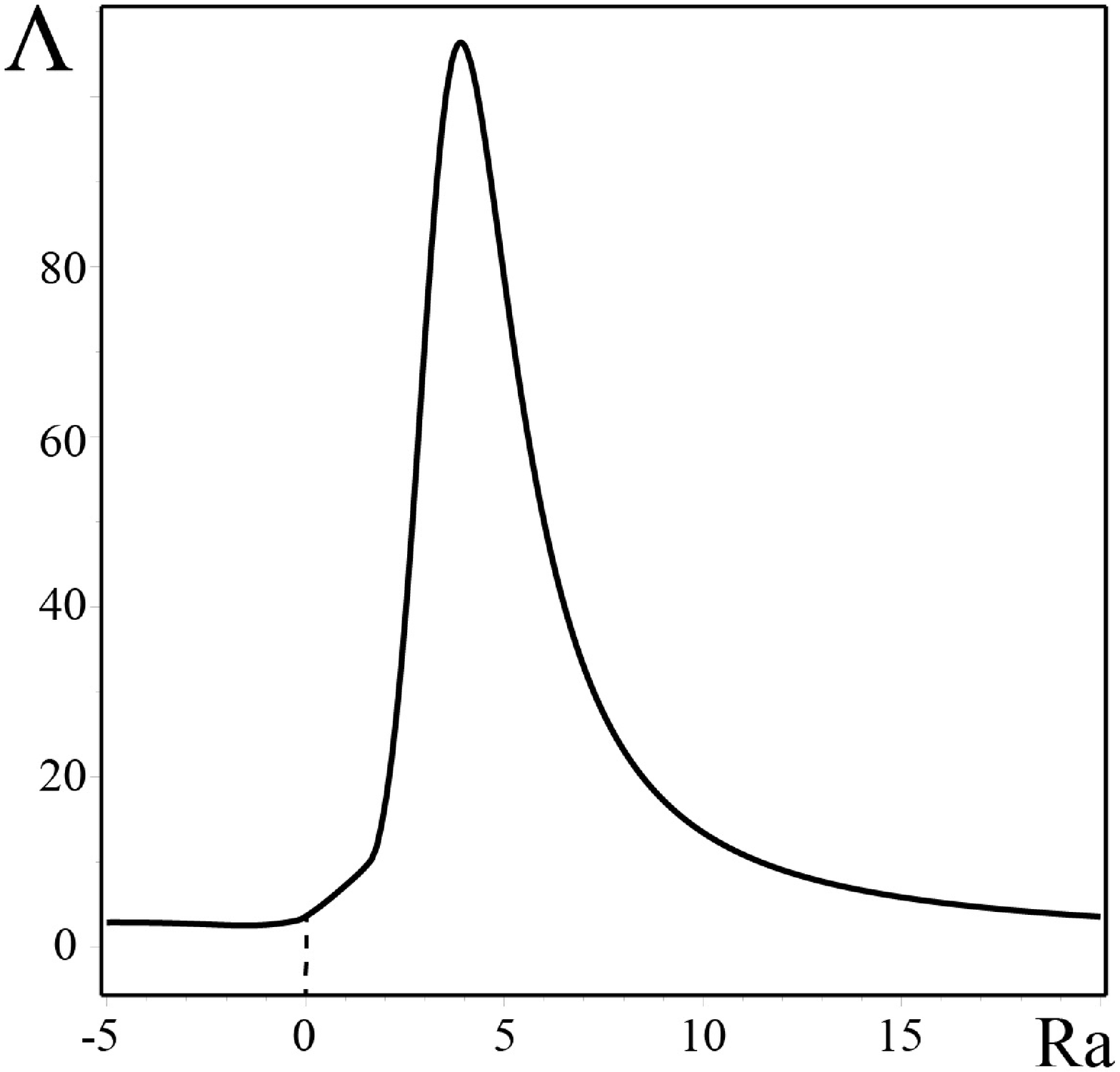}  a)
	\includegraphics[height=5.5 cm, width=5 cm]{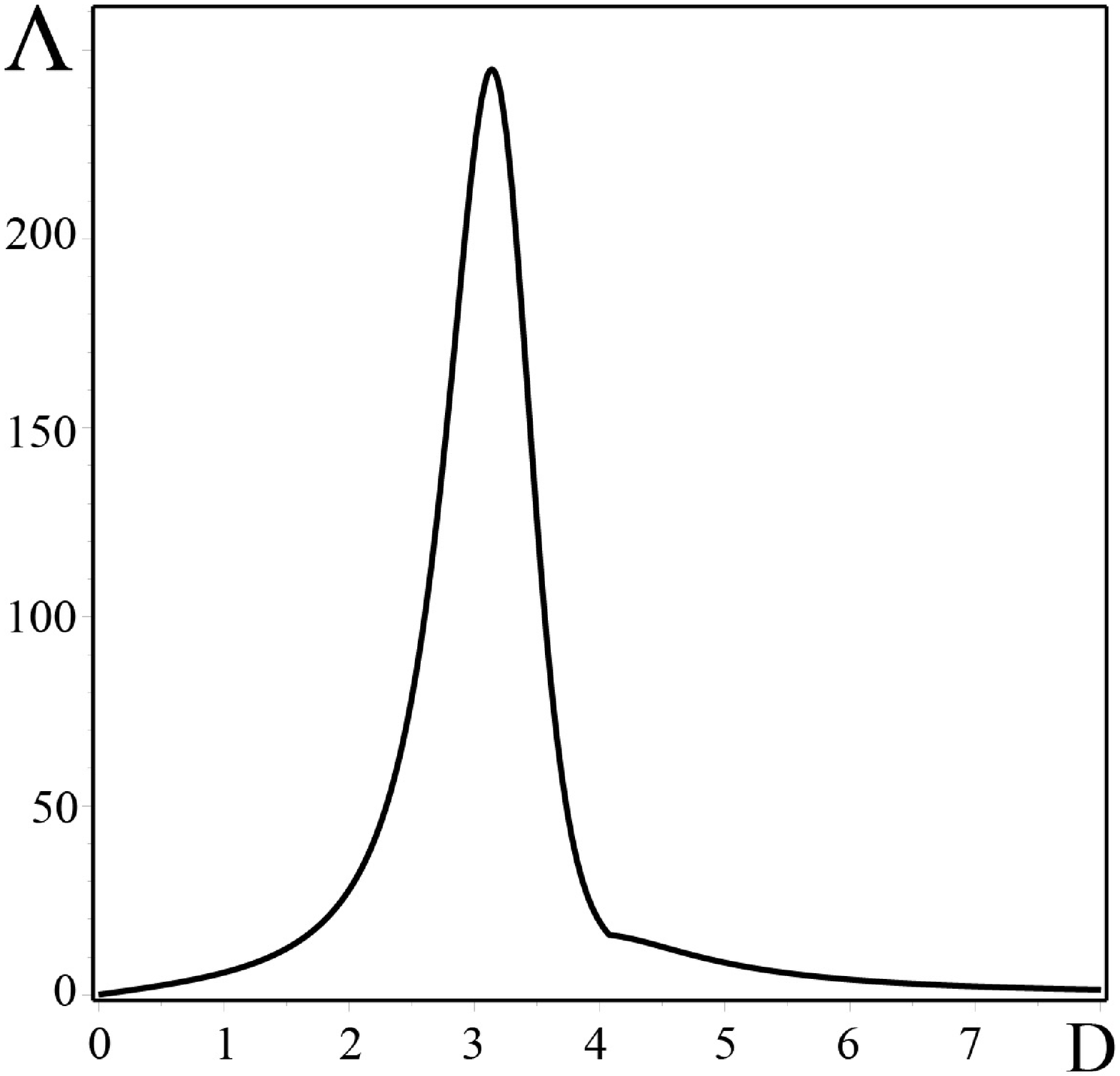}  b)
  \includegraphics[height=5.5 cm, width=5 cm ]{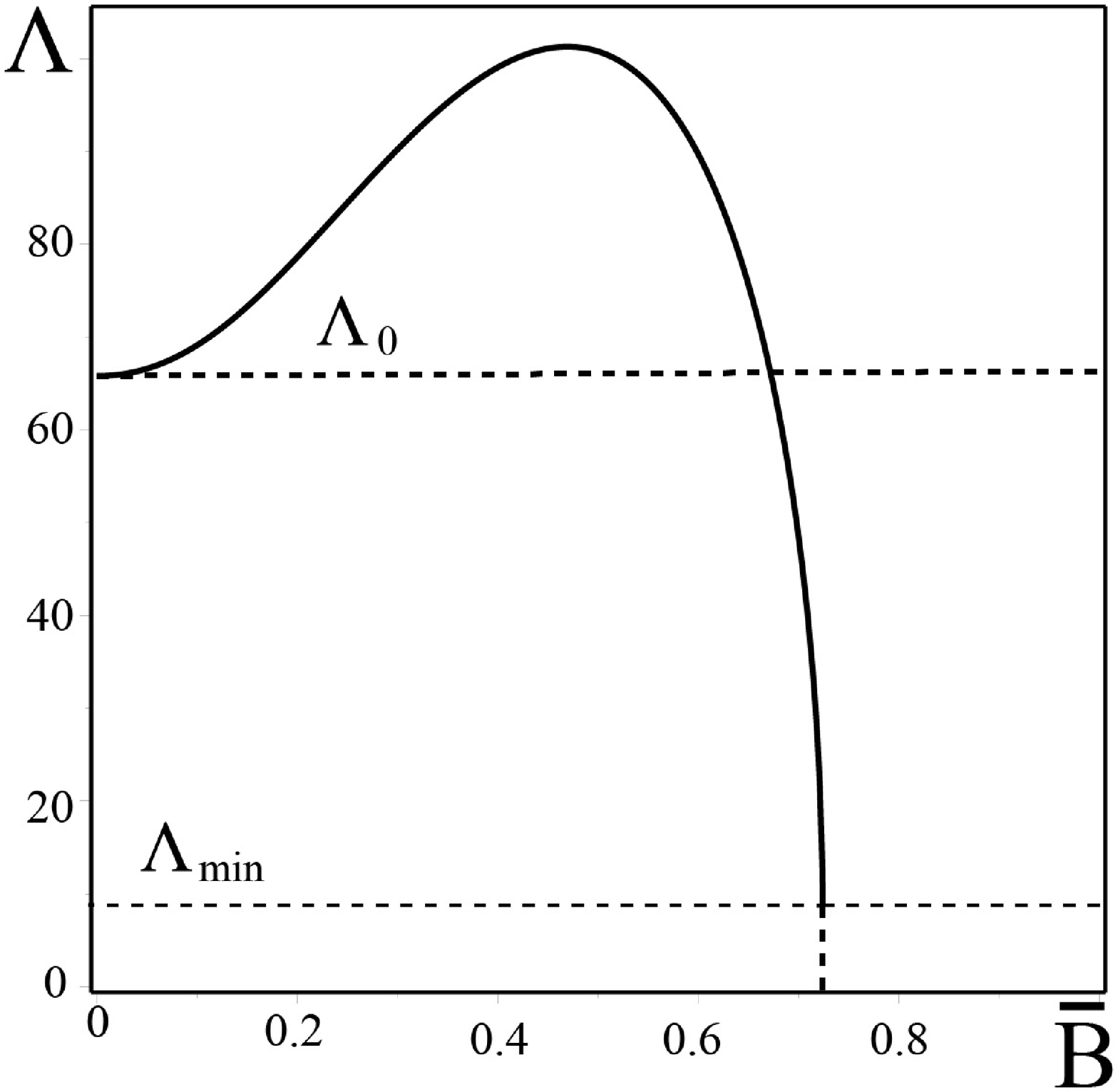} c)\\
  \caption{\small  $\textrm{a)}$  is the plot of the dependence of the  $\Lambda $-effect on the stratification parameter of the medium  $Ra$ (the Rayleigh number); $\textrm{ b)}$ is the  plot of the dependence of the  $\Lambda $-effect on the parameter of rotation of the medium $D$;  $\textrm{c)}$ is the  plot of the dependence of the   $\Lambda$-effect on the external magnetic field  $\overline{B}$. } \label{fg8}
\end{figure}
\[\alpha_{\left(2\right)} =\frac{f_{0}^{2}}{2}\cdot \frac{D_{2} q_{2} Q_{2} \left(1-W_{2} \right)^{-1}}{4
\left(1-W_{2}\right)^{2}q_2^2\widetilde{Q}_2^2  +\left[D_{2}^{2}+W_{2}\left(2-W_{2}\right)+\mu_{2}\right]^{2} +\xi_2},\]
\[\alpha_{H}^{\left(1\right)}=f_{0}^{2}\cdot \frac{D_{1} \left(1-W_{1}\right)Pm \widetilde{Q}_{1}(1+\overline{B}_1/H_1) }{\left(1+Pm^{2}\left(1-W_{1} \right)^{2} \right)\left[4\left(1-W_{1}\right)^{2}q_1^2\widetilde{Q}_1^2 +\left[D_{1}^{2} +W_{1} \left(2-W_{1} \right)+\mu_{1} \right]^{2}+\xi_1 \right]}, \]
\[\alpha_{H}^{\left(2\right)}=f_{0}^{2}\cdot \frac{D_{2}\left(1-W_{2} \right)Pm \widetilde{Q}_{2}(1+\overline{B}_2/H_2) }{\left(1+Pm^{2} \left(1-W_{2} \right)^{2} \right)\left[4\left(1-W_{2} \right)^{2}q_2^2\widetilde{Q}_2^2   +\left[D_{2}^{2} +W_{2} \left(2-W_{2}\right)+\mu _{2} \right]^{2}+\xi_2 \right]}. \]

The expressions which denote $q_{1,2} $, $Q_{1,2} $, $\widetilde{Q}_{1,2} $, $\mu_{1,2} $, $\sigma_{1,2} $, $\chi_{1,2} $, $\xi_{1,2} $ are also presented in Appendix  III.
The coefficients  $\alpha_{\left(1\right)} $, $\alpha_{\left(2\right)}$  and  $\alpha_{H}^{\left(1\right)} $, $\alpha_{H}^{\left(2\right)}  $ correspond to the nonlinear HD  $\alpha $-effect and  the nonlinear MHD  $\alpha $-effect, respectively. Thus, we have obtained the self-consistent system of nonlinear evolution equations for large-scale perturbations  of the velocity and magnetic field  that  will be further called the equations of nonlinear  magneto-vortex dynamo. It should be noted that the mechanism of dynamo works  only  due to  the effect of rotation of the medium.  If such a rotation is absent ($\Omega =0$), then there occurs conventional diffuse spreading of large-scale fields. In the absence of heating  $(\nabla \overline{T}=0)$ and external magnetic field  $(\overline{B}=0)$ Eqs.(\ref{eq27})-(\ref{eq28}) coincide with the results reported in  \cite{47s}. In the case of non-electroconductive fluid   $(\sigma =0)$ with the temperature gradient  $(\nabla \overline{T}\ne 0)$ we will have  the results obtained in \cite{50s}. In the limit of non-electroconductive  $(\sigma =0)$ and homogeneous fluid  $(\nabla \overline{T}=0)$ there will be obtained the results of \cite{46s}.  For more illustrative representation of the physical mechanism of the said  dynamo model,  it is necessary at first to consider the evolution of small perturbations and then to start studying the nonlinear effects.

\section{Large-scale instability }

Consider the behavior of small perturbations of the field of velocity  $\left(W_{1} ,W_{2} \right)$ and the magnetic field  $\left(H_{1} ,H_{2} \right)$. Then expand the nonlinear coefficients  $\alpha_{(1,2)} $ and  $\alpha_{H}^{(1,2)} $ in Eqs. (\ref{eq27})-(\ref{eq30}) into the Taylor series with respect to the small values  $\left(W_{1},W_{2} \right)$, $\left(H_{1} ,H_{2} \right)$:
\begin{equation} \label{eq31} \alpha_{(1,2)}\cdot (1-W_{1,2})\approx \alpha_0^{(1,2)}-\alpha_{1,2}^{(H)}\cdot H_{1,2}-\alpha_{1,2}^{(W)}\cdot W_{1,2},\quad \alpha_0^{(1,2)}=\textrm{const}, $$
$$ \alpha_H^{(1,2)}\cdot H_{1,2} \approx \alpha_{0H}^{(1,2)}+\widetilde{\alpha}_{H}^{(1,2)}\cdot H_{1,2}-\beta_{W}^{(1,2)}\cdot W_{1,2},\quad \alpha_{0H}^{(1,2)}=\textrm{const}.
\end{equation}
\begin{figure}
  \centering
  \includegraphics[height=7 cm, width=7 cm]{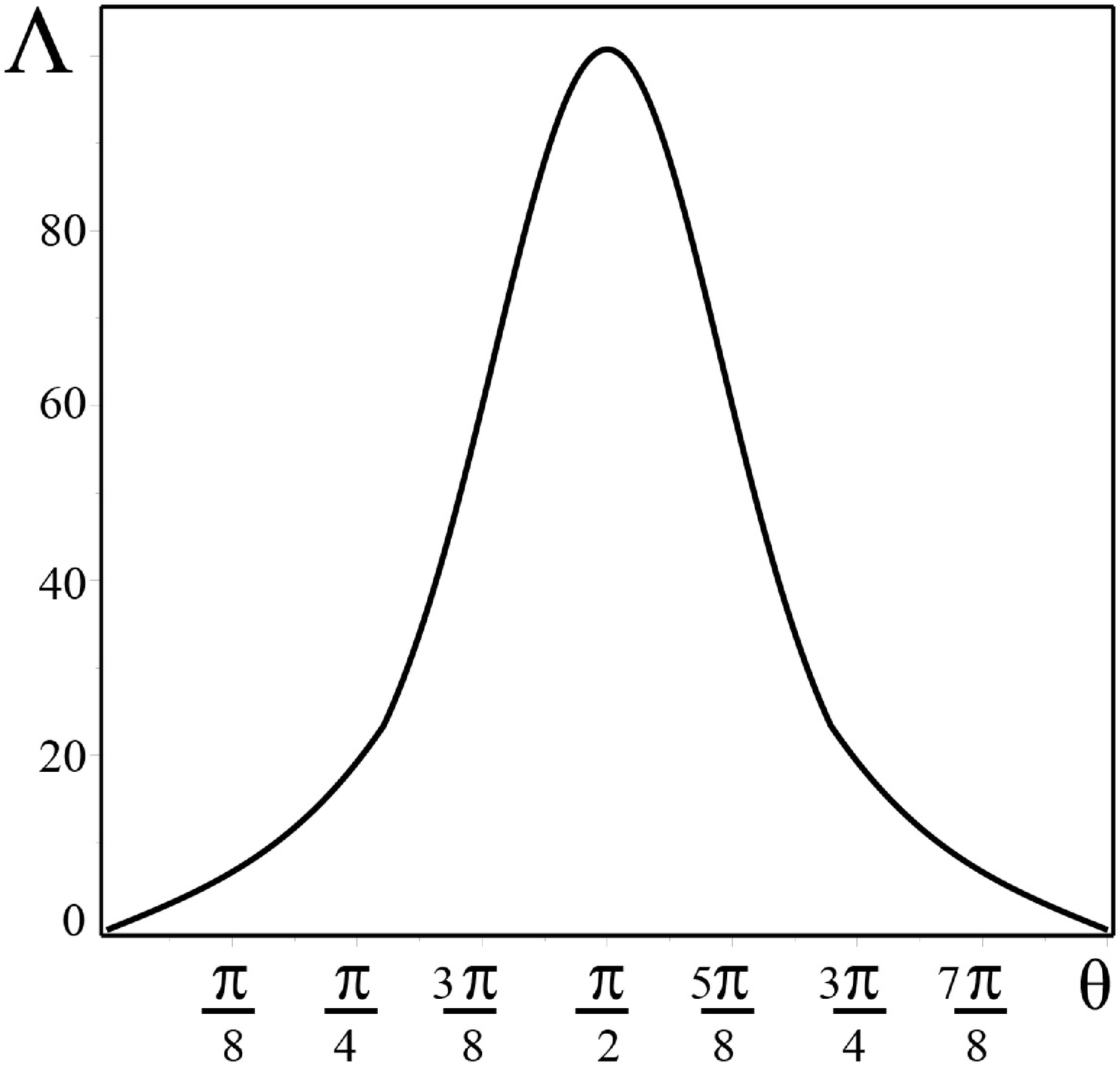}
  \includegraphics[height=7 cm, width=7 cm ]{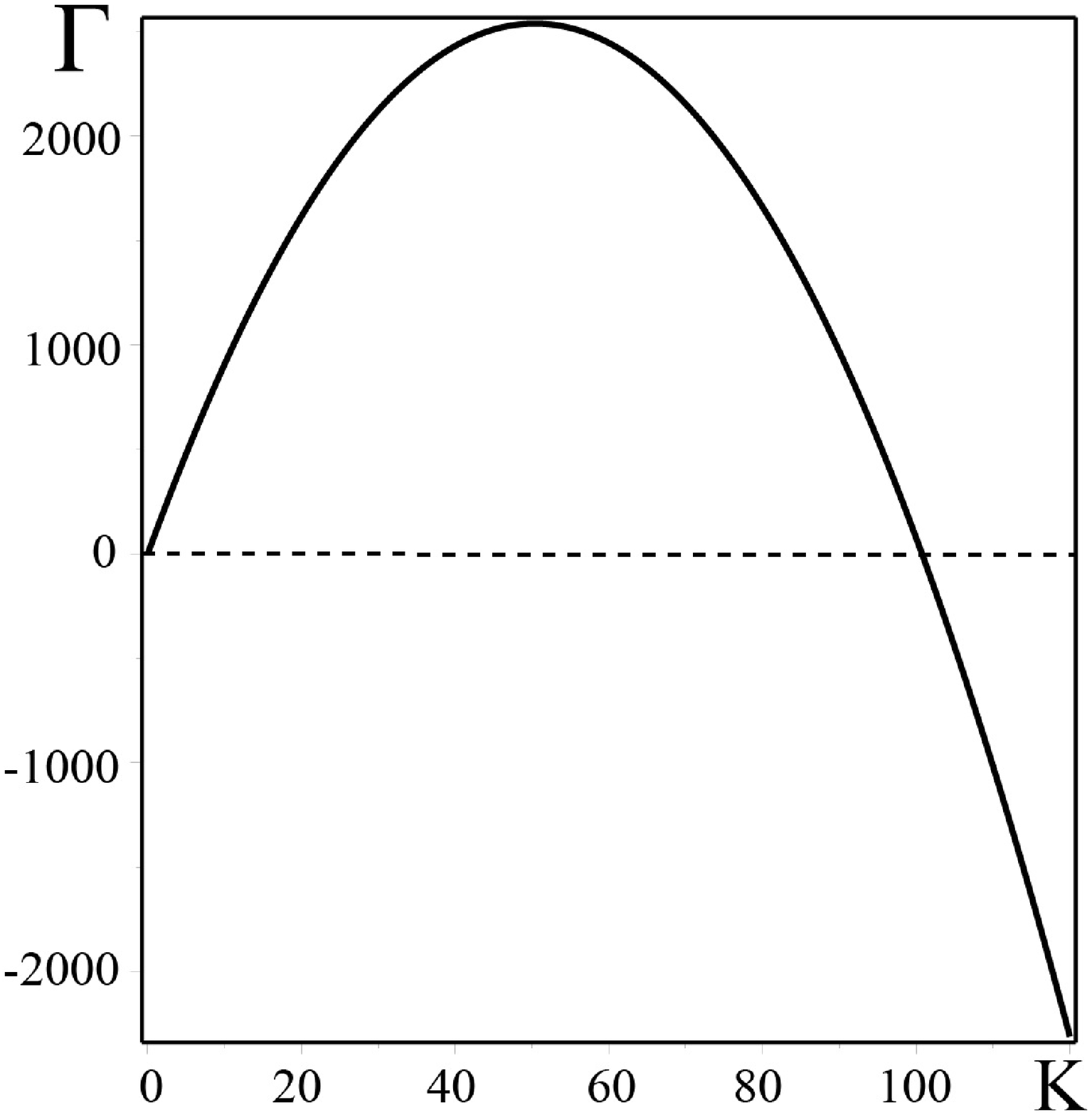}\\
  \caption{ \small   On the left  is the plot of of the dependence of the  $\Lambda$-effect on the angle of inclination $\theta$ of the angular velocity vector $\vec{\Omega}$; on the right is the plot  of the dependence of the instability increment for the   $\Lambda $-effect on the wave numbers $K$.} \label{fg9}
\end{figure}
After substituting (\ref{eq31}) into Eqs. (\ref{eq27})-(\ref{eq30}) we obtain the linearized system of equations :
\begin{equation} \label{eq32}
 \partial_{T} W_{1}-\nabla_Z^2 W_{1}-\alpha_2^{(H)}\cdot\nabla_{Z}H_2-\alpha_2^{(W)}\cdot\nabla_Z W_2=0
\end{equation}
\begin{equation}\label{eq33}
 \partial_{T} W_{2}-\nabla_Z^2 W_{2}+\alpha_1^{(H)}\cdot\nabla_{Z}H_1+\alpha_1^{(W)}\cdot\nabla_Z W_1 =0
\end{equation}
\begin{equation} \label{eq34}
 \partial_{T} H_{1}-\nabla_Z^2 H_{1} + \widetilde{\alpha}_{H}^{(2)}\cdot\nabla_{Z}H_2-\beta_{W}^{(2)}\cdot\nabla_{Z}W_2=0
\end{equation}
\begin{equation} \label{eq35}
 \partial_{T} H_{2}-\nabla_Z^2 H_{2}-\widetilde{\alpha}_{H}^{(1)}\cdot\nabla_{Z}H_1+\beta_{W}^{(1)}\cdot\nabla_{Z}W_1 =0,
\end{equation}
 where the constant coefficients $\alpha_{1,2}^{(H)} ,\alpha_{1,2}^{(W)} ,\widetilde{\alpha}_{H}^{(1,2)} ,\beta_{W}^{(1,2)} $ have the following form:
\begin{equation} \label{eq36}
\alpha_{1,2}^{(H)}=\frac{f_0^2 D_{1,2}}{2}\cdot Q\overline{B}_{1,2}\times$$
$$ \times \left[\frac{\left(2-Ra\right)(2-Q\overline{B}_{1,2}^2)\left(4(D_{1,2}^2-Ra)+(Ra+1)^2+7\right)}{4\left(4+(D_{1,2}^2-Ra)^2\right)^2}+\frac{Q\overline{B}_{1,2}^2-2(Ra-1)}{4\left(4+(D_{1,2}^2-Ra)^2\right)}\right],
\end{equation}
\begin{equation} \label{eq37}
\alpha_{1,2}^{(W)}=\frac{f_0^2 D_{1,2}}{2}\times$$
$$ \times \left[\frac{\left(2-Ra\right)(2-Q\overline{B}_{1,2}^2)\left(D_{1,2}^2-Ra-2\right)}{\left(4+(D_{1,2}^2-Ra)^2\right)^2}+\frac{Q\overline{B}_{1,2}^2+Ra\left(1-Q\overline{B}_{1,2}^2\right)}{2\left(4+(D_{1,2}^2-Ra)^2\right)}\right],
\end{equation}
\begin{equation} \label{eq38}
 \widetilde{\alpha}_{H}^{(1,2)}=\frac{f_0^2 D_{1,2}}{4}\times$$
$$ \times \left[\frac{2+Ra-Q\overline{B}_{1,2}^2+\overline{B}_{1,2}(2+Ra)}{4+(D_{1,2}^2-Ra)^2}-\frac{\overline{B}_{1,2}(2+Ra)\left(4(D_{1,2}^2-Ra)+(Ra+1)^2+7\right)}{\left(4+(D_{1,2}^2-Ra)^2\right)^2}\right],
\end{equation}
\begin{figure}
\centering
\includegraphics[width=19 cm, height=19 cm]{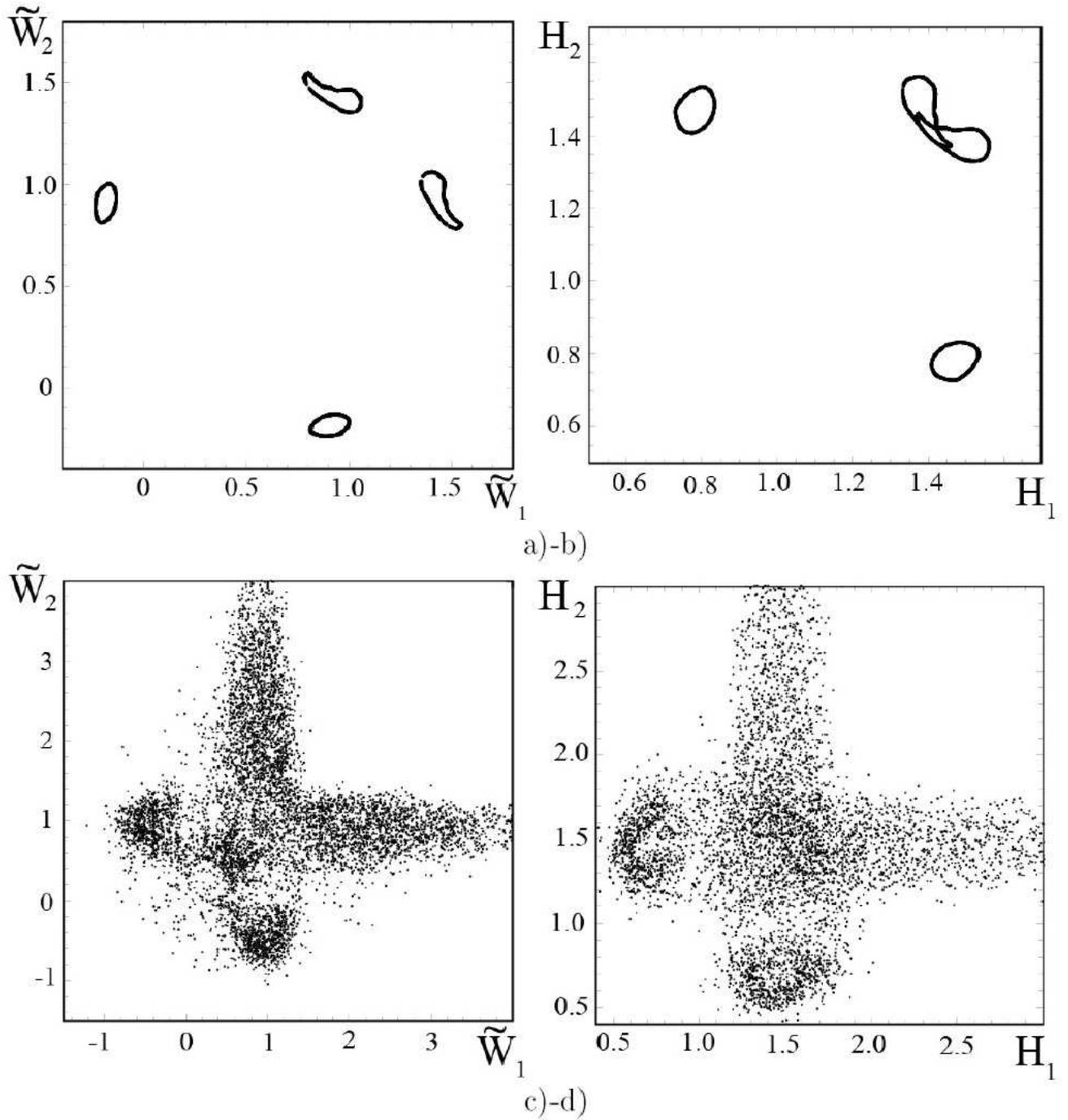} \\
    \caption{\small  In  the figures $\textrm{a)}$ and $\textrm{b)}$ shown the Poincar\'{e} sections for a trajectory with initial conditions $\widetilde W_1 (0) = 1.25$, $\widetilde W_2 (0) = 1.25$, $ H_1 (0) = 1.4$, $ H_2 (0) = 1.4$. This is a regular type of trajectory, which is wound on the tori. The figures  $\textrm{c)}$  and $\textrm{d)}$  correspond to Poincar\'{e} sections for a trajectory with initial conditions  $\widetilde W_1 (0) = 1.398$, $\widetilde W_2 (0) = 1.398$, $ H_1 (0) = 1.4$, $ H_2 (0) = 1.4$. These pictures show stochastic layers, to which the corresponding chaotic trajectory belongs. The calculations were carried out for the case $\overline{B}=0$.} \label{fg10}
\end{figure}
\begin{equation} \label{eq39}
\beta_{W}^{(1,2)}=f_0^2\cdot \frac{D_{1,2}\overline{B}_{1,2}(2+Ra)\left(D_{1,2}^2-Ra-2\right)}{\left(4+(D_{1,2}^2-Ra)^2\right)^2}-
\frac{f_0^2}{4}\cdot\frac{D_{1,2}\overline{B}_{1,2}Ra}{4+(D_{1,2}^2-Ra)^2}
\end{equation}
While deriving Eqs.(\ref{eq32})-(\ref{eq35})  we used the simplification connected with the equality: $Pr=Pm=1$. As seen from Eqs.(\ref{eq32})-(\ref{eq35}), in the presence of external magnetic field the coefficients  $\alpha_{1,2}^{(H)} $ and $\beta_{W}^{(1,2)} $ define the positive feedback in the self-consistent dynamics of the fields   $W_{1,2} $ and $H_{1,2} $. We will search for the solution of the linear system of Eqs.(\ref{eq32})-(\ref{eq35})  in the form of plane waves with the wave vector  $\vec{K} \parallel OZ$:
\begin{equation} \label{eq40}
 \left(\begin{array}{c} {W_{1,2} } \\ {H_{1,2}
} \end{array}\right)=\left(\begin{array}{c} {\hat{W}_{1,2} } \\ {\hat{H}_{1,2} }  \end{array}
\right){exp}\left(-i\omega T+iKZ\right)
\end{equation}
After substituting  (\ref{eq40}) into the system (\ref{eq32})-(\ref{eq35})  we obtain the dispersion equation:
\begin{equation} \label{eq41}
\left[\left(K^2-i\omega \right)^2-K^2\left( \alpha_1^{(W)}\alpha_2^{(W)}+\alpha_2^{(H)}\beta_W^{(1)}\right)\right]\left[\left(K^2-i\omega \right)^2-K^2\left( \widetilde{\alpha}_H^{(1)}\widetilde{\alpha}_H^{(2)}+\alpha_1^{(H)}\beta_W^{(2)}  \right)\right]+$$
$$+K^4\left(\widetilde{\alpha}_H^{(1)}\alpha_2^{(H)}-\alpha_1^{(H)}\alpha_2^{(W)}\right)\left(\alpha_1^{(W)}\beta_W^{(2)}-\widetilde{\alpha}_H^{(2)}\beta_W^{(1)} \right)=0
\end{equation}

\subsection{Analysis of dispersion equation (\ref{eq41})  in the absence of external magnetic field  $\overline{B}_{1,2} =0$ }

It is obvious that without external magnetic field $\overline{B}_{1,2} =0$ the coefficients  $\alpha _{1,2}^{(H)} $ and $\beta _{W}^{(W)} $ vanish, and Eq.(\ref{eq41})  breaks down into two independent equations:
\begin{equation} \label{eq42}
\left[\left(K^2-i\omega \right)^2-\alpha_1^{(W)}\alpha_2^{(W)}K^2\right]\left[\left(K^2-i\omega \right)^2-\widetilde{\alpha}_H^{(1)}\widetilde{\alpha}_H^{(2)}K^2 \right]=0
\end{equation}
where  the coefficients  $\alpha_{1,2}^{(W)} $, $\widetilde{\alpha }_{H}^{(1,2)} $ do not depend on  $\overline{B}_{1,2} $. Dispersion Eq.(\ref{eq42}) corresponds to the physical situation when small perturbations of vortex and magnetic fields independently gain in intensity due to development of large-scale instability such as  $\alpha $--effect. Substituting the frequency  $\omega =\omega _{0} +i\Gamma $ from Eq.(\ref{eq42}) we find:
\begin{equation} \label{eq43}
 \Gamma_{1} =Im \omega_{1} =\pm \sqrt{\alpha_{1}^{(W)} \alpha_{2}^{(W)}}K-K^{2}
\end{equation}
\begin{equation} \label{eq44}
 \Gamma_{2} =Im\omega_{2} =\pm \sqrt{\widetilde{\alpha}_{H}^{(1)}\widetilde{\alpha}_{H}^{(2)} } K- K^{2}
\end{equation}
Solutions (\ref{eq43})   testify to instability at  $\alpha_{1} \alpha_{2}>0$ for large-scale vortex   perturbations with the maximum  instability increment    $\Gamma _{1max} =\frac{\alpha _{1} \alpha _{2} }{4} $ at the wave numbers $K_{1max} =\frac{\sqrt{\alpha _{1} \alpha _{2} } }{2} $. Similarly, for magnetic perturbations the instability  increment   $\Gamma_{2max} =\frac{\widetilde{\alpha}_{H}^{(1)}\widetilde{\alpha}_{H}^{(2)}}{4}$  reaches its maximum at the wave  numbers $K_{2max} =\frac{\sqrt{\widetilde{\alpha}_{H}^{(1)}\widetilde{\alpha}_{H}^{(2)}}}{2}$.  If  $\alpha _{1} \alpha_{2} <0$ and  ${\widetilde{\alpha}_{H}^{(1)}\widetilde{\alpha}_{H}^{(2)}} <0$, then instead of instability there arise damped oscillations  with the frequences  $\omega _{01} =\sqrt{\alpha _{1} \alpha_{2} } K$  and $\omega_{02} =\sqrt{\widetilde{\alpha}_{H}^{(1)}\widetilde{\alpha}_{H}^{(2)}} K$, respectively.

It is clear that in the considered linear theory the coefficients  $\alpha_{1}^{(W)} $, $\alpha_{2}^{(W)} $, $\widetilde{\alpha}_{H}^{(1)}$, $\widetilde{\alpha}_{H}^{(2)}$  depend not on the amplitudes of the fields,  but  on the rotation parameters  $D_{1,2} $, the Rayleigh number $Ra$ and the amplitude of the external force  $f_{0} $. Now  analyze  the dependence of these coefficients on the dimensionless parameters. For simplicity let us assume that the dimensionless amplitude of the external force  $f_{0} =10$. Fixation of the level of the dimensionless force signifies the choice of a certain level of  steady background of small-scale and fast oscillations. It is convenient to replace the Cartesian projections  $D_{1} $ and $D_{2} $ in the coefficients  $\alpha _{1}^{(W)} $, $\alpha _{2}^{(W)} $, $\widetilde{\alpha}_{H}^{(1)}$, $\widetilde{\alpha}_{H}^{(2)}$ by their projections in the spherical coordinate system  $(D,\varphi ,\theta )$.
\begin{figure}
\centering
   \includegraphics[width=18 cm, height=18 cm]{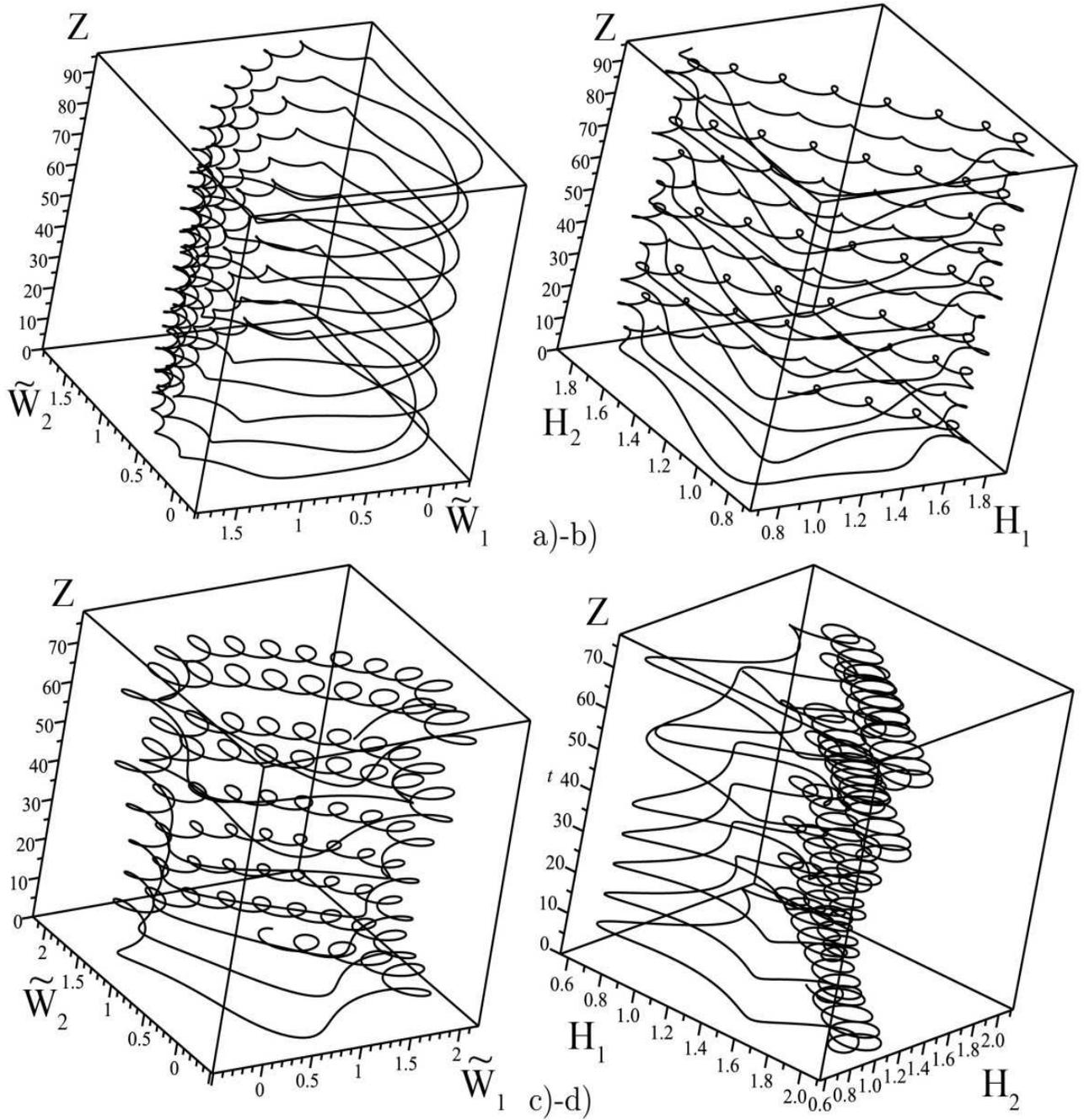}  \\
  \caption{ \small The upper part ( $\textrm{a)}$, $\textrm{b)}$ ) shows  the dependence of the velocity and magnetic field on the height  $Z$  for the numerical solution of equation system (\ref{eq54})-(\ref{eq57}) with the initial conditions $\widetilde W_1 (0) = 1.25$, $\widetilde W_2 (0) = 1.25$, $ H_1 (0) = 1.4$, $ H_2 (0) = 1.4$.   This dependence corresponds to regular motions of the Poincar\'{e} section shown on top of  Fig.   Below ($\textrm{c)}$, $\textrm{d)}$) a similar  dependence is shown  for the numerical solution of equation system (\ref{eq54})-(\ref{eq57}) with the initial conditions:  $\widetilde W_1 (0) = 1.398$, $\widetilde W_2 (0) = 1.398$, $ H_1 (0) = 1.4$, $ H_2 (0) = 1.4$. This chaotic dependence corresponds to the Poincar\'{e} sections in Fig. \ref{fg10}$\textrm{c}$-\ref{fg10}$\textrm{d}$ shown at the bottom. } \label{fg11}
\end{figure}
The coordinate surface  $D=\textrm{const}$ is a sphere: $\theta $, the latitude: $\theta \in [0,\pi ]$, $\varphi $, the longitude: $\varphi \in [0,2\pi ]$ (see Fig. \ref{fg2}). Let us analyze the dependences of the coefficients $\alpha _{1} $, $\alpha _{2} $, $\widetilde{\alpha}_{H}^{(1)}$, $\widetilde{\alpha}_{H}^{(2)}$   on the effect of rotation and stratification. For simplicity assume that   $D_{1} =D_{2} $, which   corresponds to the fixed longitude value  $\varphi =\pi /4+\pi n$, where  $n=0,1,2...k$, $k$ being an integer. In this case the coefficients for vortex and magnetic perturbations are:
\begin{equation} \label{eq45}
 \alpha =\alpha_{1}^{(W)} =\alpha_{2}^{(W)} =$$
$$=f_{0}^{2}\sqrt{2}D\sin \theta \cdot \frac{4\left(D^2\sin^{2}\theta-2Ra-4\right)(2-Ra)+\frac{Ra}{2}\left((D^2\sin^2\theta-2Ra)^2+16\right)}{\left((D^2\sin^2\theta-2Ra)^2 +16 \right)^2} , \end{equation}
\begin{equation} \label{eq46}
\alpha_{H} =\widetilde{\alpha}_{H}^{(1)} =\widetilde{\alpha}_{H}^{(2)}=\frac{f_{0}^{2}\sqrt{2}}{2}\cdot\frac{D(2+Ra)\sin \theta }{(D^2\sin^2\theta-2Ra)^2+16}
\end{equation}
 As seen from these relations, at the poles  $(\theta =0,\; \theta =\pi )$ generation of vortex and magnetic perturbations is inefficient, since $\alpha ,\alpha _{H} \to 0$, i.e. large-scale instability occurs in the case when the vector of angular rotation velocity  $\vec{\Omega }$ deviates  from the axis  $Z$. In the case of a homogeneous medium $Ra=0$, where  the generation of large-scale vortex and magnetic disturbances is due to the action of an external small-scale non-spiral force and the Coriolis force \cite{47s}. The  coefficient $\alpha $ of vortex perturbations  for a rotating  stratified electroconductive fluid coincides with the analogous  coefficient $\alpha $ for a rotating stratified non-electroconductive fluid obtained in \cite{50s}. Therefore, the conclusions made in the said paper concerning  gain of vortex perturbations may be applied to the problem considered here. The dependence of the coefficient  $\alpha $ on the parameter of fluid stratification   (the Rayleigh number  $Ra$ ) at the fixed value of latitude  $\theta =\pi /2$ and  $D=2.5$ is presented in the left part of Fig. \ref{fg3}. As is seen, the  temperature stratification  ($Ra\ne 0$) may/can give rise  to an essential increase of the coefficient  $\alpha $ and, consequently, make generation of large-scale vortex perturbations  faster in comparison with that in a homogeneous medium.  Such an effect is especially explicit  at   $Ra\to 5$. With further rise of the Rayleigh numbers  the values of the coefficient  $\alpha $are diminishing. Now let us clarify the influence of the rotation of the medium on the  coefficient  $\alpha $. For this purpose we fix the value of the Rayleigh number  $Ra=5$ at  $\theta =\pi /2$. In this case the functional dependence  $\alpha (D)$ is presented in the right part of Fig. \ref{fg3}. One can see that at a certain value of the rotation parameter $D$ the coefficient  $\alpha $ reaches its maximum  $\alpha_{\max} $. With further rise of  $D$ the coefficient  $\alpha $ smoothly tends to zero, i.e. $\alpha $-effect is being suppressed by the  rotation of the medium. Now consider the dependence of the coefficient  $\alpha _{H} $ on the parameters of stratification $Ra$ and rotation $D$  at the latitude  $\theta =\pi /2$ . The dependence of the coefficient  $\alpha _{H} $ on the stratification parameter  (the Rayleigh number $Ra$ ) at the fixed  $\theta =\pi /2$ and  $D=2.5$ is shown in the left part of Fig. \ref{fg4}. Here we also see that the presence of temperature stratification ($Ra\ne 0$)  essentially increases the coefficient  $\alpha_{H} $, and, consequently, makes generation of large-scale perturbations faster than the one in a homogeneous medium. Magnetohydrodynamic  $\alpha $-effect (or $\alpha_{H}$-effect) also  increases at \guillemotleft slow\guillemotright rotation  up to the maximum value  $\alpha_{Hmax} $. Then with the rise of the parameter  $D$ the coefficient  $\alpha _{H} $ decreases, but its sign does not change. The analysis of  the dependence  $\alpha _{H} (D)$ shows that at \guillemotleft fast\guillemotright  rotation of the medium MHD  $\alpha $-effect is also  suppressed   (see the right part of Fig. \ref{fg5}). Similar phenomenon, i.e. suppression of $\alpha $-effect by the rotation of a  turbulent medium is shown in  \cite{51s}.

Shown in Fig. \ref{fg5} is   the graph which represents   the  influence of  rotation and stratification on  $\alpha $ and $\alpha_{H} $-effects in the plane  $(D,Ra)$. Here the regions of instability  $\alpha >0,\alpha_{H} >0$ are marked with grey color.  Having fixed the values of the rotation and stratification parameters  $D$ and  $Ra$ for the latitudinal angles  $\theta =\pi /2$ we will plot the dependences of the growth rate of the vortex  $\Gamma_{1} $ and magnetic  $\Gamma_{2} $ perturbations on the wave numbers  $K$. These graphs have the form typical of  $\alpha $-effect (see Fig. \ref{fg6}).

\begin{figure}
\centering
 \includegraphics[height=5.5 cm, width=9 cm]{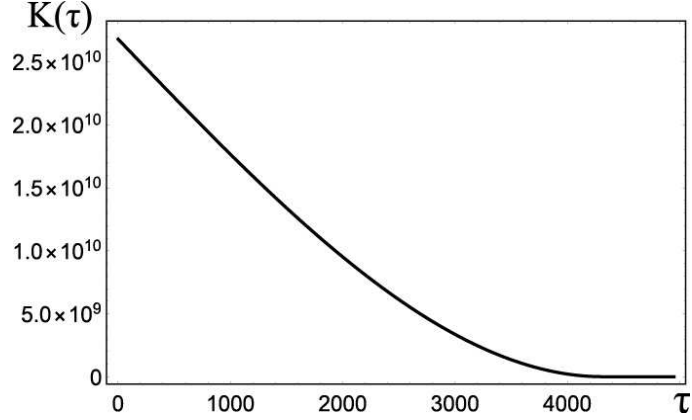}\\
\caption{\small Показан график зависимости автокорреляционной функции , от времени для траектории с начальными условиями хаотическое движение). The plot of the dependence of the autocorrelation function   ${K}_{\widetilde W_1 \widetilde W_1}$   on time  $\tau$  for a trajectory with initial conditions  $\widetilde W_1 (0) = 1.398$, $\widetilde W_2 (0) = 1.398$, $ H_1 (0) = 1.4$, $ H_2 (0) = 1.4$ (chaotic motion).}\label{fg12}
\end{figure}

\subsection{Analysis of dispersion equation (\ref{eq41})  in the presence of  external magnetic field  $\overline{B}_{1,2}\neq 0$ }

Let us study  Eq.(\ref{eq41}) at $\overline{B}_{1,2}\neq 0$. In this case it is transformed into the biquadratic equation :
\begin{equation} \label{eq47} (K^2-i\omega)^4-b(K^2-i\omega)^2+a=0,  \end{equation}
 where
\[  b=K^2\left(\alpha_1^{(W)}\alpha_2^{(W)}+\alpha_2^{(H)}\beta_W^{(1)}+\widetilde{\alpha}_H^{(1)}\widetilde{\alpha}_H^{(2)}+\alpha_1^{(H)}\beta_W^{(2)} \right)=K^2\widetilde{b} ,  \]
\[a=K^4\left(\widetilde{\alpha}_H^{(2)}\alpha_2^{(W)}\left(\alpha_1^{(W)}\widetilde{\alpha}_H^{(1)}+\alpha_1^{(H)}\beta_W^{(1)}\right)+\alpha_2^{(H)}\beta_W^{(2)}\left(\alpha_1^{(H)}\beta_W^{(1)}+\widetilde{\alpha}_H^{(1)}\alpha_1^{(W)}\right)\right)=K^4\widetilde{a}. \]
The solution of Eq.(\ref{eq47}) has the form:
\begin{equation} \label{eq48} K^2-i\omega=\pm K \sqrt{\frac{\widetilde{b}}{2}\pm\frac{1}{2}\sqrt{\widetilde{b}^2-4 \widetilde{a}}}
\end{equation}
Since we are interested in increasing   solutions,  easily find the increment of large-scale instability from Eq.(\ref{eq48}):
\begin{equation} \label{eq49} \Gamma= \textrm{Im}\,\omega=\Lambda K-K^2,
\end{equation}
where  $\Lambda=\sqrt{\frac{\widetilde{b}}{2}\pm\frac{1}{2}\sqrt{\widetilde{b}^2-4 \widetilde{a}}}$ is the  coefficient for vortex and magnetic perturbations which has a positive value at  $\widetilde{b}^{2} >4\widetilde{a}$. The maximum increment of instability  $\Gamma_{max} =\Lambda ^{2} /4$  corresponds to the wave numbers  $K_{max} =\Lambda /2$. Shown in the right part of Fig. \ref{fg9}  is  the dependence of the increment $\Gamma $  of large-scale instability  (\ref{eq49}) on the wave numbers  $K$ for fixed values of the  inclination angle   $\theta =\pi /2$, the amplitude of the extenal force  $f_{0} =10$ and the dimensionless parameters  $D=2.5$, $Ra=5$, $Q=10$, $\overline{B}=0.5$. The form of this graph is analogous to that of  graph  $\alpha $-effect  (see Fig. \ref{fg6} ).

As  in the previous Section, it is convenient to replace the Cartesian projections  $D_{1,2} $ and $\overline{B}_{1,2} $  by their projections in the spherical coordinate system   (see Fig.7). Now let us analyze the dependences of the gain coefficient  $\Lambda $ on the effects of rotation $(D)$, stratification  $(Ra)$ and the external magnetic field  $(\overline{B})$. For simplicity assume that  $D_{1} =D_{2} $ and $\overline{B}_{1} =\overline{B}_{2} $, and this  corresponds to  the fixed value of the angle  $\varphi_{\Omega } \approx \varphi _{B} =\pi /4+\pi n$, where  $n=0,1,2...k$, $k$ is an integer. In such a case the coefficients  $\alpha_{1,2}^{(W)},\alpha_{1,2}^{(H)},\widetilde{\alpha}_{H}^{(1,2)},\beta_{W}^{(1,2)} $ which enter into the gain coefficient  $\Lambda$ for the  vortex and magnetic perturbations will acquire the form:
\begin{equation} \label{eq50}
 A =\alpha_{1}^{(W)} =\alpha_{2}^{(W)} = f_{0}^{2}\sqrt{2}D\sin \theta \left[ \frac{\left(D^2\sin^{2}\theta-2Ra-4\right)(2-Ra)(4-Q\overline{B}^2)}{\left((D^2\sin^2\theta-2Ra)^2 +16 \right)^2}+\right.$$
\begin{figure}
\centering
 \includegraphics[height=4.5 cm, width=9 cm]{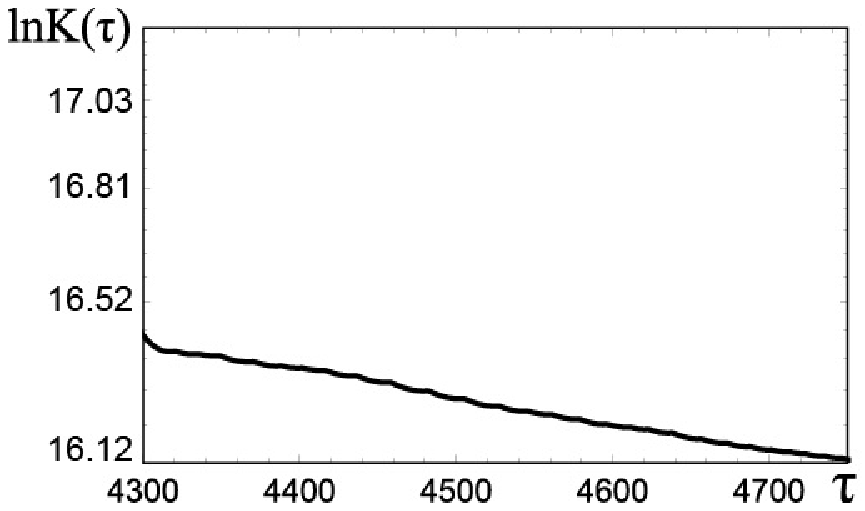}\\
	 \caption{\small  A rectilinear dependence of the autocorrelation function  ${K}_{\widetilde W_1 \widetilde W_1}$   in logarithmic scales  from the time interval  $\tau$   for strongly chaotic motion.}\label{fg13}
\end{figure}
$$\left. +\frac{Ra(2-Q\overline{B}^2)+Q\overline{B}^2}{4\left((D^2\sin^2\theta-2Ra)^2 +16 \right)} \right], \end{equation}
\begin{equation} \label{eq51}
B_{H}=\alpha_{1}^{(H)}=\alpha_{2}^{(H)}= \frac{f_0^2}{8}D Q\overline{B}\sin \theta  \times$$
$$ \times \left[\frac{4(4-Q\overline{B}^2)(2-Ra)\left(2(D^2\sin^{2}\theta-2Ra)+(Ra+1)^2+7\right)}{\left(16+(D^2\sin^{2}\theta-2Ra)^2\right)^2}+\frac{Q\overline{B}^2-4(Ra-1)}{16+(D^2\sin^{2}\theta-2Ra)^2}\right],
\end{equation}
\begin{equation} \label{eq52}
 A_H=\widetilde{\alpha}_{H}^{(1)}= \widetilde{\alpha}_{H}^{(2)}= \frac{f_{0}^{2}\sqrt{2}}{2}\cdot\frac{D\left(2+Ra-\frac{Q\overline{B}}{2}+\frac{\sqrt{2}}{2}(2+Ra)\overline{B}\right)\sin \theta }{(D^2\sin^2\theta-2Ra)^2+16}-$$
$$-f_{0}^{2}\cdot\frac{2D\overline{B}(2+Ra)\sin \theta  }{\left((D^2\sin^2\theta-2Ra)^2+16\right)^2}\cdot\left((2(D^2\sin^{2}\theta-2Ra)+(Ra+1)^2+7\right),
\end{equation}
\begin{equation} \label{eq53}
 B_W= \beta_{W}^{(1)}=\beta_{W}^{(2)}= f_0^2\cdot \frac{4D\overline{B}(2+Ra)\sin \theta}{\left((D^2\sin^2\theta-2Ra)^2 +16 \right)^2}\cdot\left(D^2\sin^{2}\theta-2Ra-4\right)-$$
$$-\frac{f_0^2}{4}\cdot\frac{D\overline{B}Ra\sin \theta }{(D^2\sin^2\theta-2Ra)^2 +16 }.
\end{equation}
Fig. 8a shows the dependence of the  coefficient  $\Lambda $ on the Rayleigh number  $Ra$ at the fixed latitude values  $\theta =\pi /2$ and the dimensionless numbers  $D=2.5$, $Q=10$, $\overline{B}=0.2$. As before, assume that the amplitude of the external force $f_{0} =10$. In Fig. \ref{fg8}$\textrm{a}$  the value of the coefficient  $\Lambda $ at  $Ra=0$ (homogeneous medium) is denoted by dashes. As one can see, with the increase of the Rayleigh number  $Ra\to 5$ the coefficient   $\Lambda $ considerably exceeds its value for a homogeneous medium, i.e. reaches its peak magnitude. Further rise of the parameter $Ra$ leads to a drop of the value of  $\Lambda $ and, consequently, to less intense   generation  of the magneto-vortex perturbations. Let us fix the Rayleigh number e.g. on the level of $Ra=5$ and find the dependence of the coefficient  $\Lambda $ on the rotation parameter $D$ at the external magnetic field  $\overline{B}=0.2$ and $Q=10$. The graph presented  in Fig. \ref{fg8}$\textrm{b}$ shows the dependence  $\Lambda (D)$. Here we observe the increase  of  $\Lambda $ to a certain maximum value   $\Lambda _{max} $ for  $D\approx 3$. With the rise of the parameter  $D$ the value of  $\Lambda $ diminishes, and generation of magneto-vortex perturbations becomes less efficient.  \guillemotleft Fast\guillemotright  rotation of the medium also suppresses the considered  $\Lambda $-effect. To clarify the influence of the homogeneous magnetic field  $\overline{B}$ on  $\Lambda $-effect, let us fix the following parameters: $D=2.5$, $Ra=5$, $Q=10$. Fig.  \ref{fg8}$\textrm{c}$  presents the dependence  $\Lambda (\overline{B})$. The upper dashed line denotes the level  $\Lambda _{0} $ corresponding to the case when the external magnetic field is absent: $\overline{B}=0$. As seen from this figure, the growth of the magnetic field value provides intensification of the magneto-vortex
\begin{figure}
      \centering
    \includegraphics[width=18 cm, height=18 cm]{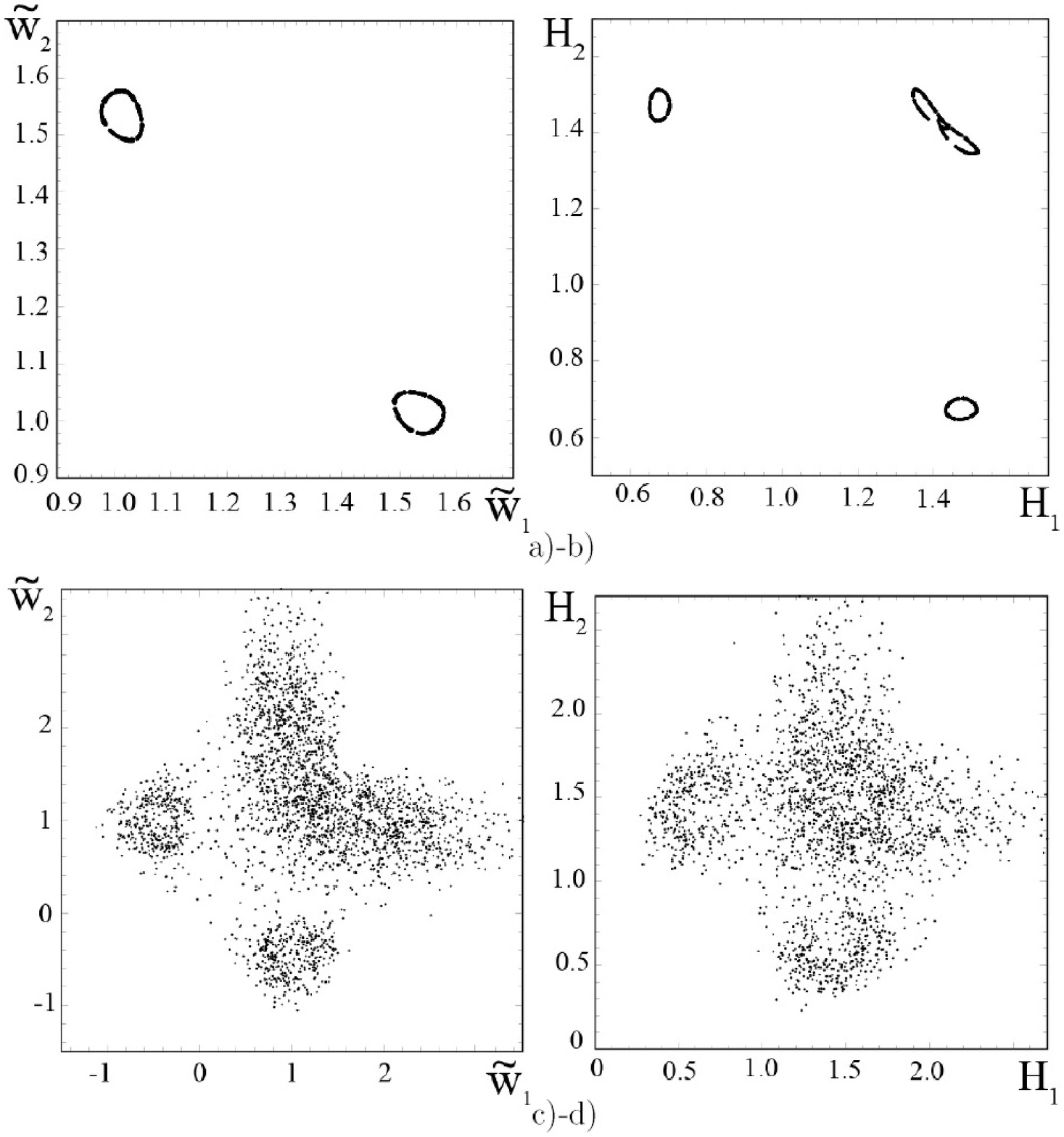}  \\
    \caption{\small In  the figures $\textrm{a)}$  and $\textrm{b)}$ shown the Poincar\'{e} sections for a trajectory with initial conditions  $\widetilde W_1 (0) = 1.31$, $\widetilde W_2 (0) = 1.31$, $ H_1 (0) = 1.4$, $ H_2 (0) = 1.4$. This is a regular type of trajectory, which is wound on the tori. The figures  $\textrm{c)}$  and $\textrm{d)}$ correspond to Poincar\'{e} sections for a trajectory with initial conditions  $\widetilde W_1 (0) = 1.31$, $\widetilde W_2 (0) = 1.31$, $ H_1 (0) = 1.8$, $ H_2 (0) = 1.8$. These pictures show stochastic layers, to which the corresponding chaotic trajectory belongs. The calculations were carried out for the case  $\overline{B}=0.1$.  } \label{fg14}
\end{figure}
pertubations up to a certain level  $\Lambda_{max} \sim 100$. The lower dashed line in Fig. 8c shows the minimum level of the  coefficient  $\Lambda_{min} \approx 9.63$ which corresponds to the value of the magnetic field   $\overline{B}\approx 0.72$ for the given parameters  $D$, $Ra$ and $Q$. From here it follows that \guillemotleft strong\guillemotright external magnetic field suppresses the considered  $\Lambda $-effect. For the fixed parameters $D=2.5$, $Ra=5$, $Q=10$ and $\overline{B}=0.5$ one can find the dependence  of the  coefficient  $\Lambda $ on the angle  $\theta $ of deviation  for  the vector of  the angular rotation velocity   $\vec{\Omega }$ from the vertical direction  $OZ$. This dependence  $\Lambda (\theta )$ is presented in the left part of Fig. \ref{fg9}. As is seen, generation of magneto-vortex perturbations does not occur  $(\Lambda \to 0)$ at  $\theta \to 0$ and $\theta \to \pi $ (the pole) , whereas at $\theta \to \pi /2$ (the equator) it is most effective.

\section{Nonlinear stationary structures}

With the growth of the amplitude of  the perturbations $W_{1,2} $ and  $H_{1,2} $  due to development of large-scale instability, the linear theory considered in the previous Section becomes inapplicable. The evolution of these perturbations will be described by the nonlinear system of (\ref{eq27})-(\ref{eq30}).  Now study the regime of instability saturation which leads to the formation of nonlinear stationary structures. To describe  such structures, let us put  $\partial_{T} W_{1} =\partial_{T} W_{2} =\partial_{T} H_{1} =\partial_{T} H_{2}=0$ in the system of Eqs. \eqref{eq27}-\eqref{eq30}, and then integrate these equation over to Z:
\begin{equation} \label{eq54}
  \frac{d{\widetilde{W}}_1}{dZ}=- f_{0}^{2} \cdot \frac{\sqrt{2}D q_{2} Q_{2}}{16\widetilde{W}_2^{2}q_2^2\widetilde{Q}_2^2 +\left[D^{2}+2(1-\widetilde{W}_{2}^2)+2\mu_{2}\right]^{2} +4\xi_2}+C_1
\end{equation}
\begin{equation} \label{eq55}
  \frac{d{\widetilde{W}}_2}{dZ}= f_{0}^{2} \cdot \frac{\sqrt{2}D q_{1} Q_{1}}{16\widetilde{W}_1^{2}q_1^2\widetilde{Q}_1^2 +\left[D^{2}+2(1-\widetilde{W}_{1}^2)+2\mu_{1}\right]^{2} +4\xi_1}+C_2
\end{equation}
\begin{equation} \label{eq56}
  \frac{dH_1}{dZ}= f_{0}^{2}\cdot \frac{\sqrt{2}D\widetilde{W}_{2}\widetilde{Q}_{2}(2H_2+ \overline{B}\sqrt{2} )}{(1+\widetilde{W}_{2}^{2})\left[16\widetilde{W}_2^{2}q_2^2\widetilde{Q}_2^2 +\left[D^{2}+2(1-\widetilde{W}_{2}^2)+2\mu_{2}\right]^{2} +4\xi_2 \right] } +C_3
\end{equation}
\begin{equation} \label{eq57}
  \frac{dH_2}{dZ}= -f_{0}^{2}\cdot \frac{\sqrt{2}D\widetilde{W}_{1}\widetilde{Q}_{1}(2H_1+ \overline{B}\sqrt{2} )}{(1+\widetilde{W}_{1}^{2})\left[16\widetilde{W}_1^{2}q_1^2\widetilde{Q}_1^2 +\left[D^{2}+2(1-\widetilde{W}_{1}^2)+2\mu_{1}\right]^{2} +4\xi_1 \right] } +C_4
\end{equation}
Here $\widetilde{W}_{1} =1-W_{1} $, $\widetilde{W}_{2} =1-W_{2} $; $C_{1} $, $C_{2} $, $C_{3} $ and  $C_{4} $ are arbitrary integration constants. While obtaining Eqs. \eqref{eq54}-\eqref{eq57} we put the Prandtl numbers  $Pr=Pm=1$, and replaced the Cartesian projections for  $D_{1,2} $ and  $\overline{B}_{1,2} $ in the coefficients $\alpha_{\left(1,2\right)} $, $\alpha _{H}^{\left(1,2\right)} $  by their projections in the spherical coordinate system  (see Fig. \ref{fg7} ). For simplicity we fixed  the values of the angles  $\varphi_{\Omega } =\varphi_{B} =\pi /4$ and  $\theta =\pi /2$. In this case the expressions for  $q_{1,2} $, $Q_{1,2} $, $\widetilde{Q}_{1,2} $, $\mu _{1,2} $, $\sigma _{1,2} $, $\chi _{1,2} $, $\xi _{1,2} $ will be also simplified:
\[q_{1,2} =1 + \frac{Q H_{1,2}(2H_{1,2}+\overline{B}\sqrt{2})}{2(1+\widetilde{W}_{1,2}^{2})}-\frac{Ra}{1+\widetilde{W}_{1,2}^2}, \quad Q_{1,2} =1-\frac{Q(2H_{1,2}+\overline{B}\sqrt{2})^2}{4(1+\widetilde{W}_{1,2}^{2})},\]
\[\mu _{1,2} = QH_{1,2}(2H_{1,2}+\overline{B}\sqrt{2})+Q^2H_{1,2}^2(2H_{1,2}+\overline{B}\sqrt{2})^2 \cdot\frac{1-\widetilde{W}_{1,2}^{2}}{4(1+\widetilde{W}_{1,2}^{2})}-$$
$$-Ra\cdot\frac{1+\widetilde{W}_{1,2}^2+QH_{1,2}(2H_{1,2}+\overline{B}\sqrt{2})\cdot \frac{1-\widetilde{W}_{1,2}^{2}}{1+\widetilde{W}_{1,2}^{2}}}{1+ \widetilde{W}_{1,2}^2}, \]
\begin{figure}
\centering
 \includegraphics[width=18 cm, height=18 cm]{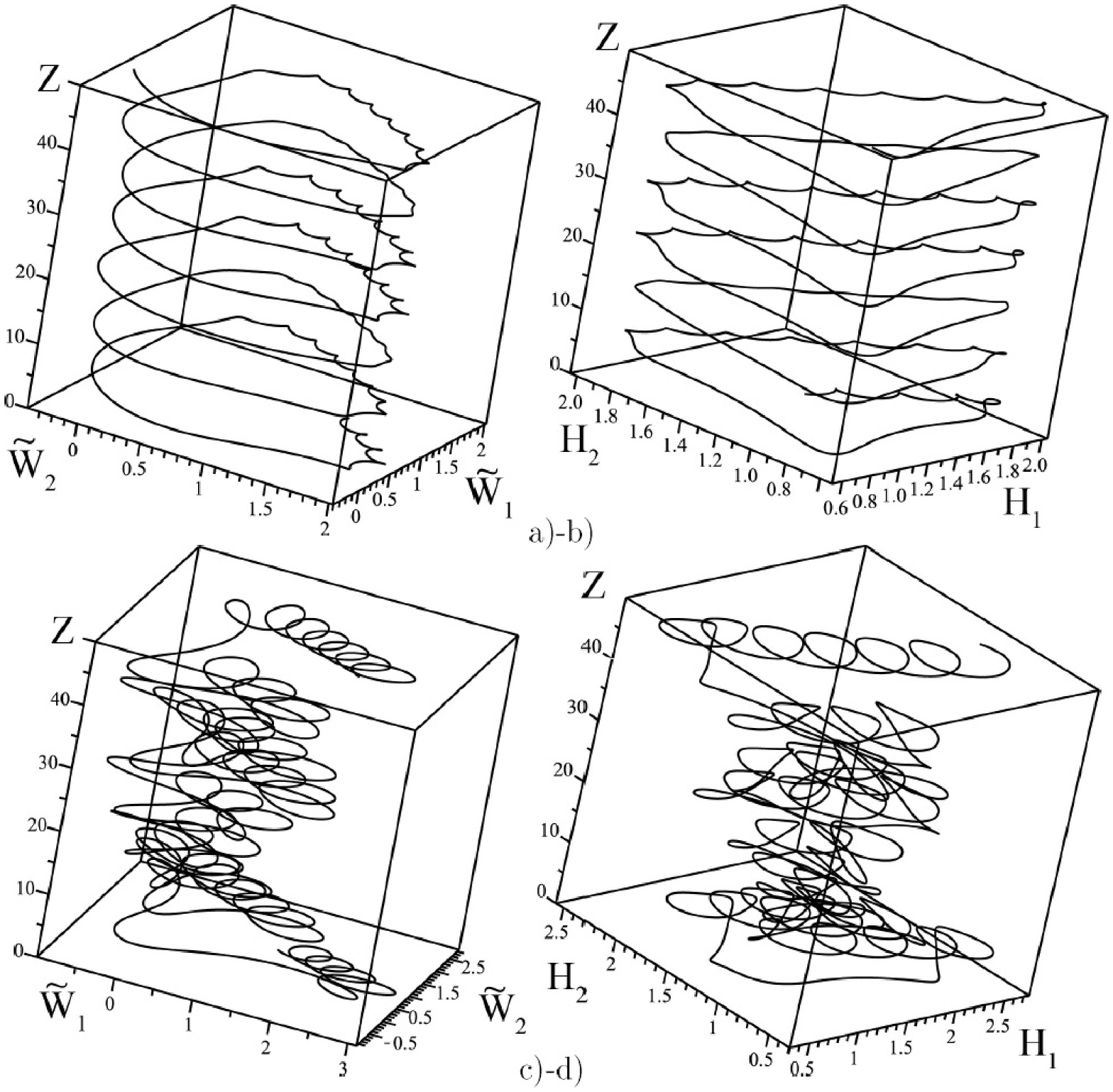} \\
  \caption{\small  The upper part  ($\textrm{a)}$, $\textrm{b)}$) shows  the dependence of the velocity and magnetic field on the height  $Z$   for the numerical solution of equation system (\ref{eq54})-(\ref{eq57}) with the initial conditions $\widetilde W_1 (0) = 1.31$, $\widetilde W_2 (0) = 1.31$, $ H_1 (0) = 1.4$, $ H_2 (0) = 1.4$.  This dependence corresponds to regular motions of the Poincar\'{e} section shown on top of  Fig. \ref{fg14}$\textrm{a}$-\ref{fg14}$\textrm{b}$. Below ($\textrm{c)}$, $\textrm{d)}$) a similar  dependence is shown  for the numerical solution of equation system (\ref{eq54})-(\ref{eq57})  with the initial conditions: $\widetilde W_1 (0) = 1.31$, $\widetilde W_2 (0) = 1.31$, $ H_1 (0) = 1.8$, $ H_2 (0) = 1.8$. This chaotic dependence corresponds to the Poincar\'{e} sections in Fig. \ref{fg14}$\textrm{c}$-\ref{fg14}$\textrm{d}$ shown at the bottom. } \label{fg15}
\end{figure}	
\[\xi_{1,2}=2\Xi_{1,2}+2\widetilde{W}_{1,2}^2\Pi_{1,2}-2\widetilde{W}_{1,2}^2(1-\widetilde{Q}_{1,2}^2)\Pi_{1,2}-2(1-{q}_{1,2}^2)\Xi_{1,2}+ \Xi_{1,2}\Pi_{1,2}+$$
$$+\chi_{1,2}\widetilde{W}_{1,2}^2+\chi_{1,2}(1+\sigma_{1,2}), \]
\[\Xi_{1,2}=-\frac{4\widetilde{W}_{1,2}^2\widetilde{Q}_{1,2}Ra}{1+\widetilde{W}_{1,2}^2}+\frac{2\widetilde{W}_{1,2}^2 Ra^2}{\left(1+\widetilde{W}_{1,2}^2\right)^2}+ Ra\cdot\frac{1+\widetilde{W}_{1,2}^2+QH_{1,2}(2H_{1,2}+\overline{B}\sqrt{2})\cdot \frac{1-\widetilde{W}_{1,2}^{2}}{1+\widetilde{W}_{1,2}^{2}}}{1+ \widetilde{W}_{1,2}^2},  \]
\[\Pi_{1,2}=\frac{4q_{1,2}Ra}{1+\widetilde{W}_{1,2}^2}+\frac{2Ra^2}{\left(1+\widetilde{W}_{1,2}^2\right)^2}-Ra\cdot\frac{1+\widetilde{W}_{1,2}^2+QH_{1,2}(2H_{1,2}+\overline{B}\sqrt{2})\cdot \frac{1-\widetilde{W}_{1,2}^{2}}{1+\widetilde{W}_{1,2}^{2}}}{1+ \widetilde{W}_{1,2}^2}, \]
\[\sigma_{1,2}=\frac{QH_{1,2}(2H_{1,2}+\overline{B}\sqrt{2})}{4(1+\widetilde{W}_{1,2}^{2})}\cdot\left[4(1+\widetilde{W}_{1,2}^{2})+QH_{1,2}(2H_{1,2}+\overline{B}\sqrt{2})\right], \]
\[ \chi_{1,2}=\frac{Ra}{1+\widetilde{W}_{1,2}^{2}}\cdot \left[Ra-\left(2(1-\widetilde{W}_{1,2}^{2})+QH_{1,2}(2H_{1,2}+\overline{B}\sqrt{2})\right)\right], \]
\[\widetilde{Q}_{1,2}=1-\frac{QH_{1,2}(2H_{1,2}+\overline{B}\sqrt{2})}{2(1+\widetilde{W}_{1,2}^{2})}+\frac{Ra}{1+\widetilde{W}_{1,2}^2}. \]
Eqs.(\ref{eq54})-(\ref{eq57}) constitute the nonlinear dynamic system in 4-dimensional  phase space in which phase flow divergence is (equal to) zero. Therefore, the system of Eqs. (\ref{eq54})-(\ref{eq57}) is conservative. Search for the Hamiltonian of this system is a very difficult task, as the integration is complicated by the dependence of the nonlinear coefficients  $\alpha_{(1,2)} $, $\alpha_{H}^{(1,2)} $ on the fields  $\vec{W}$, $\vec{H}$, that takes it beyond the class of elementary functions.  A complete qualitative analysis of this system is extremely complicated due to a high dimension of the phase space, as well as to a large number of the parameters included in the system. Proceeding from general ideas, it is to be expected that this system of conservative equations may contain  structures of resonance and non-resonance tori in the phase space and, consequently, chaotic stationary structures of hydrodynamic and magnetic fields. The considered system of nonlinear Eqs. (\ref{eq54})-(\ref{eq57} can be studied e.g. using the Poincar\'{e} cross-section method.

\section{Stationary chaotic structures  in the absence of external magnetic field $\overline{B}=0$.}

Using the standard Mathematica programs, let us build the Poincar\'{e} cross-sections for the trajectories in the phase space for the case of rotating electroconductive fluid stratified with respect to temperature  $(Ra\ne 0)$ not taking into account the external magnetic field  $\overline{B}=0$. All the numerical calculations will be performed for the following parameters: $f_{0} =10$, $D=2$, $Q=1$, $Ra=0.1$ and the constants $C_{1} =1,C_{2} =-1$, $C_{3} =-0.5,C_{4} =0.5$. For the initial conditions  $\widetilde W_1 (0) = 1.25$, $\widetilde W_2 (0) = 1.25$, $ H_1 (0) = 1.4$, $ H_2 (0) = 1.4$   the Poincar\'{e} cross-sections presented in Fig. \ref{fg10}$\textrm{a}$-\ref{fg10}$\textrm{b}$, demonstrate regular trajectories for the velocity and  magnetic fields. With the rise  of the initial perturbation velocity $\widetilde W_1 (0) = 1.398$, $\widetilde W_2 (0) = 1.398$, $ H_1 (0) = 1.4$, $ H_2 (0) = 1.4$ the regular trajectories become chaotic. They correspond to the  Poincar\'{e} cross-sections shown in Fig. \ref{fg10}$\textrm{c}$-\ref{fg10}$\textrm{d}$. Fig. \ref{fg11}$\textrm{a}$-\ref{fg11}$\textrm{d}$ present the dependence of the stationary large-scale fields on the altitude $Z$. The latter was obtained numerically for the initial conditions corresponding to the  Poincar\'{e} cross-sections presented in Fig.  \ref{fg10}$\textrm{a}$-\ref{fg10}$\textrm{d}$. These figures also show  the emergence  of stationary chaotic solutions for   magnetic and vortex fields.  To prove the emergence of chaotic regime of stationary large-scale fields, we will  use not only the Poincar\'{e} cross-sections, but also the notion of autocorrelated function.  As is known (see e.g. \cite{52s} ), the autocorrelated function  $K(\tau )$ is the value which characterizes the intensity of chaos.
\begin{figure}
\centering
 \includegraphics[height=6 cm, width=10 cm]{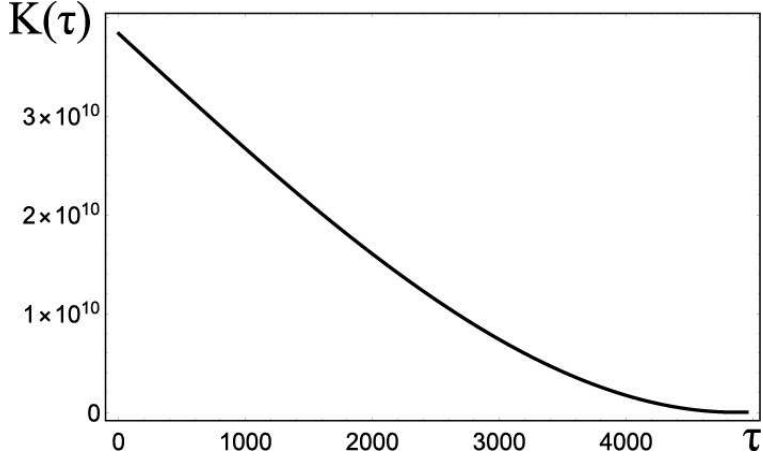}\\
\caption{\small The plot of the dependence of the autocorrelation function   ${K}_{\widetilde W_1 \widetilde W_1}$   on time   $\tau$    for a trajectory with initial conditions  $\widetilde W_1 (0) = 1.31$, $\widetilde W_2 (0) = 1.31$, $ H_1 (0) = 1.8$, $ H_2 (0) = 1.8$ (chaotic motion) on condition $\overline{B}=0.1$.}\label{fg16}
\end{figure}	
It is defined as averaging of the product of random functions $P(t)$  and $P(t+\tau )$ at the moment of time  $t$ and $t+\tau $, respectively, over \guillemotleft long \guillemotright   interval of time $\Delta t$ : $K(\tau ) = \mathop {\lim }\limits_{\Delta t \to \infty } \frac{1}{{\Delta t}}\int\limits_0^{\Delta t} {P(t)P(t + \tau )dt} $. In the case we consider the coordinate  $Z$ acts as the time $t$, whereas the product  $P(t)P(t+\tau )$ consists of 16 components:
\[ P(t)P(t + \tau ) = \left[ {\begin{array}{*{20}c}
   {\widetilde W_1 (t)}  \\
   {\widetilde W_2 (t)}  \\
   {H_1 (t)}  \\
   {H_2 (t)}  \\
\end{array}} \right]\left[ {\begin{array}{*{20}c}
   {\widetilde W_1 (t + \tau )} & {\widetilde W_2 (t + \tau )} & {H_1 (t + \tau )} & {H_2 (t + \tau )}  \\
\end{array}} \right] . \]
The plot of the dependence of the autocorrelated function for the component  $K_{\widetilde{W}_{1} \widetilde{W}_{1} } $ on the time  $\tau $ is presented in Fig. \ref{fg12}. The case of chaotic motion corresponds to the section of the trajectory with  the exponential  decay of the function  $K_{\widetilde{W}_{1} \widetilde{W}_{1} } $. It is evident that the said section in the logarithmic scale of the autocorrelated function $K_{\widetilde{W}_{1} \widetilde{W}_{1} } $ is approximated by a straight line  (see Fig. \ref{fg13}). The data presented in Fig. \ref{fg13} make it possible to determine the characteristic correlation time  $\tau_{cor} \approx 1324$ of the stationary random process  $P_{\widetilde{W}_{1} } $. If to take into account the introduced to the above definition of \guillemotleft time\guillemotright, it becomes clear that we have found the estimated value of the altitude  $Z_{cor} \approx 1324$ corresponding to the onset of chaotic motion of the large-scale fields. Shown in Fig. \ref{fg11}$\textrm{c}$-\ref{fg11}$\textrm{d}$ are the chaotic solutions for the velocity and magnetic fields of height  $Z\approx 90$ which considerably less than  $Z_{cor} $. However, even in this case one can see the start of the intricate trajectory for the large-scale fields  at  the increase of the altitude  $Z$. Therefore, such trajectories cannot be plotted. Thus,   with the increase of the altitude  $Z$ up to a critical value  $Z_{cor} $, quasi-periodic motion of the stationary large-scale fileds becomes chaotic.

\section{Stationary chaotic structures in the presence of external magnetic field $\overline{B}\neq 0$. }

Now let us  build the Poincar\'{e} cross-sections of the trajectories in the phase space for the nonlinear system of Eqs. (\ref{eq54})-(\ref{eq57}) taking into account external homogeneous magnetic field, by means of the standard Matematica programs. For this purpose all the numerical
\begin{figure}
\centering
 \includegraphics[height=6 cm, width=10 cm]{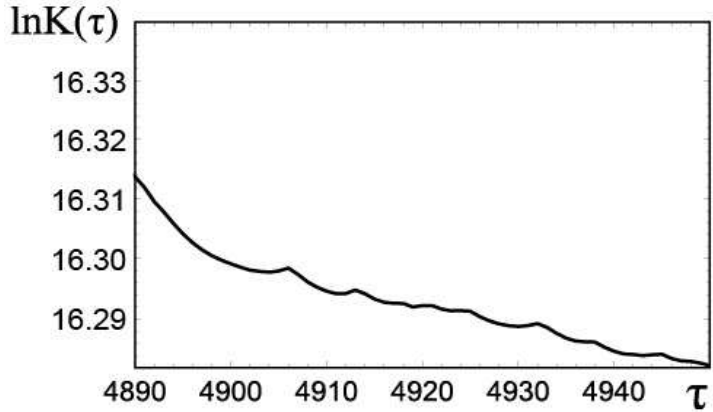}\\
	 \caption{\small A rectilinear dependence of the autocorrelation function   ${K}_{\widetilde W_1 \widetilde W_1}$  in logarithmic scales  from the time interval  $\tau$   for strongly chaotic motion on condition  $\overline{B}=0.1$.}\label{fg17}
\end{figure}
calculations will be carried out for the following dimensionless parameters: $f_{0} =10$, $D=2$, $Q=1$, $Ra=0.1$, $\overline{B}=0.1$ and the constants  $C_{1} =1,C_{2} =-1$, $C_{3} =-0.5,C_{4} =0.5$. Shown in Fig. \ref{fg14}$\textrm{a}$-\ref{fg14}$\textrm{b}$  are the regular trajectories of the velocity and magnetic fields built at the numerical solutions of Eqs. (\ref{eq54})-(\ref{eq57})  with the following initial conditions : $\widetilde W_1 (0) = 1.31$, $\widetilde W_2 (0) = 1.31$, $ H_1 (0) = 1.4$, $ H_2 (0) = 1.4$. This type of trajectories corresponds to quasi-periodic character of motion  for  large-scale perturbations of the velocity  $(\widetilde{W}_{1,2} )$ and  magnetic fields  $(H_{1,2} )$. By increasing only the amplitudes of the initial values of perturbations for the magnetic field $\widetilde W_1 (0) = 1.31$, $\widetilde W_2 (0) = 1.31$, $ H_1 (0) = 1.8$, $ H_2 (0) = 1.8$   we find that the quasi-periodic motion  transforms into chaotic. This case demonstrates the Poincare cross-sections shown in Fig. \ref{fg14}$\textrm{c}$-\ref{fg14}$\textrm{d}$. Using the initial data for the regular ($\widetilde W_1 (0) = 1.31$, $\widetilde W_2 (0) = 1.31$, $ H_1 (0) = 1.4$, $ H_2 (0) = 1.4$)  and chaotic  ($\widetilde W_1 (0) = 1.31$, $\widetilde W_2 (0) = 1.31$, $ H_1 (0) = 1.8$, $ H_2 (0) = 1.8$) trajectories, one can numerically build  the dependence of the stationary large-scale fields on the altitude  $Z$ (see Fig. \ref{fg15}$\textrm{a}$-\ref{fg15}$\textrm{d}$). The emergence of stationary chaotic solutions for the magnetic and vortex fields is also observed in Fig. \ref{fg15}$\textrm{a}$-\ref{fg15}$\textrm{d}$. To confirm the onset of chaotic regime of the stationary large-scale fields  we will plot the dependence of the autocorrelated function for the component  $K_{\widetilde{W}_{1} \widetilde{W}_{1} } $ on the time  $\tau $ (see Fig. \ref{fg16} ). The trajectories of  chaotic motion  correspond to the  to the section of the graph with the exponential decay of the function  $K_{\widetilde{W}_{1} \widetilde{W}_{1} } $ in Fig. \ref{fg16}. In the logarithmic scale of the autocorrelated function $K_{\widetilde{W}_{1} \widetilde{W}_{1} } $ this section is approximated by a  straight line (see Fig. \ref{fg17} ). Using this graph it is easy to find the estimated value of the characteristic correlation time for a stationary random process: $\tau_{cor} \approx 2000$. The obtained value of the  correlation time corresponds to  the altitude  $Z_{cor} \approx 2000$, above which there arise strongly chaotic stationary structures of the large-scale fields. Shown in Fig. \ref{fg15}$\textrm{c}$-\ref{fg15}$\textrm{d}$  are chaotic solutions for the velocity and magnetic fields at the altitude  $Z\approx 50$ which is considerably less than $Z_{cor} $. It is evident that for  $Z$ tending to a critical value  $Z_{cor} $ the motion trajectories  become more intricate and, finally, completely chaotic.

\section{Conclusion  }

There is obtained the closed system of nonlinear equations for vortex and magnetic large-scale perturbations (magneto-vortex dynamo) in an obliquely rotating stratified electroconductive fluid in external uniformly   magnetic field.  At the initial stage small amplitudes of large-scale perturbations increase due to the average  helicity $\overline{\vec{v}_{0} rot\vec{v}_{0} }\ne 0$ of small-scale motion in a rotating stratified electroconductive  fluid excited by the external non-helical force  $\vec{F}_{0} rot\vec{F}_{0} =0$. The mechanism of the amplification of the large-scale perturbations is bound up with development of large-scale instability of  $\alpha $-effect type. Thereat, in the absence of  external magnetic field  $(\overline{\vec{B}}=0)$ the linear equations of magneto-vortex dynamo are split into two subsystems: vortex and magnetic ones.  In this case the large-scale vortex and magnetic perturbations are generated owing to development of instability such as HD  $\alpha $-effect and MHD $\alpha $-effect, respectively. Both types of instability occur when the vector of angular rotation velocity  $\vec{\Omega }$ is deflected from the vertical axis  $OZ$. Unlike the case of a homogeneous medium \cite{46s}-\cite{47s} , the combined effects of rotarion and stratification of the medium  (at heating from below)  give rise to an essential amplification of the large-scale perturbations. Such a phenomenon becomes  especially noticeable at the parameters of the medium  $D\to 3$ and $Ra\to 5$ (see Fig. \ref{fg3}). In this case  there arises the regime of maximal  generation of the small-scale helical motion caused by the action of the Coriolis force and inhomogeneity  of the medium with respect to the temperature.  In the presence of the external magnetic field  $(\overline{\vec{B}}\ne 0)$ the  evolution of the vortex and magnetic perturbations is characterized by a positive feedback  due to which the rates of the growth of the vortex and magnetic large-scale perturbations coincide. Thereat, \guillemotleft weak\guillemotright external magnetic field favours generation of the said    perturbations, whereas \guillemotleft strong\guillemotright field suppresses them (see Fig. 8c). Generation of the large-scale vortex and magnetic perturbations also depends on the angle of deflection of the vector of angular rotation velocity  $\vec{\Omega }$. It is minimal at  $\theta \to 0$ or  $\theta \to \pi $ (nearby the poles) and maximal at  $\theta \to \pi /2$ (nearby the equator) (see the left part of Fig. 9).  The performed analysis of the influence of rotation on the growth of the vortex and magnetic perturbations shows that at \guillemotleft fast\guillemotright rotation they are being suppressed. With the rise of  the perturbation amplitude   the instability is stabilized and then becomes stationary. Under such conditions there arise nonlinear stationary vortex and magnetic structures. The dynamical system of equations describing  these structures is Hamiltonian in the four-dimensional phase space. The possibility of the existence of the large-scale chaotic vortex and magnetic fields in stationary regime is  proved by numerical methods. In the absence of external magnetic field $(\overline{\vec{B}}=0)$  stationary chaotic structures arise in a rotating stratified electroconductive fluid at the increase of the initial velocity of perturbations $\widetilde{W}_{1,2}(0)$.   In the presence of external magnetic field  $(\overline{\vec B}\neq 0)$ these structures are formed at the rise of the initial values of the perturbed field  $H_{1,2}(0)$.

\section{Appendix I. Multi-scale asymptotic expansions}

Let us consider the algebraic structure of the asymptotic expansion of Eqs.(\ref{eq10})-(\ref{eq13}) in different orders in  $R$ starting with the lowest  of them.  In the order  $R^{-3} $ we have only one  equation:
\begin{equation}\label{eqI1}
    \partial _{i} P_{-3} =0 \quad \Rightarrow \quad P_{-3}=P_{-3} \left( X \right)
\end{equation}
In the order  $R^{-2} $ there is the following equation:
\begin{equation}\label{eqI2}
    \partial _{i} P_{-2} =0 \quad \Rightarrow \quad P_{-2} =P_{-2} \left(X \right)
\end{equation}
Eqs. (\ref{eqI1}) and (\ref{eqI2}) are satisfied automatically, since  $P_{-3} $ and $P_{-2} $ are the functions of  slow variables only.
In the order  $R^{-1} $ we obtain the system of equations :
\[\partial _{t} W_{-1}^{i} +W_{-1}^{k} \partial _{k} W_{-1}^{i} =-\partial _{i} P_{-1}
-\nabla _{i} P_{-3} +\partial _{k}^{2} W_{-1}^{i} +\varepsilon _{ijk} W_{j} D_{k}+$$
$$+\widetilde{Q}\varepsilon _{ijk} \varepsilon _{jml} \partial _{m} B_{-1}^{l} B_{-1}^{k}+\widetilde{Q}\varepsilon _{ijk} \varepsilon _{jml} \partial _{m} B_{-1}^{l}\overline{B_k}+e_{i}\widetilde{Ra}T_{-1} \]

\begin{equation}\label{eqI3}
 \partial_{t} B_{-1}^{i} -Pm^{-1} \partial _{k}^{2} B_{-1}^{i} =\varepsilon _{ijk} \varepsilon
_{knp} \partial _{j} W_{-1}^{n} B_{-1}^{p}+ \varepsilon _{ijk} \varepsilon_{knp} \partial _{j} W_{-1}^{n}\overline{B_p} \end{equation}

\[ \partial_{t} T_{-1} -Pr^{-1} \partial _{k}^{2} T_{-1} =-W_{-1}^{k} \partial _{k} T_{-1} + W_{-1}^{z} \]

\[ \partial_{i} W_{-1}^{i} =0, \qquad \partial _{i} B_{-1}^{i}=0 \]
The averaging of Eqs. (\ref{eqI3})  over the \guillemotleft fast\guillemotright variables gives the secular equation :
\begin{equation}\label{eqI4}
-\nabla_{i} P_{-3} +\varepsilon_{ijk} W_{j} D_{k}+e_{i}\widetilde{Ra}T_{-1} =0,\quad  W_{-1}^{z}=0,  \end{equation}
which  corresponds to geostrophic equilibrium.

In the zero order in  $R$ we have the following system of equations:
\[\partial _{t} v_{0}^{i} +W_{-1}^{k} \partial _{k} v_{0}^{i} +v_{0}^{k} \partial
_{k} W_{-1}^{i} =-\partial _{i} P_{0} -\nabla _{i} P_{-2} +\partial _{k}^{2} v_{0}^{i}
+\varepsilon _{ijk} v_{0}^{j} D_{k} + \]

\[+ \widetilde{Q}\varepsilon _{ijk} \varepsilon
_{jml} \left(\partial _{m} B_{-1}^{l} B_{0}^{k} +\partial _{m} B_{0}^{l} B_{-1}^{k}
\right)+\widetilde{Q}\varepsilon _{ijk} \varepsilon_{jml}\partial _{m} B_{0}^{l}\overline{B_k}+e_{i}\widetilde{Ra}T_{0}+F_{0}^{i}\]
\begin{equation}\label{eqI5}
\partial _{t} B_{0}^{i} -Pm^{-1} \partial _{k}^{2} B_{0}^{i} =\varepsilon _{ijk}\varepsilon _{knp}\left(\partial _{j} W_{-1}^{n} B_{0}^{p} +\partial _{j} v_{0}^{n}B_{-1}^{p} \right)+ \varepsilon _{ijk}\varepsilon _{knp}\partial _{j} v_{0}^{n}\overline{B_p} \end{equation}

\[\partial _{t} T_{0} -Pr^{-1} \partial _{k}^{2} T_{0} =-W_{-1}^{k} \partial _{k}T_{0}-\partial _{k}(v_{0}^{k} T_{-1}) +v_{0}^{z} \]

\[ \partial _{i} v_{0}^{i} =0, \quad \partial _{i} B_{0}^{i} =0 \]

These equations give only one secular term:
\[\nabla P_{-2} =0 \quad \Rightarrow \quad P_{-2} =const\]
Now consider the first-order approximation  $R^{1} $ :
\[\partial _{t} v_{1}^{i} +W_{-1}^{k} \partial _{k} v_{1}^{i} + v_{0}^{k}
\partial _{k} v_{0}^{i} + v_{1}^{k} \partial_{k} W_{-1}^{i} +W_{-1}^{k} \nabla _{k}
W_{-1}^{i} =-\nabla _{i} P_{-1} -$$
$$- \partial _{i} \left(P_{1} +\overline{P}_{1} \right)+
\partial _{k}^{2} v_{1}^{i} + 2\partial _{k} \nabla _{k}
W_{-1}^{i} +{\varepsilon}_{ijk} {v}_{1}^{j} {D}_{k}+
\]
\begin{equation}\label{eqI6}
    +\widetilde{Q}\varepsilon _{ijk} \varepsilon _{jml} \left(\partial _{m} B_{-1}^{l} B_{1}^{k}
+\partial _{m} B_{0}^{l} B_{0}^{k} +\partial _{m} B_{1}^{l} B_{-1}^{k} +\nabla
_{m} B_{-1}^{l} B_{-1}^{k} \right)+$$
$$+\widetilde{Q}\varepsilon_{ijk} \varepsilon_{jml} \left(\partial_{m} B_{1}^{l}\overline{B_k}+\nabla_{m} B_{-1}^{l}\overline{B_k} \right)+e_{i}\widetilde{Ra}T_{1}
\end{equation}
\[\partial _{t} B_{1}^{i} -Pm^{-1} \partial _{k}^{2} B_{1}^{i} -Pm^{-1} 2\partial
_{k} \nabla _{k} B_{-1}^{i} =\varepsilon _{ijk} \varepsilon _{knp} \left(\partial _{j}
W_{-1}^{n} B_{1}^{p} +\partial _{j} v_{0}^{n} B_{0}^{p} +\partial _{j} v_{1}^{n}
B_{-1}^{p} +\right.$$
$$\left. + \nabla_{j} {W}_{-1}^{n} {B}_{-1}^{p} \right)+\varepsilon_{ijk} \varepsilon_{knp}\left( \partial _{j} v_{1}^{n}
\overline{B_p}+\nabla_{j} {W}_{-1}^{n}\overline{B_p} \right)   \]
\[\partial_{t} T_{1} -Pr^{-1} \partial _{k}^{2} T_{1} -Pr^{-1}2\partial _{k} \nabla _{k} T_{-1}
=-W_{-1}^{k} \partial _{k} T_{1} -W_{-1}^{k} \nabla _{k} T_{-1}
-v_{0}^{k} \partial _{k} T_{0} -v_{1}^{k} \partial _{k} T_{-1} +v_{1}^{z} \]
\[\partial_{i} v_{1}^{i} +\nabla _{i} W_{-1}^{i} =0, \quad \partial_{i} B_{1}^{i} +\nabla _{i} B_{-1}^{i} =0 \]
This system yields the following secular equations:
\begin{equation}\label{eqI7}
    W_{-1}^{k} \nabla _{k} W_{-1}^{i} =-\nabla _{i} P_{-1} +\widetilde{Q}\varepsilon_{ijk} \varepsilon _{jml}\left( \nabla _{m} B_{-1}^{l} B_{-1}^{k}+\nabla _{m} B_{-1}^{l} \overline{B_k} \right)
\end{equation}
\begin{equation}\label{eqI8}
    \varepsilon_{ijk} \varepsilon_{knp} \left(\nabla _{j} W_{-1}^{n} B_{-1}^{p}+\nabla _{j} W_{-1}^{n}\overline{B_p} \right)=0
\end{equation}
\begin{equation}\label{eqI9}
 W_{-1}^{k} \nabla _{k} T_{-1} =0,\quad \nabla_{i} W_{-1}^{i} =0, \quad \nabla _{i} B_{-1}^{i} =0
\end{equation}
For the second order  $R^{2} $ we obtain the equations:
\[\partial _{t} v_{2}^{i} +W_{-1}^{k} \partial _{k} v_{2}^{i} +v_{0}^{k}
\partial _{k} v_{1}^{i} +W_{-1}^{k} \nabla _{k} v_{0}^{i} +v_{0}^{k} \nabla _{k}
W_{-1}^{i} +v_{1}^{k} \partial _{k} v_{0}^{i} +v_{2}^{k} \partial _{k} W_{-1}^{i}
=\]
\[= -\nabla _{i} P_{2} -\nabla _{i} P_{0} +\partial _{k}^{2} v_{2}^{i} +2\partial _{k}
\nabla _{k} v_{0}^{i} +{\varepsilon }_{ijk} {v}_{2}^{j}
{D}_{k} +\]
\[+\widetilde{Q}\varepsilon_{ijk} \varepsilon_{jml} \left(\partial
_{m} B_{-1}^{l} B_{2}^{k} +\partial_{m} B_{0}^{l} B_{1}^{k} +\partial_{m} B_{1}^{l}
B_{0}^{k} +\partial _{m} B_{2}^{l} B_{-1}^{k} +\nabla_{m} B_{-1}^{l} B_{0}^{k} +
\nabla _{m} B_{0}^{l} B_{-1}^{k} \right)+$$
$$+ \widetilde{Q}\varepsilon_{ijk} \varepsilon_{jml} \left(\partial_{m} B_{2}^{l} \overline{B_k}+\nabla_{m} B_{0}^{l} \overline{B_k} \right) +e_{i}\widetilde{Ra}T_{2}\]
\begin{equation}\label{eqI10}
\partial_{t} B_{2}^{i} -Pm^{-1} \partial_{k}^{2} B_{2}^{i} -Pm^{-1}
2\partial_{k} \nabla_{k} B_{0}^{i} = \varepsilon_{ijk} \varepsilon _{knp}
\left(\partial_{j} W_{-1}^{n} B_{2}^{p} +\partial_{j} v_{0}^{n} B_{1}^{p} +\partial_{j}v_{1}^{n} B_{0}^{p}+\right.$$
$$\left.+\partial_{j}v_{2}^{n}B_{-1}^{p}+\nabla_{j} W_{-1}^{n} B_{0}^{p} +\nabla_{j} v_{0}^{n}B_{-1}^{p} \right)+\varepsilon_{ijk} \varepsilon_{knp} \left( \partial_{j}v_{2}^{n}\overline{B_p}+\nabla_{j}v_{0}^{n} \overline{B_p} \right)  \end{equation}
\[\partial_{t} T_{2} -Pr^{-1} \partial_{k}^{2} T_{2} -Pr^{-1} 2\partial _{k} \nabla_{k} T_{0} =-W_{-1}^{k} \partial_{k} T_{2}-W_{-1}^{k} \nabla _{k} T_{0}-v_{0}^{k}\partial_{k} T_{1} -v_{0}^{k} \nabla _{k} T_{-1}-$$
 $$-v_{1}^{k} \partial_{k}T_{0}-v_{2}^{k} \partial _{k} T_{-1} +v_{2}^{z} \]
 \[ \partial_{i}  v_{2}^{i} +\nabla_{i} v_{0}^{i} =0, \quad \partial_{i} B_{2}^{i} +\nabla_{i}  B_{0}^{i} =0 \]
As seen after the averaging of the system of Eqs. (\ref{eqI10})  over the \guillemotleft fast\guillemotright variables, in the order  $R^{2} $ secular terms are absent.

Finally, let us consider the most significant order $R^{3} $. Here the equations have the following form:
\[\partial_{t} v_{3}^{i} +\partial_{T} W_{-1}^{i} +W_{-1}^{k}
\partial _{k} v_{3}^{i} +v_{0}^{k} \partial _{k} v_{2}^{i} +W_{-1}^{k} \nabla _{k}
v_{1}^{i} +v_{0}^{k} \nabla_{k} v_{0}^{i} +v_{1}^{k} \partial_{k} v_{1}^{i} +v_{1}^{k}
\nabla _{k} W_{-1}^{i} + v_{2}^{k} \partial _{k} W_{-1}^{i} =\]
\[=-\partial_{i} P_{3}
- \nabla _{i} \left(P_{1} +\overline{P}_{1} \right)+\partial_{k}^{2} v_{3}^{i}
+2\partial _{k} \nabla _{k} v_{1}^{i} +\Delta W_{-1}^{i}+{\varepsilon}_{ijk} {v}_{3}^{j} {D}_{k} +\]
\[+\widetilde{Q}\varepsilon_{ijk}\varepsilon_{jml} \left(\partial _{m} B_{-1}^{l} B_{3}^{k} +\partial _{m} B_{0}^{l}
B_{2}^{k} +\partial_{m} B_{1}^{l} B_{1}^{k} +\partial _{m} B_{2}^{l} B_{0}^{k} +\partial _{m} B_{3}^{l} B_{-1}^{k}+\right.$$
$$\left. +\nabla_{m} B_{-1}^{l} B_{1}^{k}+\nabla_{m} B_{0}^{l} B_{0}^{k}+\nabla_{m} B_{1}^{l} B_{-1}^{k} \right)+ \widetilde{Q}\varepsilon_{ijk}\varepsilon_{jml} \left(\nabla_{m} B_{3}^{l} \overline{B_k}+\nabla_{m} B_{1}^{l}\overline{B_k}\right)+e_{i}\widetilde{Ra}T_{3}\]
\begin{equation}\label{eqI11}
\partial_{t} B_{3}^{i} +\partial_{T} B_{-1}^{i} -Pm^{-1} \partial _{k}^{2} B_{3}^{i}-2Pm^{-1}\partial _{k} \nabla_{k} B_{1}^{i} -Pm^{-1} \Delta B_{-1}^{i} =$$
$$=\varepsilon_{ijk} \varepsilon_{knp} \left(\partial_{j} W_{-1}^{n} B_{3}^{p}+\partial_{j}v_{0}^{n} B_{2}^{p} +\partial_{j} v_{1}^{n} B_{1}^{p}+\partial_{j}v_{2}^{n} B_{0}^{p}+\partial_{j} v_{3}^{n} B_{-1}^{p} +\nabla _{j} W_{-1}^{n} B_{1}^{p} +\right.$$
 $$\left.+\nabla_{j}v_{0}^{n} B_{0}^{p}+\nabla_{j}v_{1}^{n} B_{-1}^{p} \right)+\varepsilon_{ijk}\varepsilon_{knp} \left(\partial_{j} v_{3}^{n} \overline{B_p}+\nabla_{j}v_{1}^{n} \overline{B_p}\right) \end{equation}
\[ \partial_{t} T_{3} +\partial_{T} T_{-1} -Pr^{-1} \partial_{k}^{2}T_{3}-Pr^{-1} 2\partial_{k} \nabla_{k} T_{1}-Pr^{-1} \Delta T_{-1} =-W_{-1}^{k}\partial_{k} T_{3} -W_{-1}^{k} \nabla_{k} T_{1}-$$
$$-v_{0}^{k} \partial_{k} T_{2} -v_{0}^{k} \nabla_{k} T_{0} -v_{1}^{k}\nabla_{k}T_{1}-v_{1}^{k} \nabla_{k} T_{-1} -v_{2}^{k} \partial _{k} T_{0}-v_{3}^{k} \partial _{k} T_{-1} +v_{3}^{z} \]
\[\partial_{i} v_{3}^{i} +\nabla _{i} v_{1}^{i} =0, \quad \partial_{i} B_{3}^{i} +\nabla_{i} B_{1}^{i} =0 \]
By averaging this system of equations over the \guillemotleft fast\guillemotright variables we obtain the basic secular equations which describe the evolution of the large-scale perturbations:
\begin{equation}\label{eqI12}
    \partial _{T} W_{-1}^{i} -\Delta W_{-1}^{i} +\nabla _{k} \left(\overline{v_{0}^{k}
v_{0}^{i} }\right)=-\nabla _{i} \overline{P}_{1} +\widetilde{Q}\varepsilon _{ijk}
\varepsilon _{jml} \nabla _{m} \left(\overline{B_{0}^{l} B_{0}^{k} }\right)
\end{equation}
\begin{equation}\label{eqI13}
    \partial_{T} B_{-1}^{i} -Pm^{-1} \Delta B_{-1}^{i} =\varepsilon _{ijk} \varepsilon _{knp}
\nabla _{j} (\overline{v_{0}^{n} B_{0}^{p} })
\end{equation}
\begin{equation}\label{eqI14}\partial_{T} T_{-1} -Pr^{-1} \Delta T_{-1} =-\nabla _{k} \left(\overline{v_{0}^{k} T_{0}
}\right) \end{equation}

\section{Appendix II. Small-scale fields }

In Appendix  I we have obtained the equations of asymptotic expansion in the zero-order approximation. Taking into account the new denotations $\vec{W}=\vec{W}_{-1},\;  \vec{H}=\vec{B}_{-1} $ they can be written in the following form:
\begin{equation} \label{eqII1}
 \widehat{D}_{W} v_{0}^{i} =-\partial_{i} P_{0}
+\varepsilon_{ijk} v_{0}^{j} D_{k} +\widetilde {Q}H_{k} \left(\partial_{k} B_{0}^{i}-\partial_{i} B_{0}^{k}\right)+\widetilde{Q} \overline{B}_k \left(\partial_{k} B_{0}^{i}-\partial_{i} B_{0}^{k} \right)+e_{i}\widetilde{Ra}T_{0}+F_{0}^{i}
\end{equation}
\begin{equation} \label{eqII2}
\widehat{D}_{H} B_{0}^{i} =\left(H_{p}\partial_{p}+\overline{B}_k\partial_k \right)v_{0}^{i}
\end{equation}
\begin{equation}
\label{eqII3} \widehat{D}_{\theta } T_{0} =e_{k} v_{0}^{k}  \end{equation}
\begin{equation} \label{eqII4}
 \partial_{i} v_{0}^{i} =\partial_{k} B_{0}^{k}= \partial_{i} F_{0}^{i}=0
\end{equation}
where the operators are denoted as :
\[\widehat{D}_{W} =\partial_{t} -\partial_{k}^{2} +W_{k} \partial_{k} ,\;\widehat{D}_{_{H}} =\partial_{t} -Pm^{-1} \partial_{k}^{2} +W_{k} \partial_{k},\; \widehat{D}_{\theta } =\partial_{t}-Pr^{-1}\partial ^{2}+W_{k} \partial_{k}. \]
The small-scale oscillations of the magnetic field and the temperature are easily found from Eqs. (\ref{eqII2})-(\ref{eqII3}):
\begin{equation}\label{eqII5}
 B_{0}^{i} =\frac{\left(H_{p}+\overline{B}_p\right)\partial _{p}v_{0}^{i}}{\widehat{D}_{H}}, \quad T_{0}=\frac{v_{0}^{z} }{\widehat{D}_{\theta} }
\end{equation}
Now we substitute  (\ref{eqII5}) into (\ref{eqII1}) and find the pressure $P_{0} $ using the condition of field solenoidality   (\ref{eqII4}) :
\begin{equation}\label{eqII6}
P_0=\varepsilon_{ijk}\frac{\partial_i v_0^{j}}{\partial^2}D_k+e_i e_k\widetilde{Ra}\frac{\partial_i v_0^{k}}{\partial^2 \widehat{D}_{\theta}}-\frac{\widetilde{Q}(H_p+\overline{B}_p)}{\partial^2\widehat{D}_H}\partial_p(H_k \partial^2 v_0^{k})
\end{equation}
Using the above-derived  formula (\ref{eqII6}) we exclude the pressure from Eq. (\ref{eqII1}) and obtain the equation for the velocity field of the zeroth-order approximation:
\begin{equation}\label{eqII7} \left[\left(\widehat{D}_W-\frac{\widetilde{Q}((H_k\partial_k)^2+H_k\overline{B}_l\partial_k\partial_l)}{\widehat{D}_H}\right)\delta_{ij}-\left(\widetilde{Ra}\frac{e_je_p}{\widehat{D}_{\theta}}+\varepsilon_{ijk}D_k\right)\widehat{P}_{ip}\right]v_0^{j}=F_0^{i},
\end{equation}
where $\widehat{P}_{ip} =\delta_{ip}-\frac{\partial_{i} \partial_{p} }{\partial^{2} } $ is the projection operator. For finding the small-scale field  $\vec{v}_{0} $  it is convenient to present Eq.(\ref{eqII7}) in the coordinate form:
\begin{equation}\label{eqII8}
\left\{\begin{array}{c} {\widehat{d}_{11} v_{0}^{x}+\widehat{d}_{12} v_{0}^{y} +\widehat{d}_{13} v_{0}^{z} =\widehat{F}_{0}^{x} } \\
{\widehat{d}_{21} v_{0}^{x} +\widehat{d}_{22} v_{0}^{y} +\widehat{d}_{23} v_{0}^{z}=\widehat{F}_{0}^{y} } \\
{\widehat{d}_{31} v_{0}^{x} +\widehat{d}_{32} v_{0}^{y}+\widehat{d}_{33} v_{0}^{z} =\widehat{F}_{0}^{z} } \end{array}\right.  \end{equation}
The components of the tensor  $\widehat{d}_{ij} $ have the following form:
\[\widehat{d}_{11}=\widehat{D}_W-\frac{\widetilde{Q}((H_k\partial_k)^2+H_k\overline{B}_l\partial_k\partial_l)}{\widehat{D}_H}+\frac{D_{2} \partial_{x}
\partial_{z} -D_{3} \partial_{x} \partial_{y} }{\partial ^{2} },\; \widehat{d}_{12}=\frac{D_{3} \partial^{2}_{x}-D_{1} \partial_{x} \partial_{z} }{\partial ^{2}}-D_{3}, \]
\[\widehat{d}_{13}=D_2+\frac{D_{1} \partial_{x}\partial _{y}-D_{2} \partial^{2}_{x} }{\partial^{2}}+\widehat{Ra}\frac{\partial_x\partial_z}{\partial^2\widehat{D}_{\theta}},\; \widehat{d}_{21}= D_3+\frac{D_{2} \partial_{y} \partial_{z}-D_{3} \partial ^{2}_{y} }{\partial^{2} },  \]
\[\widehat{d}_{22}=\widehat{D}_W-\frac{\widetilde{Q}((H_k\partial_k)^2+H_k\overline{B}_l\partial_k\partial_l)}{\widehat{D}_H}+\frac{D_{3}\partial
_{y} \partial_{x}-D_{1} \partial_{y} \partial_{z} }{\partial^{2}}, \]
\[\widehat{d}_{23}=\widehat{Ra}\frac{\partial_y\partial_z}{\partial^2\widehat{D}_{\theta}}+\frac{D_{1} \partial^{2}_{y}-D_{2} \partial_{y} \partial_{x} }{\partial^{2}}-D_1,\; \widehat{d}_{31}= \frac{D_{2}\partial^{2}_{z}-D_{3} \partial_{z} \partial_{y} }{\partial^{2}}-D_2, \]
\[\widehat{d}_{32}= \frac{D_{3} \partial_{z} \partial_{x}-D_{1} \partial^{2}_{z}}{\partial^{2} }+D_1,\; \widehat{d}_{33}=\widehat{D}_W-\frac{\widetilde{Q}((H_k\partial_k)^2+H_k\overline{B}_l\partial_k\partial_l)}{\widehat{D}_H}+$$
$$+\frac{D_{1} \partial_{z} \partial_{y}-D_{2} \partial_{z} \partial_{x} }{\partial^{2}}-\frac{\widetilde{Ra}}{\widehat{D}_{\theta}}+ \widetilde{Ra}\frac{\partial_z^2}{\partial^2\widehat{D}_{\theta}}.  \]
As is known, the solution for the system  (\ref{eqII8}) can  be found by means of the Cramer rule:
\begin{equation} \label{eqII9} v_{0}^{x} =u_{0} =\frac{1}{\Delta } \left\{
\left(\widehat{d}_{22} \widehat{d}_{33} -\widehat{d}_{32} \widehat{d}_{23} \right)F_{0}^{x}
+\left(\widehat{d}_{13} \widehat{d}_{32} -\widehat{d}_{12} \widehat{d}_{33} \right)F_{0}^{y}
+\left(\widehat{d}_{12} \widehat{d}_{23} -\widehat{d}_{13} \widehat{d}_{22} \right)F_{0}^{z}
\right\} \end{equation}
\begin{equation} \label{eqII10} v_{0}^{y} =v_{0} =\frac{1}{\Delta } \left\{
\left(\widehat{d}_{23} \widehat{d}_{31} -\widehat{d}_{21} \widehat{d}_{33} \right)F_{0}^{x}
+\left(\widehat{d}_{11} \widehat{d}_{33} -\widehat{d}_{13} \widehat{d}_{31} \right)F_{0}^{y}
+\left(\widehat{d}_{13} \widehat{d}_{21} -\widehat{d}_{11} \widehat{d}_{23} \right)F_{0}^{z}
\right\} \end{equation}
\begin{equation} \label{eqII11} v_{0}^{z} =w_{0} =\frac{1}{\Delta } \left\{
\left(\widehat{d}_{21} \widehat{d}_{32} -\widehat{d}_{22} \widehat{d}_{31} \right)F_{0}^{x}
+\left(\widehat{d}_{12} \widehat{d}_{31} -\widehat{d}_{11} \widehat{d}_{32} \right)F_{0}^{y}
+\left(\widehat{d}_{11} \widehat{d}_{22} -\widehat{d}_{12} \widehat{d}_{21} \right)F_{0}^{z}
\right\} \end{equation}
Here  $\Delta $ is the determinant of the system of equations (\ref{eqII8}) which in the open form is the following:
\begin{equation} \label{eqII12} \Delta =\widehat{d}_{11} \widehat{d}_{22} \widehat{d}_{33}
+\widehat{d}_{21} \widehat{d}_{32} \widehat{d}_{13} +\widehat{d}_{12} \widehat{d}_{23}
\widehat{d}_{31} -\widehat{d}_{13} \widehat{d}_{22} \widehat{d}_{31} -\widehat{d}_{32}
\widehat{d}_{23} \widehat{d}_{11} -\widehat{d}_{21} \widehat{d}_{12} \widehat{d}_{33}  \end{equation}
Now  let us  present the external force  $\vec{F}_{0} $ in the complex form:
\begin{equation} \label{eqII13} \vec{F}_{0} =\vec{i}\frac{f_{0} }{2} \; e^{i\phi _{2} } +\vec{j}\frac{f_{0} }{2} e^{i\phi _{1} } +c.c. \end{equation}
Then all the operators contained in  (\ref{eqII9})-(\ref{eqII12}) act on the eigenfunctions from the left:
\[\widehat{D}_{W,H,\theta} e^{i\phi _{1} } =e^{i\phi_{1} } \widehat{D}_{W,H,\theta} \left(\vec{\kappa }_{1} , -\omega _{0} \right), \quad \widehat{D}_{W,H,\theta}e^{i\phi _{2} } =e^{i\phi _{2} } \widehat{D}_{W,H,\theta} \left(\vec{\kappa }_{2} ,- \omega _{0} \right), \]
\begin{equation} \label{eqII14}
 \Delta e^{i\phi _{1}} =e^{i\phi _{1} } \Delta \left(\vec{\kappa}_{1} ,\; -\omega _{0} \right),\quad \Delta e^{i\phi _{2} } =e^{i\phi _{2} } \Delta \left(\vec{\kappa }_{2} ,\; -\omega _{0} \right)
\end{equation}
To simplify the formulae, assume that $\kappa_{0} =1$, $\omega_{0} =1$ and introduce the new denotations :
\[ \widehat{D}_{W} \left(\vec{\kappa}_{1}, -\omega _{0} \right)=\widehat{D}_{W_{1} }^{*} =1-i\left(1-W_{1} \right),\quad \widehat{D}_{W} \left(\vec{\kappa}_{2} ,-\omega_{0} \right)=\widehat{D}_{W_{2} }^{*} =1-i\left(1-W_{2} \right) \]
\begin{equation}\label{eqII15}
\widehat{D}_{H}\left(\vec{\kappa}_{1},-\omega_{0} \right)=\widehat{D}_{H_{1} }^{*} =Pm^{-1}
-i\left(1-W_{1} \right),\quad \widehat{D}_{H} \left(\vec{\kappa }_{2} , -\omega_{0}
\right)=\widehat{D}_{H_{2} }^{*} =Pm^{-1} -i\left(1-W_{2} \right)  \end{equation}
\[\widehat{D}_{\theta}\left(\vec{\kappa }_{1},-\omega_{0} \right)=\widehat{D}_{\theta_{1} }^{*} =Pr^{-1}
-i\left(1-W_{1} \right),\quad \widehat{D}_{\theta} \left(\vec{\kappa}_{2} ,-\omega_{0}
\right)=\widehat{D}_{\theta_{2}}^{*}=Pr^{-1} -i\left(1-W_{2} \right)\]
Here and further the complex-conjugate terms are marked by asterisk. In subsequent  calculations some components in the tensors  $\widehat{d}_{ij} \left(\vec{\kappa }_{1} ,-\omega _{0} \right)$ and $\widehat{d}_{ij} \left(\vec{\kappa }_{2} ,-\omega _{0} \right)$ vanish, and there remain the non-zero components:
\[\widehat{d}_{11} \left(\vec{\kappa}_{1},-\omega_0 \right)=\widehat{D}_{W_1}^*+\frac{\widetilde{Q}H_1(H_1+\overline{B}_1)}{\widehat{D}_{H_1}^*},\;\widehat{d}_{12} \left(\vec{\kappa}_{1},-\omega_0 \right)=0,\; \widehat{d}_{13} \left(\vec{\kappa}_{1},-\omega_0 \right)=0,  \]
\begin{equation}\label{eqII16}
\widehat{d}_{21} \left(\vec{\kappa}_{1},-\omega_0 \right)=D_3, \;\widehat{d}_{22} \left(\vec{\kappa}_{1},-\omega_0 \right)=\widehat{d}_{11} \left(\vec{\kappa}_{1},-\omega_0 \right), \;\widehat{d}_{23} \left(\vec{\kappa}_{1},-\omega_0 \right)=-D_1, \end{equation}
\[\widehat{d}_{31} \left(\vec{\kappa}_{1},-\omega_0 \right)=-D_2,\;\widehat{d}_{32} \left(\vec{\kappa}_{1},-\omega_0 \right)=D_1,\;\widehat{d}_{33} \left(\vec{\kappa}_{1},-\omega_0 \right)=\widehat{d}_{22} \left(\vec{\kappa}_{1},-\omega_0 \right)-\frac{\widetilde{Ra}}{\widehat{D}_{\theta_1}^*} \]
\[\widehat{d}_{11} \left(\vec{\kappa}_{2},-\omega_0 \right)=\widehat{D}_{W_2}^*+\frac{\widetilde{Q}H_2(H_2+\overline{B}_2)}{\widehat{D}_{H_2}^*},\;\widehat{d}_{12} \left(\vec{\kappa}_{2},-\omega_0 \right)=-D_3,\; \widehat{d}_{13} \left(\vec{\kappa}_{2},-\omega_0 \right)=D_2,  \]
\begin{equation}\label{eqII17}
\widehat{d}_{21} \left(\vec{\kappa}_{2},-\omega_0 \right)=0, \;\widehat{d}_{22} \left(\vec{\kappa}_{2},-\omega_0 \right)=\widehat{d}_{11} \left(\vec{\kappa}_{2},-\omega_0 \right), \;\widehat{d}_{23} \left(\vec{\kappa}_{2},-\omega_0 \right)=0, \end{equation}
\[\widehat{d}_{31} \left(\vec{\kappa}_{2},-\omega_0 \right)=-D_2,\;\widehat{d}_{32} \left(\vec{\kappa}_{1},-\omega_0 \right)=D_1,\;\widehat{d}_{33} \left(\vec{\kappa}_{2},-\omega_0 \right)=\widehat{d}_{22} \left(\vec{\kappa}_{2},-\omega_0 \right)-\frac{\widetilde{Ra}}{\widehat{D}_{\theta_2}^*} \]
Taking into account the expressions (\ref{eqII16})-(\ref{eqII17})  we find the velocity fields in the zero-order  approximation:
\begin{equation}\label{eqII18}
 u_{0} =\frac{f_{0} }{2} \frac{\widehat{B}_{2}^* }{\widehat{A}_{2}^{*}\widehat{B}_{2}^{*}+D_{2}^{2} }e^{i\phi_{2} } +c.c.=u_{03} +u_{04}
\end{equation}
\begin{equation} \label{eqII19}
 v_{0} =\frac{f_{0} }{2} \frac{\widehat{B}_{1}}{\widehat{A}_{1}^{*}\widehat{B}_{1}^{*}+D_{1}^{2}} e^{i\phi_{1} } +c.c.=v_{01} +v_{02}
\end{equation}
\begin{equation} \label{eqII20}
 w_{0} =-\frac{f_{0}}{2} \frac{D_{1}}{\widehat{A}_{1}^{*}\widehat{B}_{1}^{*}+D_{1}^{2} }e^{i\phi_{1} } +\frac{f_{0}}{2} \frac{D_{2}}{\widehat{A}_{2}^{*}\widehat{B}_{2}^{*}+D_{2}^{2}} e^{i\phi_{2} }+c.c.= w_{01} +w_{02} +w_{03} +w_{04}
\end{equation}
where
\begin{equation} \label{eqII21}
\widehat{A}_{1,2}^* =\widehat{D}_{W_{1,2}}^{*} +\frac{\widetilde{Q} H_{1,2} }{\widehat{D}_{H_{1,2}}^{*}}(H_{1,2}+\overline{B}_{1,2}) , \; \widehat{B}_{1,2}^* = \widehat{A}_{1,2}^* -\frac{\widetilde{Ra}}{\widehat{D}_{\theta_{1,2}}^*}. \end{equation}
The velocity components satisfy the following relations: $w_{02} =\left(w_{01} \right)^{*} $, $w_{04} =\left(w_{03} \right)^{*} $, $v_{02} =\left(v_{01} \right)^{*} ,v_{04} =\left(v_{03} \right)^{*} $, $u_{02} =\left(u_{01} \right)^{*} $, $u_{04} =\left(u_{03} \right)^{*} $. In the limiting case of non-electroconductive fluid  $(\sigma =0)$, in the absence of temperature gradient  $(\nabla \overline{T}=0)$ and  external magnetic field  $(\overline{B}_{1,2} =0)$ the  formulae (\ref{eqII18})-(\ref{eqII20}) coincide with the results obtained in  \cite{41s}. Now we will calculate the small-scale oscillations of the magnetic field  $\vec{B}_{0} $ using  the expressions   (\ref{eqII5})  and  (\ref{eqII18})-(\ref{eqII20}):
\begin{equation} \label{eqII22}
 B_{0}^{x} =\widetilde{u}_{0} =\frac{f_{0}}{2}\frac{i(H_2+\overline{B}_2)\widehat{B}_{2}^* }{\widehat{D}_{H_{2}}^{*}(\widehat{A}_{2}^{*}\widehat{B}_{2}^{*}+D_{2}^{2})}e^{i\phi_{2}}+c.c.=\widetilde{u}_{03}+\widetilde{u}_{04}
\end{equation}
\begin{equation} \label{eqII23}
 B_{0}^{y} =\widetilde{v}_{0} =\frac{f_{0}}{2}\frac{i(H_1+\overline{B}_1)\widehat{B}_{1}^* }{\widehat{D}_{H_{1}}^{*}(\widehat{A}_{1}^{*}\widehat{B}_{1}^{*}+D_{1}^{2})}e^{i\phi_{1}}+c.c.=\widetilde{u}_{03}+\widetilde{u}_{04}
\end{equation}
\begin{equation} \label{eqII24}
 B_{0}^{z} =\widetilde{w}_{0} =-\frac{f_{0}}{2} \frac{i(H_1+\overline{B}_1)D_{1}}{ \widehat{D}_{H_{1}}^{*}(\widehat{A}_{1}^{*}\widehat{B}_{1}^{*}+D_{1}^{2}) }e^{i\phi_{1}}+\frac{f_{0}}{2} \frac{i(H_2+\overline{B}_2)D_{2}}{\widehat{D}_{H_{2}}^{*} (\widehat{A}_{2}^{*}\widehat{B}_{2}^{*}+D_{2}^{2})} e^{i\phi_{2}}+c.c.= $$
 $$= w_{01} +w_{02} +w_{03} +w_{04}
\end{equation}
In the expressions for the small-scale oscillations  $(\vec{v}_{0} ,\vec{B}_{0} ,T_{0} )$  the component of the angular velocity  $D_{3} $ is absent due to the choice of the external force.  Further Eqs. (\ref{eqII18})-(\ref{eqII24})  will be used while  calculating the correlation functions.

\section{Appendix  III.  Calculation of the Reynolds stresses, Maxwell stresses and turbulent e.m.f. }

To close  the system of Eqs. (\ref{eq17})-(\ref{eq20}) which describe the evolution of the large-scale fields, it is necessary to calculate the correlators of the types
\begin{equation}\label{eqIII1}
 T^{31} =\overline{w_{0} u_{0} }=\overline{w_{03} \left(u_{03}\right)^{*} }+\overline{\left(w_{03} \right)^{*} u_{03}}
\end{equation}

\begin{equation} \label{eqIII2}
 T^{32} =\overline{w_{0} v_{0} }=\overline{w_{01}\left(v_{01} \right)^{*} }+\overline{\left(w_{01} \right)^{*} v_{01}}
\end{equation}

\begin{equation}\label{eqIII3}
 S^{31} =\overline{\widetilde{w}_{0}\widetilde{u}_{0}}=\overline{\widetilde{w}_{03} \left(\widetilde{u}_{03}\right)^{*}}+\overline{\left(\widetilde{w}_{03} \right)^{*} \widetilde{u}_{03}}
\end{equation}

\begin{equation}\label{eqIII4}
 S^{32} =\overline{\widetilde{w}_{0} \widetilde{v}_{0}}=\overline{\widetilde{w}_{01} \left(\widetilde{v}_{01} \right)^{*}}+\overline{\left(\widetilde{w}_{01}\right)^{*} \widetilde{v}_{01}}
\end{equation}

\begin{equation}\label{eqIII5}
 G^{13} =\overline{u_{0} \widetilde{w}_{0} }=
\overline{u_{03} \left(\widetilde{w}_{03} \right)^{*}}+\overline{\left(u_{03}\right)^{*} \widetilde{w}_{03}}
\end{equation}

\begin{equation} \label{eqIII6}
 G^{31} =\overline{w_{0} \widetilde{u}_{0} }=
\overline{w_{03} \left(\widetilde{u}_{03} \right)^{*} }+\overline{\left(w_{03} \right)^{*} \widetilde{u}_{03} }
\end{equation}

\begin{equation} \label{eqIII7}
 G^{23} =\overline{v_{0} \widetilde{w}_{0} }=
\overline{v_{01} \left(\widetilde{w}_{01} \right)^{*} }+\overline{\left(v_{01} \right)^{*}
\widetilde{w}_{01}}
\end{equation}

\begin{equation} \label{eqIII8}
 G^{32} =\overline{w_{0} \widetilde{v}_{0} }=
\overline{w_{01} \left(\widetilde{v}_{01} \right)^{*} }+\overline{\left(w_{01} \right)^{*}
\widetilde{v}_{01} }
\end{equation}
At first let us calculate the Reynolds stresses (\ref{eqIII1})-(\ref{eqIII2}).   For this purpose we will use the expressions for the small-scale velocity fields  (\ref{eqII18})-(\ref{eqII20}). Their  substitution into  (\ref{eqIII1})-(\ref{eqIII2})  gives:
\begin{equation} \label{eqIII9}
 T^{31} =\frac{f_{0}^{2} }{2} \frac{D_{2}q_{2}}{\left|\widehat{A}_{2}\widehat{B}_{2} +D_{2}^{2} \right|^{2} },
\end{equation}
\begin{equation}\label{eqIII10}
    T^{32} =-\frac{f_{0}^{2} }{2} \frac{D_{1} q_{1} }{\left|\widehat{A}_{1}\widehat{B}_{1} +D_{1}^{2} \right|^{2}},
\end{equation}
where  \[q_{1,2} =1 + \frac{Q H_{1,2}(H_{1,2}+\overline{B}_{1,2})}{1+Pm^{2} \left(1-W_{1,2} \right)^{2}}-\frac{Ra}{1+Pr^{2} \left(1-W_{1,2} \right)^{2}}.  \]
 To calculate the correlators of the magnetic field or the Maxwell stresses  $S^{31} $ and $S^{32} $, we will use the expressions  (\ref{eqII21})-(\ref{eqII23}). By substituting   (\ref{eqII21})-(\ref{eqII23}) into (\ref{eqIII3}), (\ref{eqIII4})  obtain :
\begin{equation}\label{eqIII11}
    S^{31} =\frac{H_{2}^{2} }{\left|\widehat{D}_{H_{2} } \right|^{2}
} T^{31}, \quad S^{32} =\frac{H_{1}^{2} }{\left|\widehat{D}_{H_{1}
} \right|^{2} } T^{32}
\end{equation}
 The differences   $T^{31}-\widetilde{Q}S^{31} $ and $T^{32}-\widetilde{Q}S^{32} $ contained in the right sides of Eqs. (\ref{eq17})-(\ref{eq18})  can be easily found using the expressions (\ref{eqIII10})-(\ref{eqIII11}):
\begin{equation} \label{eqIII12}
 T^{31} -\widetilde{Q}S^{31} =T^{31} \left(1-\frac{Q(H_{2}+\overline{B}_2)^{2} }{Pm\left|\widehat{D}_{H_{2} } \right|^{2} } \right)=T^{31} Q_{2}
\end{equation}

\begin{equation}\label{eqIII13}
 T^{32} -\widetilde{Q}S^{32} =T^{32} \left(1-\frac{Q(H_{1}+\overline{B}_1)^{2}}{Pm\left|\widehat{D}_{H_{1} } \right|^{2} } \right)=T^{32} Q_{1}
\end{equation}
To calculate the group of oscillators (\ref{eqIII5})-(\ref{eqIII8}) we will use the expressions for the small-scale velocity field  (\ref{eqII18})-(\ref{eqII20}) and  the magnetic field (\ref{eqII21})-(\ref{eqII23}) . Simple mathematical operations yield:
\begin{equation} \label{eqIII14}
 G^{13} =\frac{f_{0}^{2} }{4} \frac{i(H_{2}+\overline{B}_2) D_{2}
}{\left|\widehat{A}_{2}\widehat{B}_{2} +D_{2}^{2} \right|^{2} }\cdot \left(\frac{ \widehat{B}_{2}}{\widehat{D}_{H_{2}
}^{*} } -\frac{\widehat{B}_{2}^* }{\widehat{D}_{H_{2}}} \right)
\end{equation}

\begin{equation} \label{eqIII15}
 G^{31} =\frac{f_{0}^{2} }{4} \frac{i(H_{2}+\overline{B}_2) D_{2}
}{\left|\widehat{A}_{2}\widehat{B}_{2} +D_{2}^{2} \right|^{2}} \cdot\left(\frac{\widehat{B}_{2}^* }{\widehat{D}_{H_{2}}^{*}}-\frac{ \widehat{B}_{2}}{\widehat{D}_{H_{2} } } \right)
\end{equation}

\begin{equation} \label{eqIII16}
 G^{23} =\frac{f_{0}^{2} }{4} \frac{i(H_{1}+\overline{B}_1) D_{1}}{\left|\widehat{A}_{1}\widehat{B}_{1} +D_{1}^{2} \right|^{2}} \cdot\left(\frac{ \widehat{ B}_{1}^*}{\widehat{D}_{H_{1}}} -\frac{ \widehat{B}_{1}}{\widehat{D}_{H_{1}}^{*}} \right)
\end{equation}

\begin{equation} \label{eqIII17}
 G^{32} =\frac{f_{0}^{2} }{4} \frac{i(H_{1}+\overline{B}_1)D_{1}}{ \left|\widehat{A}_{1}\widehat{B}_{1} +D_{1}^{2} \right|^{2}} \cdot\left(\frac{ \widehat{B}_{1} }{\widehat{D}_{H_{1}}}-\frac{ \widehat{B}_{1}^*}{\widehat{D}_{H_{1}}^{*}} \right)
\end{equation}
To close the equations for the large-scale magnetic field (\ref{eq24}), (\ref{eq25}) , it is necessary to calculate the differences  $G^{13} -G^{31} $ and  $G^{23} -G^{32} $ corresponding to the  turbulent e.m.f. components  ${\mathcal{E}}_{2}={\mathcal{E}}_{y}$ и ${\mathcal{E}}_{1}={\mathcal{E}}_{x}$. In view of the expressions (\ref{eqIII14})-(\ref{eqIII17}) we obtain:
\begin{equation} \label{eqIII18}
 {\mathcal{E}}_{2}=G^{13} -G^{31} =
\frac{f_{0}^{2} }{4} \frac{i(H_{2}+\overline{B}_2)D_{2}}{\left|\widehat{A}_{2}\widehat{B}_{2} +D_{2}^{2} \right|^{2}}\cdot \frac{(\widehat B_{2}-\widehat{B}_{2}^* )}{\left|\widehat{D}_{H_{2}}\right|^{2}}\cdot\left(\widehat{D}_{H_{2} }^{*} +\widehat{D}_{H_{2} } \right)
\end{equation}
\begin{equation} \label{eqIII19}
 {\mathcal{E}}_{1}  =G^{23} -G^{32} =-\frac{f_{0}^{2}}{4} \frac{i(H_{1}+\overline{B}_1) D_{1}}{\left|\widehat{A}_{1}\widehat{B}_{1} +D_{1}^{2} \right|^{2}}\cdot\frac{(\widehat{B}_{1}-\widehat{B}_{1}^* )}{\left|\widehat{D}_{H_{1} } \right|^{2} }\cdot\left(\widehat{D}_{H_{1} }^{*} +\widehat{D}_{H_{1} } \right)
\end{equation}
Using the expressions (\ref{eqII15}) and (\ref{eqII21}) let us write  several relations :
\[\left|\widehat{A}_{1,2}\widehat{B}_{1,2}+D_{1,2}^2\right|^2=\left|\widehat{A}_{1,2}\right|^2 \cdot\left|\widehat{B}_{1,2}\right|^2 +D_{1,2}^2\cdot(\widehat{A}_{1,2}^*\widehat{B}_{1,2}^*+\widehat{A}_{1,2}\widehat{B}_{1,2})+D_{1,2}^4, \]
\[ \left|\widehat{A}_{1,2}\right|^2=\left|\widehat{D}_{W_{1,2}}\right|^2+\widetilde{Q}H_{1,2}(H_{1,2}+\overline{B}_{1,2})\left(\frac{\widehat{D}_{W_{1,2}}}{\widehat{D}_{H_{1,2}}^*}+\frac{\widehat{D}_{W_{1,2}}^*}{\widehat{D}_{H_{1,2}}}\right)+\frac{\widetilde{Q}^2H_{1,2}^2(H_{1,2}+\overline{B}_{1,2})^2}{\left|\widehat{D}_{H_{1,2}}\right|^2}, \]
\[ \left|\widehat{B}_{1,2}\right|^2=\left|\widehat{A}_{{1,2}}\right|^2-\widetilde{Ra}\left(\frac{\widehat{A}_{{1,2}}}{\widehat{D}_{\theta_{1,2}}^*}+\frac{\widehat{A}_{{1,2}}^*}{\widehat{D}_{\theta_{1,2}}}\right)+\frac{\widetilde{Ra}^2}{\left|\widehat{D}_{\theta_{1,2}}\right|^2}, \]
\begin{equation} \label{eqIII20}
 \widehat{A}_{1,2}^*\widehat{B}_{1,2}^*+\widehat{A}_{1,2}\widehat{B}_{1,2}= \left(\widehat{A}_{1,2}^*\right)^2+\left(\widehat{A}_{1,2}\right)^2-\widetilde{Ra}\left(\frac{\widehat{A}_{{1,2}}^*}{\widehat{D}_{\theta_{1,2}}^*}+\frac{\widehat{A}_{{1,2}}}{\widehat{D}_{\theta_{1,2}}}\right), \end{equation}
\[\left|\widehat{D}_{W_{1,2}}\right|^2=1+(1-W_{1,2})^2, \; \left|\widehat{D}_{H_{1,2}}\right|^2=Pm^{-2}+(1-W_{1,2})^2,\; \left|\widehat{D}_{\theta_{1,2}}\right|^2=Pr^{-2}+(1-W_{1,2})^2, \]
\[ \widehat{D}_{H_{1,2}}+\widehat{D}_{H_{1,2}}^*=2Pm^{-1}, \; \widehat{D}_{W_{1,2}}\widehat{D}_{H_{1,2}}+\widehat{D}_{W_{1,2}}^*\widehat{D}_{H_{1,2}}^*=2(Pm^{-1}-(1-W_{1,2})^2), \]
\[\widehat{D}_{\theta_{1,2}}\widehat{D}_{H_{1,2}}^*+\widehat{D}_{H_{1,2}}\widehat{D}_{\theta_{1,2}}^*=2(Pm^{-1}Pr^{-1}+(1-W_{1,2})^2), \;\left(\widehat{D}_{W_{1,2}}\right)^2+\left(\widehat{D}_{W_{1,2}}^*\right)^2=2(1-(1-W_{1,2})^2), \]
\[ \widehat{D}_{W_{1,2}}^*\widehat{D}_{\theta_{1,2}}+\widehat{D}_{W_{1,2}}\widehat{D}_{\theta_{1,2}}^*=2(Pr^{-1}+(1-W_{1,2})^2),\;\widehat{D}_{W_{1,2}}^*\widehat{D}_{H_{1,2}}+\widehat{D}_{W_{1,2}}\widehat{D}_{H_{1,2}}^*=2(Pm^{-1}+(1-W_{1,2})^2). \]

These relations substituted  into (\ref{eqIII12})-(\ref{eqIII13}) make it possible to obtain  the difference of the Reynolds and Maxwell stresses:
\begin{equation} \label{eqIII21}
 T^{31} -\widetilde{Q}S^{31} =\frac{f_{0}^{2}}{2}\cdot \frac{D_{2} q_{2} Q_{2}}{4\left(1-W_{2} \right)^{2}q_2^2\widetilde{Q}_2^2  +\left[D_{2}^{2}+W_{2}\left(2-W_{2}\right)+\mu_{2}\right]^{2} +\xi_2}, $$
$$  T^{32} -\widetilde{Q}S^{32} =\frac{f_{0}^{2} }{2} \cdot\frac{D_{1} q_{1} Q_{1}}{4\left(1-W_{1}\right)^{2}q_1^2\widetilde{Q}_1^2  +\left[D_{1}^{2} +W_{1}\left(2-W_{1}\right)+\mu_{1} \right]^{2}+\xi_1 },
\end{equation}
with the following denotations:
\[q_{1,2} =1 + \frac{Q H_{1,2}(H_{1,2}+\overline{B}_{1,2})}{1+Pm^{2} \left(1-W_{1,2}\right)^{2}}-\frac{Ra}{1+Pr^2(1-W_{1,2})^2}, \quad Q_{1,2} =1 - \frac{QPm (H_{1,2}+\overline{B}_{1,2})^{2}}{1+Pm^{2} \left(1-W_{1,2} \right)^{2}},\]
\[\widetilde{Q}_{1,2}=1-\frac{QPm H_{1,2}(H_{1,2}+\overline{B}_{1,2})}{1+Pm^{2}\left(1-W_{1,2}\right)^{2}}+\frac{RaPr}{1+Pr^2(1-W_{1,2})^2}, \]
\[\mu _{1,2} = 2QH_{1,2}(H_{1,2}+\overline{B}_{1,2})\cdot\frac{1+Pm\left(1-W_{1,2}\right)^{2}}{1+Pm^2\left(1-W_{1,2} \right)^{2}}+Q^2H_{1,2}^2(H_{1,2}+\overline{B}_{1,2})^2 \cdot\frac{1-Pm^2\left(1-W_{1,2}\right)^{2}}{\left(1+Pm^2\left(1-W_{1,2}\right)^{2}\right)^2}-$$
$$-Ra\cdot\frac{1+Pr(1-W_{1,2})^2+2QH_{1,2}(H_{1,2}+\overline{B}_{1,2})\cdot \frac{1-PrPm\left(1-W_{1,2}\right)^{2}}{1+Pm^2\left(1-W_{1,2}\right)^{2}}}{1+Pr^2(1-W_{1,2})^2}, \]
\[\xi_{1,2}=2\Xi_{1,2}+2\left(1-W_{1,2}\right)^2\Pi_{1,2}-2\left(1-W_{1,2}\right)^2(1-\widetilde{Q}_{1,2}^2)\Pi_{1,2}-2(1-{q}_{1,2}^2)\Xi_{1,2}+ \Xi_{1,2}\Pi_{1,2}+$$
$$+\chi_{1,2}\left(1-W_{1,2}\right)^2+\chi_{1,2}(1+\sigma_{1,2}), \]
\[\Xi_{1,2}=-\frac{4\left(1-W_{1,2}\right)^2\widetilde{Q}_{1,2}RaPr}{1+Pr^{2} \left(1-W_{1,2}\right)^{2}}+\frac{2\left(1-W_{1,2}\right)^2Ra^2Pr^2}{\left(1+Pr^{2} \left(1-W_{1,2}\right)^{2}\right)^2}+ $$
 $$+Ra\cdot\frac{1+Pr(1-W_{1,2})^2+2QH_{1,2}(H_{1,2}+\overline{B}_{1,2})\cdot \frac{1-PrPm\left(1-W_{1,2}\right)^{2}}{1+Pm^2\left(1-W_{1,2}\right)^{2}}}{1+Pr^2(1-W_{1,2})^2},  \]
\[\Pi_{1,2}=\frac{4q_{1,2}Ra}{1+Pr^{2} \left(1-W_{1,2}\right)^{2}}+\frac{2Ra^2}{\left(1+Pr^{2} \left(1-W_{1,2}\right)^{2}\right)^2}-$$
$$-Ra\cdot\frac{1+Pr(1-W_{1,2})^2+2QH_{1,2}(H_{1,2}+\overline{B}_{1,2})\cdot \frac{1-PrPm\left(1-W_{1,2}\right)^{2}}{1+Pm^2\left(1-W_{1,2}\right)^{2}}}{1+Pr^2(1-W_{1,2})^2}, \]
\[\sigma_{1,2}=\frac{QH_{1,2}(H_{1,2}+\overline{B}_{1,2})}{1+Pm^2\left(1-W_{1,2}\right)^{2}}\cdot\left[2\left(1+Pm^2\left(1-W_{1,2}\right)^{2}\right)+QH_{1,2}(H_{1,2}+\overline{B}_{1,2})\right], \]
\[ \chi_{1,2}=\frac{2Ra}{1+Pr^2\left(1-W_{1,2}\right)^{2}}\cdot \left[\frac{Ra}{2}-\left(1-Pr\left(1-W_{1,2}\right)^{2}+\right. \right.$$
$$\left.\left.+\frac{QH_{1,2}(H_{1,2}+\overline{B}_{1,2})(1+PrPm\left(1-W_{1,2}\right)^{2})}{1+Pm^2\left(1-W_{1,2}\right)^{2}}\right)\right]. \]
By substituting the relations  (\ref{eqIII20})  into  (\ref{eqIII18})-(\ref{eqIII19}) we find the expressions for a turbulent e.m.f.  $\mathcal{E}_{1,2}$   in the explicit form:
\begin{equation} \label{eqIII22}
{\mathcal{E}}_{1}=f_{0}^{2}\cdot \frac{D_{1} \left(1-W_{1}\right)Pm \widetilde{Q}_{1}(H_1+\overline{B}_1) }{\left(1+Pm^{2}\left(1-W_{1} \right)^{2} \right)\left[4\left(1-W_{1}\right)^{2}q_1^2\widetilde{Q}_1^2 +\left[D_{1}^{2} +W_{1} \left(2-W_{1} \right)+\mu_{1} \right]^{2}+\xi_1 \right]},$$
$$ {\mathcal{E}}_{2}=f_{0}^{2}\cdot \frac{D_{2}\left(1-W_{2} \right)Pm \widetilde{Q}_{2}(H_2 +\overline{B}_2) }{\left(1+Pm^{2} \left(1-W_{2} \right)^{2} \right)\left[4\left(1-W_{2} \right)^{2}q_2^2\widetilde{Q}_2^2  +\left[D_{2}^{2} +W_{2} \left(2-W_{2}\right)+\mu _{2} \right]^{2}+\xi_2 \right]}.
\end{equation}

\end{document}